\documentclass[acmsmall]{acmart}

\usepackage{float}
\usepackage[caption = false]{subfig}
\usepackage{multirow}
\usepackage{makecell}
\usepackage{glossaries}
\usepackage{booktabs}
\usepackage{cases}
\usepackage{pgfplots}
\usepackage{amsmath}
\usepackage{graphicx}
\usepackage{tikzexternal}
\usepackage[absolute,showboxes]{textpos}

\usepgfplotslibrary{groupplots,colormaps,colorbrewer,statistics}

\pgfplotscreateplotcyclelist{exotic_bars}{thick,teal,fill=teal\\thick,orange,fill=orange\\thick,cyan!60!black,fill=cyan!60!black\\thick,red!70!white,fill=red!70!white\\thick,lime!80!black,fill=lime!80!black\\thick,red,fill=red\\}

\usepackage{pifont}

\usepackage{xspace}

\newcommand{\etal}{\textit{et al.\ }}
\newcommand{\eg}{\textit{e.g.,}~}
\newcommand{\ie}{\textit{i.e.,}~}
\newcommand{\etc}{\textit{etc.}~}
\newcommand{\one}{({\em i})\xspace}
\newcommand{\two}{({\em ii})\xspace}
\newcommand{\three}{({\em iii})\xspace}
\newcommand{\four}{({\em iv})\xspace}

\let\orgautoref\autoref
\renewcommand{\autoref}
{\def\sectionautorefname{Section}\def\subsectionautorefname{Section}\def\subsubsectionautorefname{Section}\orgautoref}

\makeatletter
\renewcommand{\paragraph}[1]{\vspace*{0.03in}\noindent{\bf #1.}\hspace{0.25ex \@plus1ex \@minus.2ex}}
\newcommand{\paragraphS}[1]{\vspace*{0.03in}\noindent{\bf #1}\hspace{0.25ex \@plus1ex \@minus.2ex}}
\makeatother

\AtBeginDocument{\providecommand\BibTeX{{\normalfont B\kern-0.5em{\scshape i\kern-0.25em b}\kern-0.8em\TeX}}}

\setcopyright{acmcopyright}
\copyrightyear{2022}
\acmYear{2022}
\acmDOI{xx.yyyy/zzzzzzz.aaaaaaa}

\acmJournal{TOSN}
\acmVolume{1}
\acmNumber{1}
\acmArticle{}
\acmMonth{1}

\makeglossaries

\newacronym{cap}{CAP}{Contention Access Period}
\newacronym{cfp}{CFP}{Contention Free Period}
\newacronym{bs}{BS}{Beacon Slot}
\newacronym{cad}{CAD}{Channel Activity Detection}
\newacronym{psdu}{PSDU}{Physical Service Data Unit}
\newacronym{csma}{CSMA/CA}{Carrier Sense Multiple Access/Collision Avoidance}
\newacronym{cca}{CCA}{Clear Channel Assessment}
\newacronym{pan}{PAN}{Personal Area Network}
\newacronym{gts}{GTS}{Guaranteed Time Slot}
\newacronym{so}{SO}{Superframe Order}
\newacronym{mo}{MO}{Multisuperframe Order}
\newacronym{bo}{BO}{Beacon Order}
\newacronym{mac}{MAC}{Media Access Control}
\newacronym{phy}{PHY}{Physical Layer}
\newacronym{dsme}{DSME}{Deterministic Synchronous Multichannel Extension}
\newacronym{tsch}{TSCH}{Time Slotted Channel Hopping}
\newacronym{6tisch}{6TiSCH}{IPv6 over the TSCH mode of IEEE 802.15.4e}
\newacronym{crc}{CRC}{Cyclic Redundancy Check}
\newacronym{rssi}{RSSI}{Received Signal Strength Indicator}
\newacronym{der}{DER}{Data Extraction Rate}
\newacronym{fhss}{FHSS}{Frequency-hopping spread spectrum}

\newacronym{lpwan}{LPWAN}{Low Power Wide Area Network}

\newacronym{prr}{PRR}{Packet Reception Ratio}

\tikzsetexternalprefix{figures/}

\begin{document}

\title{Towards long range peer to peer communication for the constrained IoT}
\title{DSME-LoRa: Seamless Long Range Communication Between Arbitrary Nodes in the Constrained IoT}

\author{Jos{\'e} {\'A}lamos}
\email{jose.alamos@haw-hamburg.de}

\author{Peter Kietzmann}
\email{peter.kietzmann@haw-hamburg.de}

\author{Thomas C. Schmidt}
\email{t.schmidt@haw-hamburg.de}

\affiliation{\department{Department Informatik}
  \institution{HAW Hamburg}
  \streetaddress{Berliner Tor 7}
  \city{Hamburg}
  \postcode{20099}
  \country{Germany}
}

\author{Matthias W{\"a}hlisch}
\email{m.waehlisch@fu-berlin.de}

\affiliation{\department{Institut f{\"u}r Informatik}
  \institution{Freie Universit{\"a}t Berlin}
  \streetaddress{Takustr. 9}
  \city{Berlin}
  \postcode{14195}
  \country{Germany}
}

\renewcommand{\shortauthors}{{\'A}lamos~\etal}

\begin{abstract}
Long range radio communication is preferred in many IoT deployments as it avoids the complexity of multi-hop wireless networks. LoRa is a popular, energy-efficient  wireless modulation but its networking substrate LoRaWAN introduces severe limitations to its users.  
    In this paper, we present and thoroughly analyze \acrshort{dsme}-LoRa, a system design of LoRa with  IEEE 802.15.4 \gls{dsme} as a \acrshort{mac} layer. \gls{dsme}-LoRa offers the advantage of seamless client-to-client communication beyond the pure gateway-centric transmission of LoRaWAN. We evaluate its feasibility via a full-stack implementation on the popular RIOT operating system, assess its steady-state packet flows in an analytical stochastic Markov model, and quantify its scalability in massive communication scenarios using large scale network simulations. Our findings indicate that \gls{dsme}-LoRa is indeed a powerful approach that opens LoRa to standard network layers and outperforms LoRaWAN in many dimensions. 
\end{abstract}



\maketitle

\setlength{\TPHorizModule}{\textwidth}
\setlength{\TPVertModule}{\paperheight}
\TPMargin{5pt}
\begin{textblock}{1}(.1,0.01)
\noindent
\footnotesize
If you cite this paper, please use the TOSN reference:
J. Alamos, P. Kietzmann, T. C. Schmidt, M. W{\"a}hlisch.\\
DSME-LoRa: Seamless Long Range Communication Between Arbitrary Nodes in the Constrained IoT.
\emph{ACM Trans. Sen. Netw.}, ACM, 2022. \url{https://doi.org/10.1145/3552432}
\end{textblock}

\section{Introduction}\label{sec:intro}

LoRa is a popular wireless modulation for the IoT that is robust against
interference and Doppler effect. It uses narrowband Chirp Spread Spectrum
modulation to achieve long range transmission (km) with low power consumption
(mW). LoRa operates in unlicensed spectra and therefore  is subject to
regional regulations that shall prevent a saturation of the spectrum.
In Europe, the ETSI EN300.220 standard~\cite{ETSI-en3002001-211} limits the transmission duty cycle to 0.1\%, 1\% or 10\% depending on the sub-band. 

LoRaWAN  was designed as an upper layer for LoRa that provides 
Media Access Control  and Internet communication between LoRa end devices and
end user applications.
LoRaWAN is a cloud-based \gls{mac} layer for LoRa that organizes \gls{phy} configurations
and \gls{mac} schedules, and routes traffic between end devices and end user applications.
The LoRaWAN architecture consist of three components: an Application Server,
which contains the application logic; a Network Server, which
coordinates access to the media between nodes and routes traffic between
the Application Server and End Devices; and Gateways, which act as the backbone
of the LoRaWAN network. The architecture prevents peer to peer communication
between end devices, which operate without network layer.

LoRaWAN defines three operational classes (modes) that show a trade-off between
downlink delay and power consumption. With class A, the reception of downlink
packets is only possible during a short interval after an uplink transmission.
Consequently, class A devices exhibit a high downlink latency, but the highest
power efficiency.
With class C, the end devices are always listening. The downlink latency
is consequently  lowest at the cost of a high energy consumption.
In class B, beacon-synchronized end devices wake up periodically in order to be
able to receive data. This class provides a good trade-off between downlink latency
and power consumption.

LoRaWAN imposes a series of limitations,  which make it impractical for scenarios with heterogeneous communication patterns. 
To increase the LoRa versatility as well as its efficiency, we propose the usage of IEEE 802.15.4 \gls{dsme} as a \gls{mac} layer for LoRa. \gls{dsme} is a
flexible \gls{mac} layer introduced in the 802.15.4e revision (2012) that provides
communication in contention-access, as well as contention-free
(time/frequency slots). 

In this paper, we want to answer the research question of how LoRa can be integrated with such a flexible \gls{mac} layer and how this stack performs in various settings. In particular, we want to show how LoRa end nodes can be opened up for hosting various network layers such as standard IP or data-centric adaptations~\cite{kaksw-liiel-22}. 
The contributions of this article are as follows.
\begin{enumerate}
\item We present \gls{dsme}-LoRa, a system design of LoRa with  IEEE 802.15.4 Deterministic Synchronous Multichannel Extension
(\gls{dsme}) as a \gls{mac} layer.
\item We evaluate in this work the performance of \gls{dsme}-LoRa on real hardware,
based on a \gls{dsme}-LoRa implementation~\cite{aksw-edmls-22} on the popular IoT operating System RIOT~\cite{bghkl-rosos-18}.
\item We propose a novel analytical stochastic model to predict transmission delay and throughput
    for \gls{dsme} slotted transmission.
\item We perform a large-scale simulation of \gls{dsme}-LoRa nodes to assess the scaling behavior of our proposed solution.
\item Based on the evaluation and model results, we derive preferred 
mappings for implementing different transmission patterns, with a balance trade-off
of energy consumption and transmission delay.
\end{enumerate}

The remainder of this paper is structured as follows. We outline the shortcomings of the current LoRaWAN system  along with a  problem statement in~\autoref{sec:problem_statement}. The relevant background on low power radio communication is summarized  in~\autoref{sec:background}. \autoref{sec:mappings} presents our \gls{dsme}-LoRa system design, which we evaluate on real hardware in~\autoref{sec:eval_hw}. We develop an analytical stochastic model in~\autoref{sec:analytics}, from which we predict the per packet performance for the slotted transmission. Peer-wise communication in large ensembles of LoRa nodes is subject to a simulation study in~\autoref{sec:simulation}. In~\autoref{sec:discussion} we discuss design decisions and options for optimization. 
Finally, we review related work in~\autoref{sec:related} and give a conclusion and outlook in~\autoref{sec:conclusions}.
The Appendix provides a supplementary figure (\autoref{sec:pi_calculation}) and lists a table of abbreviations (\autoref{sec:glossary}) which we use
throughout this article.

 \section{Problem statement}\label{sec:problem_statement}

\begin{figure}
    \tikzsetnextfilename{motivation}

\definecolor{color0}{rgb}{0.905882352941176,0.16078431372549,0.541176470588235}
\definecolor{color1}{rgb}{0.850980392156863,0.372549019607843,0.00784313725490196}
\definecolor{color2}{rgb}{0.458823529411765,0.43921568627451,0.701960784313725}
\definecolor{color3}{rgb}{0.105882352941176,0.619607843137255,0.466666666666667}
\definecolor{color4}{rgb}{0.4,0.650980392156863,0.117647058823529}
\begin{tikzpicture}[>={Latex}]
  \tikzset{
    sensor/.style = {draw, fill=color1, circle, inner sep=4pt },
    controller/.style = {draw, fill=color3, rectangle, inner sep=4pt,scale=1.5},
    graybox/.style = {draw, fill=lightgray, rectangle, inner sep=4pt, minimum width=.85cm},
    tls/.style = { double=Set2-A,double distance=4pt,line width=0.1cm },
    actuator/.style = {draw, fill=color2, isosceles triangle, isosceles triangle apex angle=60, rotate=90,scale=0.8},
  }

\node[inner sep=0pt] (lorawan) at (-3.5,0) {\includegraphics[width=4cm]{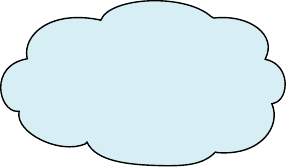}};

  \node[graybox, inner sep=4pt] (gw2) at ([xshift=0cm,yshift=-1.0cm]lorawan.south) {GW};
  \node[graybox] (NS) at ([shift={(-90:0.75cm)}]lorawan.center) {NS};
  \node[controller] (AS) at ([yshift=1cm]NS.north) {};

  \node[sensor] (lw_0) at ([shift={(-30:2cm)}]gw2.center) {};
  \node[actuator] (lw_2) at ([shift={(-150:2cm)}]gw2.center) {};

  \draw[->,thick, dashed]
    (lw_0) -- (gw2.east)
    ([xshift=0.175cm]gw2.north) -- ([xshift=0.175cm]NS.south)
    ([xshift=0.175cm]NS.north) -- ([xshift=0.175cm]AS.south)
    ;
  \draw[->,thick, dashed]
    ([xshift=-0.175  cm]AS.south) -- ([xshift=-0.175cm]NS.north)
    ([xshift=-0.175cm]NS.south) -- ([xshift=-0.175cm]gw2.north)
    (gw2.west) -- (lw_2)
    ;

  \draw[black, double=gray,double distance=2pt,line width=0.05cm]
    (gw2.north) -- (lorawan.south)
    node[pos=0.5] (lw_middle) {}
    ;

\node[inner sep=0pt] (dsme) at (3.5,0) {\includegraphics[width=4cm]{figures/cloud}};
  \node[controller, inner sep=4pt] (gw) at ([xshift=0cm,yshift=-1.0cm]dsme.south) {};

  \node[graybox] (opt) at (dsme.center) {Optional App. Logic};

  \node[sensor] (dsme_0) at ([shift={(-30:2cm)}]gw.center) {};
  \node[actuator] (dsme_2) at ([shift={(-150:2cm)}]gw.center) {};

  \draw[black, double=gray,double distance=2pt,line width=0.05cm]
    (gw.north) -- (dsme.south)
    node[pos=0.5] (dsme_middle) {}
    ;

  \draw[->, thick, dashed]
    (dsme_0) -- (gw.south)
    (gw.south) -- (dsme_2)
     ;

 \path[]
    (lw_2) -- (lw_0) coordinate[yshift=-1cm,pos=0.5](A)
    (dsme_2) -- (dsme_0) coordinate[yshift=-1cm,pos=0.5](B)
    ;
 \node[] (figa) at (A) {(a) LoRaWAN};
 \node[] (figb) at (B) {(b) Direct communication};
    \path[]
    (gw2) -- (gw) coordinate[pos=0.5,yshift=1.4cm](legend);
 
 \node[actuator,scale=0.8] (legend_a) at ([xshift=-1cm]legend) {};
 \node[anchor=north west] at ([xshift=0.25cm,yshift=0.03cm]legend_a.east) {Actuator};

 \node[sensor,anchor=south west,scale=0.8] (legend_s) at ([yshift=0.3cm]legend_a.north) {};
 \node[anchor=west] at ([xshift=0.1cm]legend_s.east) {Sensor};

 \node[controller,scale=0.8] (legend_c) at ([yshift=0.3cm]legend_s.north) {};
 \node[anchor=west] at ([xshift=0.1cm]legend_c.east) {Controller};

 \draw []
 ([xshift=-1.3cm,yshift=-0.25cm] legend) coordinate(bl) -- ([yshift=1.4cm]bl) coordinate(ul)
 -- ([xshift=2.5cm]ul) -- ([xshift=2.5cm]bl) -- (bl);
 ;

\end{tikzpicture}
     \caption{Control traffic of smart lightning scenario in LoRaWAN (left) through a gateway (GW), and direct communication (right).}
    \label{fig:motivation}
\end{figure}

The common LoRaWAN architecture adds rigid constraints to long-range
networking, which hinder many IoT deployments. Its centralized design
facilitates uplink-oriented applications, but challenges data sharing and the creation of distributed applications.

We argue that direct communication between LoRa devices overcomes these limitations,
while it still enables reliable communication in long-range deployments or harsh environments.
To further motivate the need for direct communication, we analyze an IoT control scenario for
smart lightning.

Sanchez-Sutil~\etal~\cite{sc-srees-22} design a LoRa system for
smart regulation of street lights (see~\autoref{fig:motivation} (b)) and propose an architecture
with illumination level devices (sensors), which transmit sensor data every minute.
A gateway for street lights system (controller), which acquire
illumination data from sensors, transmit control messages to actuators and
send measurement data from street lights to the cloud. Operating and
monitoring devices for street lights (actuators) control light level and transmit
electrical measurements to the controllers. The authors deploy several scenarios (up
to 64 actuators) with all devices in LoRa wireless reach.

A LoRaWAN implementation of such system may move controller logic to the cloud application and use
LoRaWAN to transmit data between sensors and actuators (see~\autoref{fig:motivation} (a)). However, this approach has the following disadvantages:

\one Traffic between controller and actuator is forced through LoRaWAN gateways.
If controllers transmit unicast control data every minute to all actuators,
nearby gateways will forward 64 downlink packets per minute.
Even if the LoRa devices and the LoRaWAN Network Server agree on the fastest
downlink data rate, a single gateway scenario will render 7\% duty cycle.
Because LoRaWAN gateways are half-duplex, packets received during downlink transmission are lost.
Therefore, such a deployment requires at least two dedicated gateways to enable a \gls{der} $\geq 99$\%.
In regions with duty cycle regulations such as EU868, even
more gateways are required to prevent additional packet losses as a result of downlink budget depletion.
Adding more gateways addresses these problems, but it increases deployment costs
and it is not always practical.

\two The LoRaWAN infrastructure prevents the deployment of edge devices and blindly forwards sensor data to the
cloud infrastructure. To further motivate the usage of edge devices, consider
a deployment in the city of London ($\approx 2.8$
million street lights). If sensors are on par with actuators and transmit every minute,
the cloud infrastructure receives ~1.5 trillion LoRaWAN messages per year, which artificially leads to a cost explosion in cloud infrastructure.

\three Devices with poor LoRaWAN wireless coverage increase transmission time on air 
 to improve link budget. This increases energy consumption \cite{lffb-rslmn-21}, which reduces
life cycle of nodes.

\four In many remote areas, cellular networks are the only uplink options for LoRaWAN gateways.
Poor Internet connectivity will lead to packet loss at the gateway, which threatens the versatility of the control system.

The proposed smart lightning topology (see~\autoref{fig:motivation} (b)) overcomes the limitations
of the LoRaWAN architecture. Instead of using a centralized controller in the cloud, the system implements low-cost controllers that run the control logic in a distributed way.
Therefore, downlink traffic is distributed between many controller devices instead
of aggregating at a few gateways. This effectively reduces downlink stress.
Because sensors and controllers are likely in wireless reach, sensors can transmit
using a fast data rate, which facilitates battery-powered operation.
Controllers can transmit preprocessed data at a lower rate
, which effectively reduces cloud transmissions and infrastructure costs. The system does not interrupt
control operation in case of intermittent connectivity at controllers. In addition, controllers are free to implement caching strategies to reduce packet loss on
intermittent Internet uplinks.

In the light of this use case, we argue that a \gls{dsme} \gls{mac} should perform better than the proposed system for two
reasons:
\one \gls{dsme} enables multichannel time slotted communication, in contrast with the
single channel approach of the system. This enables concurrent collision-free communication
without special hardware requirements (\eg LoRa concentrator). Therefore, controllers may be implemented with low
cost components, while still maintaining high \gls{prr}.
\two \gls{dsme} offers powerful built-in features such as device discovery and security mechanisms,
which facilitate deployment and secure operation.

We follow the \gls{dsme}-LoRa direction in the remainder of this paper to foster flexible long-range node-to-node communication.

 \section{Background on Low Power Radios}\label{sec:background}

\subsection{IEEE 802.15.4 with \gls{dsme} \acrshort{mac}}\label{sec:back_dsme}
The Deterministic and Synchronous Multichannel Extension (\gls{dsme}) initiates
a beacon-synchronized superframe structure that consists of a beacon slot,
a \gls{cap} and a \gls{cfp}. End devices
can choose to communicate during \gls{cap} or \gls{cfp}. During \gls{cap}, devices transmit using \gls{csma} in a common channel. During \gls{cfp}, end devices transmit in a dedicated time-frequency slot. The \gls{cfp} is divided in the time domain into seven multichannel slots, namely \gls{gts}. Each \gls{gts} is divided in the frequency domain into
the number of available channels in the channel page (usually 16). 
\gls{dsme} supports both peer to peer and cluster tree topologies. Similar to
traditional IEEE 802.15.4, there are three device roles: \gls{pan} coordinator,
regular coordinators, and child devices. Devices can transmit confirmed messages, where the \acrshort{mac} layer retransmits frames in case of missed ACK frame. We summarize the configuration parameters in~\autoref{tab:mac_params} and introduce in the reminder of this section.

\paragraph{Network formation}\label{sec:network_formation}
The \gls{pan} coordinator is the device in charge of defining the superframe structure.
For this purpose the device will transmit \textit{enhanced beacons} periodically. The
transmission of \textit{enhanced beacons} always occurs during a beacon slot and the
period is a multiple of the superframe duration. Devices that want to join the
\gls{dsme} network perform a scanning procedure to detect \textit{enhanced beacons}. When the
scanning procedure succeeds, the joining device sends an association request
to the coordinator. The association finishes when the coordinator acknowledges with a positive association reply.

The \gls{dsme} network can be extended natively by adding more coordinators. In such
case the coordinator will emit \textit{enhanced beacons} using the same period as the
\gls{pan} coordinator but in a different beacon slot offset. This ensures multiple
coordinators can share the same area without risk of beacon collisions.
In the event of beacon collisions (\ie two coordinators started emitting
beacons in the same slot offset), \gls{dsme} provides a native mechanism to resolve
collisions.
To be able to switch the common channel and \gls{phy} properties, the standard defines the \acrshort{phy}-OP-SWITCH mechanism in which neighbour devices are
instructed to switch to a different \gls{phy} configuration on reception of a dedicated \gls{mac} command. This allows dynamic switching between data rates, modulations and
frequency bands.

\begin{figure}
    \centering
    \tikzsetnextfilename{dsme_msf}

\begin{tikzpicture}[>={Latex}]
    \definecolor{col_beacon}{RGB}{252,141,98}
    \definecolor{col_cap}{RGB}{102,194,165}
    \definecolor{col_cfp}{RGB}{141,160,203}
    \definecolor{col_b}{RGB}{255,217,47}
  \tikzset{
      sbeacon/.style = {fill=col_beacon},
    scap/.style = {fill=col_cap},
    scfp/.style = {fill=col_cfp},
  }

    \pgfmathsetmacro{\sd}{0.2}
    \pgfmathsetmacro{\height}{\sd*5}
    \foreach \sfi in {0,1,2,3} {
\pgfmathsetmacro{\bs}{\sfi*\sd*16}
        \pgfmathsetmacro{\nbs}{(\sfi+1)*\sd*16}
        \pgfmathsetmacro{\caps}{\bs+\sd}
        \pgfmathsetmacro{\cfps}{\caps+8*\sd}
        \draw[sbeacon]
        (\bs,0) node[] (b{\sfi}) {} rectangle (\caps,\height) node[pos=.5,yshift=0.7cm] {\footnotesize \acrshort{bs}};
        \draw[scap]
        (\caps,0) node[] (c{\sfi}) {} rectangle (\cfps,\height) node[pos=.5,yshift=0.7cm] {\footnotesize \acrshort{cap}};
        \draw[scfp]
        (\cfps,0) node[] (cf{\sfi}) {} rectangle (\nbs,\height) node[pos=.5,yshift=0.7cm] {\footnotesize \acrshort{cfp}};

        \foreach \x in {1,...,6} {
            \pgfmathsetmacro{\start}{\cfps+\x*\sd}
        \draw[]
        (\start,0) -- (\start,\height);
        }
        \foreach \y in {1,...,4} {
            \pgfmathsetmacro{\xstart}{\cfps}
            \pgfmathsetmacro{\xend}{\nbs}
            \pgfmathsetmacro{\ystart}{\y*\sd}
            \pgfmathsetmacro{\yend}{(\y+1)*\sd}
        \draw[]
        (\xstart,\ystart) -- (\xend,\ystart);
        }
        \coordinate (end) at (\nbs,-1.5cm);
        \coordinate (lb0) at (\nbs,0);
        \coordinate (lb1) at (\nbs+\sd,\height);
    }
\path [] (cf{2}) -- (b{3}) coordinate [pos=0.2] (A) coordinate [pos=0.8] (B) coordinate[pos=0.5,yshift=-0.7cm] (C);
    \draw[->] (C) -- (A);
    \draw[->] (C) -- (B);
    \node[anchor=north] at (C) {\acrshort{gts}};

    \draw[sbeacon]
    (lb0) node[] {} rectangle (lb1) node[pos=.5,yshift=0.7cm] {\footnotesize \acrshort{bs}};
    \path[]
    (b{0}) -- ([yshift=-0.5cm]b{0}) node [] (bs0) {}
    -- ([yshift=-1cm]b{0}) node [] (bm0) {}
    -- ([yshift=-1.5cm]b{0}) node [] (bb0) {};
    \path[]
    (b{1}) -- ([yshift=-0.5cm]b{1}) node [] (bs1) {}
    -- ([yshift=-1cm]b{1}) node [] (bm1) {};
    \path[]
    (b{2}) -- ([yshift=-1cm]b{2}) node [] (bm2) {}
    -- ([yshift=-1.5cm]b{2}) node [] (bb2) {};
    \draw[dashed] (b{0}) -- (bb0); 
    \draw[dashed] (b{1}) -- (bs1); 
    \draw[dashed] (b{2}) -- (bm2); 
    \draw[dashed] (lb0) -- (end); 
    \draw[<->] (bs0) -- (bs1) node[pos=0.5,above] {Superframe};
    \draw[<->] (bm0) -- (bm2) node[pos=0.5,above] {Multisuperframe};
    \draw[<->] (bb0) -- (end) node[pos=0.5,above,yshift=0.05cm] {Beacon interval};

\end{tikzpicture}
     \caption{Overview of the \gls{dsme} Multisuperframe transmission resources.}
    \label{fig:dsme_fundamentals}
\end{figure}
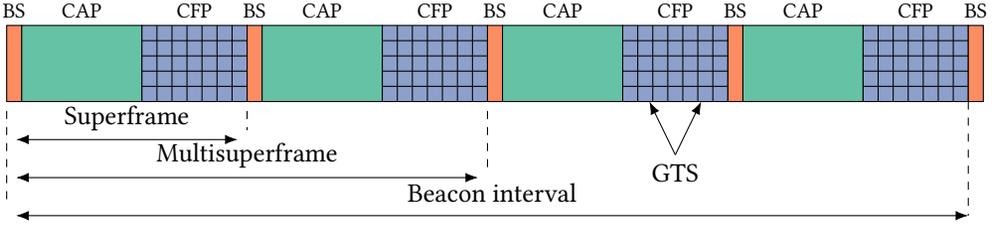

\paragraph{Superframe structure} \label{sec:superframe_structure}
Superframes merge into a multisuperframe structure as visualized
in~\autoref{fig:dsme_fundamentals}.
\gls{dsme} supports a \gls{cap} reduction mode in the multisuperframe
structure (\texttt{macCapReduction}), in which the \gls{cap} period
is replaced by 8 \gls{cfp} additional slots in all superframes except the first.
For example, a configuration with four superframes per multisuperframe exposes 28
\gls{gts} (448 unique time-frequency slots). With \gls{cap} reduction, the same structure
exposes 52 \gls{gts} (832 unique time-frequency slots).

\begin{table}
    \centering
    \caption{List of \acrshort{mac} configuration parameters for the \acrshort{dsme} \acrshort{mac}.}
    \label{tab:mac_params}
    \begin{tabular}{ll}
        \toprule
        \textbf{Parameter} & \textbf{Description} \\
        \midrule
        \texttt{macMinBE} & Minimum backoff exponent\\
        \texttt{macMaxBE} & Maximum backoff exponent \\
        \texttt{macMaxCsmaBackoff} & No. of access attempts before declaring channel failure \\
        \texttt{macMaxFrameRetries} & No. of frame retransmissions.\\
        \texttt{macSuperframeOrder (\acrshort{so})} & Describes the length of the superframe slot\\
        \texttt{macMultisuperframeOrder (\acrshort{mo})}& Describes the length of the multisuperframe \\
        \texttt{macBeaconOrder (\acrshort{bo})} & Describes the length of the beacon interval \\
        \texttt{macCapReduction} & Opt. replace \acrshort{cap} with \acrshort{cfp} in all superframes except first \\
        \texttt{macRxOnWhenIdle} & Opt. keep receiver on during \acrshort{cap} \\
        \bottomrule
    \end{tabular}
    \end{table}

\gls{dsme} defines three parameters to describe the superframe structure, namely \gls{so},
\gls{mo}, and \gls{bo} (compare~\autoref{tab:mac_params}).
The superframe order defines the slot duration as: $\texttt{aBaseSuperframeDuration} \cdot T_{Symbol} \cdot 2^{\acrshort{so}}$, where \texttt{aBaseSuperframeDuration} = 60 symbols, as per standard.
A small superframe order, which leads to a shorter superframe duration, offers
shorter latencies at the cost of higher energy consumption and smaller payload.
\gls{so}=3 enables
the transmission of standard 127 bytes 802.15.4 frames.
The multisuperframe order, together with the superframe order, define the number
of superframes per multisuperframe as $2^{(\acrshort{mo}-\acrshort{so})}$. Higher multisuperframe orders
lead to higher \gls{gts} resources with the cost of higher latencies. Finally, the beacon
order sets the beacon interval to $2^{(\acrshort{bo}-\acrshort{mo})}$ multisuperframes. Higher beacon
orders lead to higher beacon intervals, which extend the number of potential coordinator devices
at the cost of longer association time.
These three parameters must comply with $0 \leq \acrshort{so} \leq \acrshort{mo} \leq \gls{bo} \leq 14$.
We summarize the number of available \gls{gts} and the
multisuperframe duration for different multisuperframe orders for the case \acrshort{so}=3 in~\autoref{tab:msf_duration}.

\begin{table}
\centering
\caption{Number of available \gls{gts} with and without \gls{cap} reduction, and multisuperframe duration ($T_{msf}$) for varying multisuperframe orders (\acrshort{mo}), with a superframe order (\acrshort{so}) of 3 and symbol time 1\, ms.}
    \label{tab:msf_duration}
\begin{tabular}{cccr}
    \toprule
    \makecell[c]{\textbf{Multisuperframe}\\\textbf{order (\acrshort{mo}) }} & \makecell[c]{\textbf{Number of \acrshort{gts}}\\\textbf{(w/o \acrshort{cap} reduct.) [\#]}} & \makecell[c]{\textbf{Number of \acrshort{gts}}\\\textbf{(w/ \acrshort{cap} reduct.) [\#]}} & \makecell[c]{\textbf{Duration}\\\boldmath{$T_{msf} [s]$}}\\
    \midrule
    3 & 7 & 7 & 7.68\\
    4 & 14 & 22 & 15.36\\
    5 & 28 & 52 & 30.72\\
    6 & 56 & 112 & 61.44\\
    7 & 112 & 232 & 122.88\\
    \bottomrule
    \end{tabular}
\end{table}

\paragraph{\gls{csma} transmissions} \label{sec:cap_transmission}
On schedule of a \gls{csma} transmission, the \acrshort{mac} queues the packet in the \gls{cap} queue
and performs slotted \gls{csma}, aiming to avoid collisions while
accessing the common channel.
The \gls{csma} algorithm requires four parameters displayed in the first four rows in~\autoref{tab:mac_params}.
On transmission the \acrshort{mac} aligns to the backoff period, which occurs every 20 symbols since the start
of the \gls{cap}, and waits a random number of backoff periods between 0 and
$2^{\texttt{macMinBE}}$. In case the duration of the remaining portion of the \gls{cap} is
shorter than the required backoff periods, the \acrshort{mac} waits for the next \gls{cap}
period and continues its countdown accordingly. The \acrshort{mac} then performs a series
of clear channel assessments (at least two), each one at the beginning of a
backoff period. On failure, the \acrshort{mac} doubles the backoff period (below $2^{\texttt{macMaxBE}}$) and the \gls{csma} algorithm retries until it succeeds or the \acrshort{mac}
runs out of \gls{csma} attempts (\texttt{macMaxCsmaBackoff}). When successfully accessing the channel, the \acrshort{mac} transmits the frame and (optionally) waits
for an ACK frame. If the \acrshort{mac} expects an ACK frame and does not receive it, the
\acrshort{mac} repeats the \gls{csma} procedure until it runs out of \gls{csma} attempts or retransmissions (\texttt{macMaxFrameRetries}).

During \gls{cap} the \acrshort{mac} can transmit both unicast and broadcast frames.
In order to minimize the energy consumption on constrained devices, the \acrshort{mac}
offers the \texttt{macRxOnWhenIdle} configuration parameter to turn off the
receiver during \gls{cap}. This does not affect outgoing transmissions, but
prevents the \acrshort{mac} from receiving frames. To transmit frames to
these constrained devices, the standard defines the \textit{indirect transmission} mechanism.
A coordinator queues frames scheduled with \textit{indirect transmission} and appends
the target address to the next beacon. A constrained device
that finds its address in the beacon polls the coordinator with a data request
command, and waits for an ACK frame with the subsequent data frame.

\paragraph{\gls{gts} transmission} \label{sec:slot_allocation}
End devices that require communication with other devices during \gls{cap} need to
negotiate one or more \gls{gts} with the target device. \gls{dsme} provides a native
mechanism to negotiate slots---in contrast to \gls{tsch}. \gls{dsme} \gls{gts} are unidirectional
(RX or TX) and only support unicast frames.
When a device \textbf{A} wants to allocate one or several slots with device \textbf{B} (coordinator or child), it sends a \gls{dsme}-\gls{gts} request frame during \gls{cap} to device \textbf{B}. In case the
device accepts the slot, it replies with a \gls{dsme}-\gls{gts} response frame indicating
success. Finally, device \textbf{A} broadcasts a \gls{dsme}-\gls{gts} notify frame to indicate the
other node in reach about the new slot allocation.
Alternatively, a device can allocate a slot during the association procedure, by
sending a \gls{dsme} Association Request command.
On schedule of \gls{gts} transmission, the \acrshort{mac} queues the packet in the \gls{cfp} queue, which divides into multiple FIFO queues, one for each destination device among the allocated \gls{gts} resources.
\gls{gts} transmissions support two channel diversity modes, namely channel adaptation
and channel hopping. In channel adaptation mode, a source device may allocate \gls{gts}
in a single channel or in different channels based on the knowledge of the channel
quality. The source device requests channel quality information to a destination
device using the \gls{dsme} Link Report \gls{mac} command. Thereby devices agree on a different
channel if the channel quality is poor.
In channel hopping mode, each \gls{gts} hops over a predefined sequence of channels.

\gls{dsme} supports message priority for \gls{gts} transmissions. On the occurrence of a valid
\gls{gts}, the \acrshort{mac} layer transmits first the frames with high priority and then regular frames, providing a class-based service differentiation.

The 802.15.4e amendment introduces the group ACK feature, in which a coordinator
 receiving data from multiple senders transmits one group ACK frame to
all nodes in a single slot of a multisuperframe. The latest versions of the standard
do not include this feature, but we discuss its potential use cases for reducing
time on air in~\autoref{sec:regulations}.

\subsection{LoRa modulation}\label{sec:lora_modulation}

The LoRa modulation utilizes the chirp spread spectrum technique to transmit
data over the wireless channel. This technique defines a linear frequency modulated
symbol, namely chirp, which utilizes the entire allocated bandwidth
spectrum. As a result, the LoRa signal is robust against interference and multi-path
fading, and enables transmission ranges of kilometers depending on the \gls{phy}
configuration. An interesting property of the LoRa modulation, namely the capture
effect, allows to successfully decode a frame under collision if the power
difference with the colliding frames is large enough.

LoRa relies on two \gls{phy} parameters, namely bandwidth and spreading factor,
which define the symbol duration. A higher symbol duration renders better receiver
sensitivity, which increases transmission range at the cost of higher time on air and lower \gls{phy} bit rate.
A third parameter, code rate, defines the redundancy bits encoded in the
LoRa transmission. Similarly, the code rate trades-off transmission range with
time on air.

The LoRa \gls{phy} frame consist of a preamble, used to synchronize the transceiver to
the frame; an optional LoRa \gls{phy} header, which encodes payload length,
forward error correction code rate and the presence of a payload \acrshort{crc} at the end
of the \gls{phy} packet; a payload, which contains the \acrshort{psdu}; and an optional payload \gls{crc}.
The LoRa preamble defines a sync word at the end, with the purpose of isolating
networks of LoRa devices. For example, LoRaWAN sets the sync word to 0x34 for public networks and 0x12 for private networks.

LoRa devices are subject to regional Sub-GHz regulations that impose
restrictions on the transmission of LoRa frames.
These restrictions can be categorized in \one duty cycle restrictions, in which
a transmitter may not exceed a maximum time over an observation period (usually 1\% of time over an hour); \two dwell time restrictions, in which the transceiver may not
exceed a maximum time on a single channel and \three channel restriction,
in which the device must switch channels on consecutive transmissions or transmit
over a minimum number of channels.

LoRa transceivers can decode signals below the noise floor, which renders
energy detection mechanisms such as \acrshort{rssi} impractical for detecting the presence
of signals on the air. To circumvent this problem, common LoRa transceivers implement a
\gls{cad} mechanism to note the presence of a LoRa preamble signal.

An interesting feature of LoRa transceivers, which has not been exploited by LoRaWAN,
is \gls{fhss} transmission.
This feature allows to repeatedly switch carrier frequencies during radio
transmission, aiming to reduce interference and avoid interception.
 \section{DSME-LoRa system design}\label{sec:mappings}

\begin{figure}
    \tikzsetnextfilename{system_arch}

\begin{tikzpicture}[node distance=1.2cm and 0.3cm]
\definecolor{opendsme}{RGB}{255,255,179}
\definecolor{flora}{RGB}{141,211,199}
\definecolor{inet}{RGB}{190,186,218}
\definecolor{own}{RGB}{251,128,114} \tikzstyle{process} = [rectangle, minimum width=10cm, minimum height=0.7cm, text centered, draw=black, fill=orange!30]
\tikzstyle{small} = [rectangle, minimum width=2.2cm, minimum height=0.7cm, text centered, draw=black, fill=orange!30]
\tikzstyle{line} = [thick,-,>=stealth]
\tikzstyle{arrow} = [thick,->,>=stealth]

\node (mac) [process,fill=opendsme, anchor=north] {802.15.4 \acrshort{dsme} MAC};
\node (dsme_lora) [small,fill=own,below=of mac.south west,anchor=north west] {Frame mapping};
\node (chan_map) [small,fill=own,below=of mac.south,anchor=north] {Channel mapping};
\node (cad_map) [small,fill=own,below=of mac.south east,anchor=north east] {\acrshort{cca} mapping};
\node (transceiver) [process,fill=flora,below=of dsme_lora.south west, anchor=north west] {LoRa Radio};

\draw [<->,>={[scale=1]Latex}]
    ([xshift=1.1cm]mac.south west) -- ([xshift=1.1cm]dsme_lora.north west)
    node [pos=0.5,fill=white] {802.15.4 frame};
\draw [<->,>={[scale=1]Latex}]
    (mac.south) -- (chan_map.north)
    node [pos=0.5,fill=white] {802.15.4 channel};
\draw [<->,>={[scale=1]Latex}]
    ([xshift=-1.1cm]mac.south east) -- ([xshift=-1.1cm]cad_map.north east)
    node [pos=0.5,fill=white] {\acrshort{cca}};

\draw [<->,>={[scale=1]Latex}]
    ([xshift=1.1cm]transceiver.north west) -- ([xshift=1.1cm]dsme_lora.south west)
    node [pos=0.5,fill=white] {LoRa frame};
\draw [<->,>={[scale=1]Latex}]
    (transceiver.north) -- (chan_map.south)
    node [pos=0.5,fill=white] {LoRa PHY channel};
\draw [<->,>={[scale=1]Latex}]
    ([xshift=-1.1cm]transceiver.north east) -- ([xshift=-1.1cm]cad_map.south east)
    node [pos=0.5,fill=white] {LoRa \acrshort{cad}};

\draw [dashed, thick]
    ([shift={(135:0.5cm)}]dsme_lora.north west) -- ([shift={(45:0.5cm)}]cad_map.north east)
    -- ([shift={(-45:0.5cm)}]cad_map.south east) -- ([shift={(-135:0.5cm)}]dsme_lora.south west) -- cycle node [pos=0.5,rotate=90,anchor=south, yshift=0.2cm,font=\bf] {\acrshort{dsme}-LoRa};

\end{tikzpicture}
     \caption{\gls{dsme}-LoRa architecture overview.}
    \label{fig:system_arch}
\end{figure}
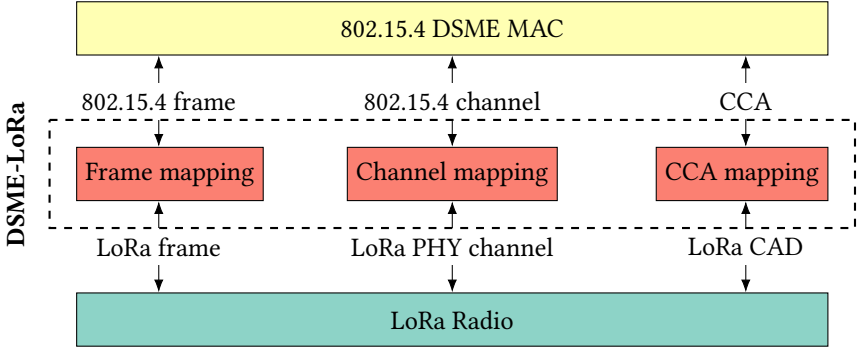

To operate LoRa radios below \gls{dsme} we define a \gls{dsme}-LoRa adaptation
, based on our original work~\cite{aksw-dfml-21}, that maps 802.15.4 \gls{mac} operations to LoRa operations. The adaptation layer
performs three tasks:
\one mapping of 802.15.4 channels to LoRa \gls{phy} channels.
\two conversion of 802.15.4 frames to LoRa frames.
\three implementation of 802.15.4 \gls{cca} on top of the LoRa device. \autoref{fig:system_arch} depicts the
system architecture.

\paragraph{Channel mapping}\label{sec:channel_mapping}
The adaptation layer maps 802.15.4 channels to LoRa \gls{phy} channels.
For this work, we define a channel page with sixteen LoRa channels (see~\autoref{tab:frequencies}) in the EU868~\cite{ETSI-en3002001-211} region. Note that the channel page may define more than sixteen
channels, as long as the channel information fits in the \gls{mac} frames that
control \gls{gts} allocation.

In the EU868 region the duty cycle of a band limits the time on air of a transmitting device
to a percentage within one hour observation period.
For 1\% and 10\% bands the duty cycle limits the cumulative time on air
36\,s and 360\,s respectively. For example, devices in a 1\% band
cannot transmit a frame if the active send time exceeds 36\,s during the last hour.
Note that the duty cycle is measured per band and not
per channel. If a device transmits in multiple channels on a same band, the
duty cycle of each channel adds up to the duty cycle of the band.

All channels utilize the same \acrshort{phy} configurations: spreading factor 7, bandwidth 125\,kHz
and code rate 4/5, which results in a \acrshort{phy} bit rate $\approx$ 5.5 kbps
and a symbol time of $\approx$ 1 ms. We choose these settings to provide a balanced
trade-off between transmission range, time on air, and throughput.

Note that different sets of LoRa \gls{phy} settings can be encoded using different
channel pages. Thereby devices can agree on different channel pages, using the \acrshort{phy}-OP-SWITCH
feature~\autoref{sec:back_dsme}, to increase transmission range or increase the channels for concurrent \gls{phy} communication.
We will investigate the feasibility of this proposal in future work.

Defining LoRa \gls{phy} channels for other
regions is viable, albeit challenging. We further discuss this situation in~\autoref{sec:regulations}.

We define one channel inside the \textit{g3} band (10\% duty cycle)
and fifteen channels inside the \textit{g} band (1\% duty cycle) with 200 kHz
channel spacing.
In order to relax duty cycle restrictions, we utilize the 10\% band channel for beacon transmissions, \gls{cap} channel, and
\gls{gts} transmissions. The remaining channels are used exclusively for \gls{gts}
transmissions.

Since the proposed channels overlap with LoRaWAN channels, we define the synchronization word of the preamble to 0x17 in order to avoid decoding of LoRaWAN frames. Furthermore, we include
the LoRa \acrshort{phy} header and payload \acrshort{crc} described in~\autoref{sec:lora_modulation}.

\begin{table}
\centering
    \caption{\gls{dsme}-LoRa \acrshort{phy} channel definition, EU868 frequency band information, and purpose.}
    \label{tab:frequencies}
    \begin{tabular}{ccccl}
    \toprule
    \textbf{Channel} & \textbf{Band} & \textbf{Frequency [MHz]} & \textbf{Duty Cycle [\%]} & \textbf{Purpose}\\
    \midrule
    11-25 & \textit{g} & 863.00 \textendash{} 868.00 & 1 & \gls{gts} transmission \\
    \midrule
    26 & \textit{g3} & 869.40 \textendash{} 869.65 & 10 & \makecell[l]{Beacon transmission,\\\gls{csma} transmission,\\\gls{gts} transmission}\\
    \bottomrule
\end{tabular}
\end{table}

\paragraph{Frame mapping}
On frame transmission the adaptation layer calculates and appends a checksum to the
\gls{mac} frame and passes the frame to the LoRa transceiver.
On frame reception, the layer receives the LoRa frame from the transceiver and
calculates the frame checksum. On success, the layer dispatches the frame to the
\gls{mac} layer.
In order to transmit full 127 bytes 802.15.4 frames, we set the superframe order
to 3. The adaptation layer defines the \gls{mac} symbol time to
1 ms, which is in line with the LoRa symbol time for the channel configuration.
With this superframe order configuration and symbol time, the superframe slot duration resolves
to 0.48\,s. Hence, the superframe duration (16 superframe slots) is 7.68\,s.
We leave the multisuperframe and beacon order configuration to the application.

\paragraph{\acrshort{cca} mapping}
On \gls{cca} requests from the \gls{mac} layer, the adaptation layer
maps to the LoRa \gls{cad} feature, which detects the presence of a LoRa preamble on the
air. On successful detection, the layer reports channel busy to the
\acrshort{mac} layer. Otherwise, the layer assumes the channel is free and
reports clear channel.

\subsection{\gls{dsme}-LoRa implementation}\label{sec:implementation}

\begin{figure}
    \centering
    \includegraphics[width=0.5\textwidth]{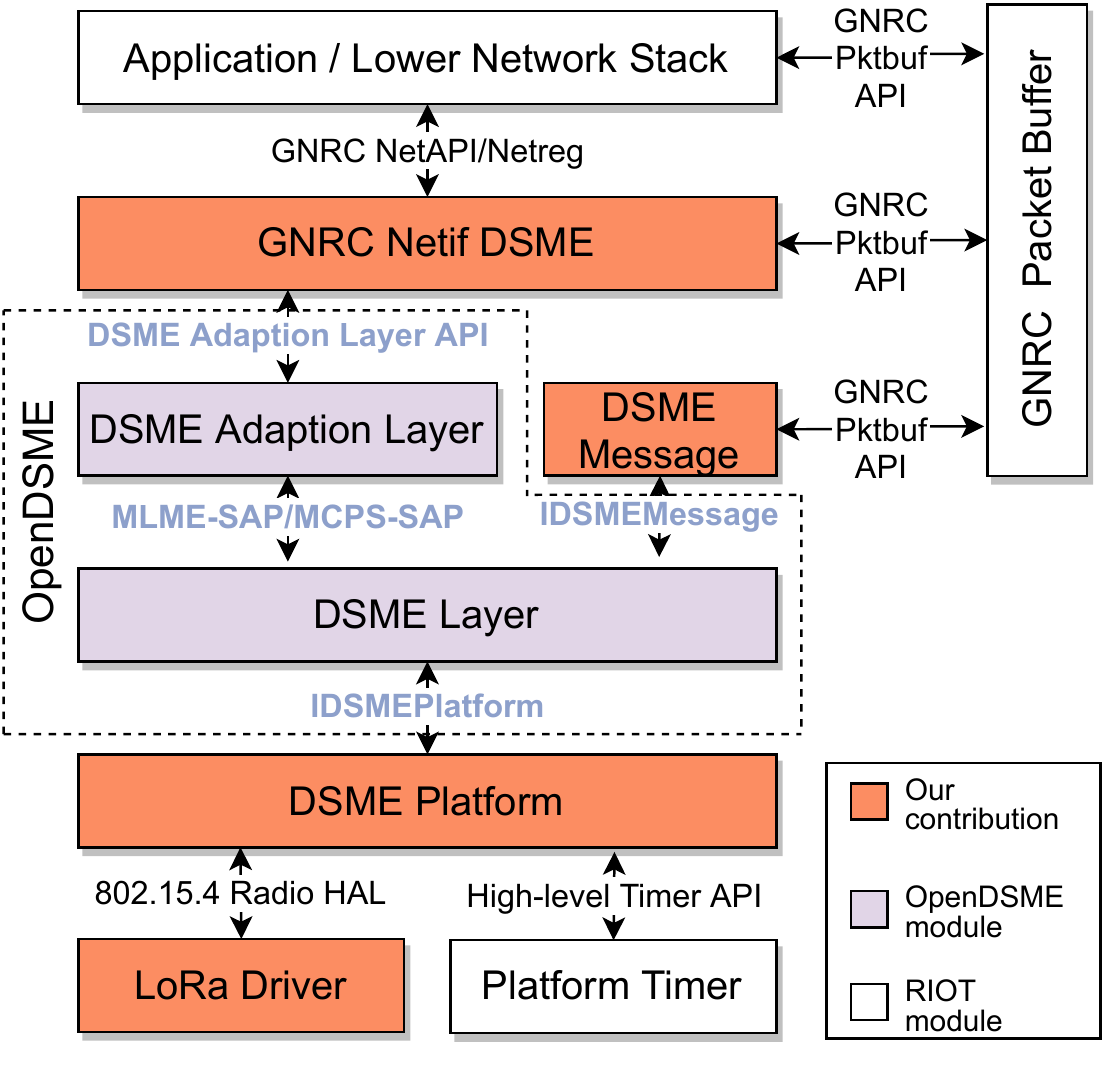}
    \caption{\gls{dsme}-LoRa integration into the networking subsystem of RIOT.}
    \label{fig:dsme_lora_integration}
\end{figure}

The integration of \gls{dsme}-LoRa on real hardware imposes a series of challenges.
\one long time on air of LoRa requires a long superframe slot duration, which
results in long beacon intervals. IoT devices are prone to clock drift from cheap crystals, which
increases the chances of desynchronization between child devices and coordinators.
\two common LoRa transceivers do not add a mechanism to timestamp frame reception,
which is required to synchronize time between neighbour.
\three \gls{dsme} accesses the transceiver based on interrupts during critical operations. This faces concurrent access with
hardware serial peripheral buses (\eg SPI) and limits the responsiveness on real
time operating systems.
\four common LoRa hardware platforms are constrained and have
low memory resources. This serves a common LoRaWAN stack. In contrast, \gls{dsme}
requires additional memory due to the complexity of the \gls{mac}.

We integrate \textit{openDSME} into the RIOT network stack (GNRC), which provides
a generic messaging interface (GNRC Netapi), a centralized packet buffer (GNRC Pktbuf), and a packet dispatch registry (GNRC Netreg). RIOT provides a high level
platform timer API and a hardware abstraction layer for 802.15.4 devices.
We further extend \textit{openDSME} to support the \textit{macRxOnWhenIdle} mode (\autoref{tab:mac_params}), in order to turn the transceiver off during \gls{cap} and save
energy, when \gls{cap} is not used.
\autoref{fig:dsme_lora_integration} presents the system integration of \gls{dsme}-LoRa in RIOT and our
contributions.

\paragraph{GNRC Netif \gls{dsme}} implements the GNRC network interface for \gls{dsme}. This
allows, via the GNRC Netapi:
\one transmission of \gls{dsme} frames.
\two configuration of the
\gls{dsme} \gls{mac} (device role, static slot allocation).
\three scanning and
association procedures. We use the \gls{dsme} Adaptation Layer, a convenience API provided by \textit{openDSME}, to implement the \gls{mac} layer logic below the network interface.
The interface dispatches the incoming frames via the GNRC Netreg. In order to
minimize memory consumption, we utilize the callback extension of GNRC Netreg as
an alternative to the default IPC implementation. This saves an additional thread
for reception, which heavily reduces RAM utilization.

\paragraph{\gls{dsme} Message} implements the \gls{dsme} Message interface (IDSMEMessage) which
abstracts the packet representation. We use the GNRC Pktbuf to implement the
packet representation. This approach has advantages on memory consumption.
\one centralized storage prevents data duplication.
\two the scattered packet representation of GNRC Pktbuf allows appending chunks of data to a packet without memory reallocation.
\three GNRC Pktbuf supports allocation with \textit{malloc}. This facilitates the
operation of \gls{dsme} in the same memory pool as \textit{openDSME}, which bases on
heap.

\paragraph{\gls{dsme} Platform} implements the \gls{dsme} Platform interface (IDSMEPlatform) which
defines the platform abstraction layer of \textit{openDSME}. The
interface implements the access to the LoRa transceiver on top of the 802.15.4
Radio HAL. It further implements the access to timer functionalities
of the operating system. Thereby, we configure the high-level timer to use the real-time timer peripheral, aiming to mitigate the effect of clock drift due to long beacon intervals.
We delegate the processing of transceiver interrupts and system timers to the
RIOT scheduler, in order to avoid concurrent access to the system bus between the transceiver and operating system.
The implementation reconfigures the symbol time of the \gls{mac} layer to 1 ms (LoRa) in compliance
with~\autoref{sec:mappings}.

\paragraph{LoRa Driver} implements a 802.15.4 compatible driver for the LoRa transceiver
(SX1272/SX1276). The driver implements the three components of the \gls{dsme}-LoRa Adaptation Layer (\autoref{sec:mappings}), namely channel mapping, frame mapping and \gls{cca} mapping. To timestamp frame reception, we calculate the time difference between the
packet reception interrupt (\textit{RxDone}) and the valid header interrupt (\textit{ValidHeader}). We use this time difference to calculate the exact reception timestamp of the frame.

As a result of these design decisions, our \gls{dsme}-LoRa implementation consumes $\approx$ 108\,kB of ROM and $\approx$ 12\,kB of RAM on ARM Cortex-M0 CPU, which is enough for common LoRa hardware platforms.
 \section{Evaluation on real hardware}\label{sec:eval_hw}

\begin{figure}
    \tikzsetnextfilename{source_sink_network}

\definecolor{color1}{rgb}{0.850980392156863,0.372549019607843,0.00784313725490196}
\definecolor{color2}{rgb}{0.458823529411765,0.43921568627451,0.701960784313725}
\begin{tikzpicture}[>={Latex}]
  \tikzset{
    source/.style = {draw, fill=color1, circle, inner sep=5pt },
    sink/.style = {draw, fill=color2, circle, inner sep=5pt },
  }

  \foreach \y in {0,1,2,3} {
    \node[source] (source_\y) at (0,-\y*1cm) {};
    \node[anchor=south east] at (source_\y.north west) {\includegraphics[width=0.5cm]{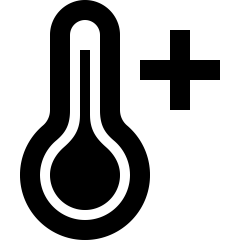}};
  }

  \foreach \y in {0,1,2} {
    \node[sink] (sink_\y) at (3cm,-0.5cm-\y*1cm) {};
    \node[anchor=south west] at (sink_\y.north east) {\includegraphics[width=0.5cm]{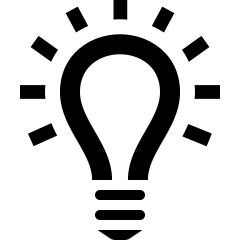}};
  }

  \draw [->,thick] (source_0) -- (sink_0);
  \draw [->,thick] (source_1) -- (sink_1);
  \draw [->,thick] (source_2) -- (sink_1);
  \draw [->,thick] (source_3) -- (sink_2);

    \node [yshift=1cm] at (source_0.north) {Source devices};
    \node [yshift=1.5cm] at (sink_0.north) {Sink devices};

\end{tikzpicture}
     \caption{Topology of a \gls{dsme}-LoRa network with source and sink devices. One sink device can receive from multiple sources.}
    \label{fig:source_sink_network}
\end{figure}
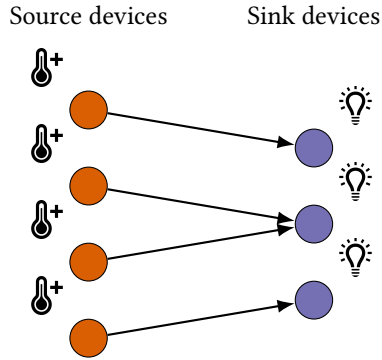

We evaluate the \gls{dsme}-LoRa implementation (see~\autoref{sec:implementation}) in a peer to peer topology
with source devices (TX-only) and sink devices (RX-only), as depicted
in~\autoref{fig:source_sink_network}.
During our experiments, each source device transmits data with exponentially distributed
interarrival times to a single sink. We vary the number of source devices (N) and the average transmission interval.

Our results include the transmission delay (time between packet schedule and
successful reception), time on air and energy consumption for
transmissions during \gls{cap} and \gls{cfp}. For the \gls{cap}, we further analyze the impact of \gls{csma} with \gls{cad} using
different backoff parameters. We also evaluate the impact of cross-traffic between
coexistent \gls{dsme}-LoRa and LoRaWAN networks and the effect of interference on
\gls{dsme}-LoRa.

\subsection{Experiment setup}\label{sec:methodology}

\begin{figure}
    \includegraphics[width=0.75\textwidth,trim=70 0 70 70,clip]{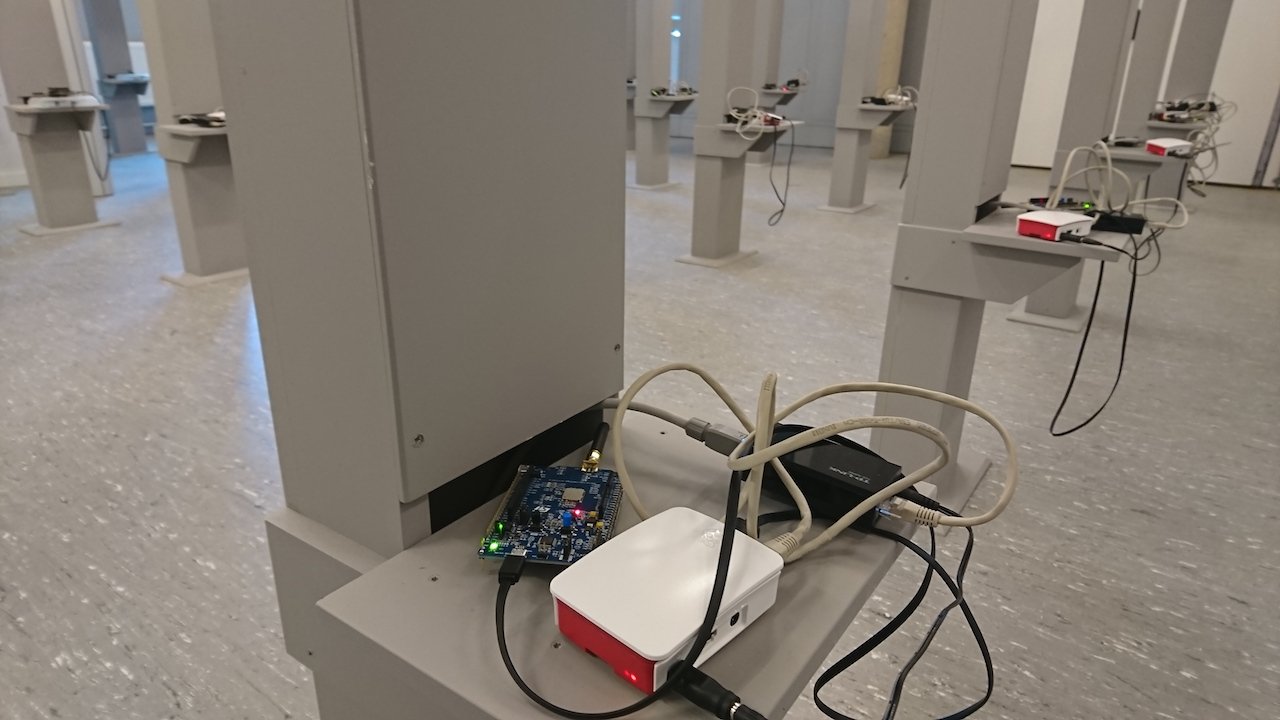}
    \caption{Deployment of LoRa platforms (B-L072z-lrwan1) on the FIT IoT-LAB testbed~\cite{abfhm-filso-15}.}
    \label{fig:iotlab}
\end{figure}

\paragraph{Testbed deployment}
We conduct our experiments in the \textit{Saclay} site of the FIT IoT-LAB testbed,  which
supplies 25 LoRa boards (B-L072z-lrwan1). These are distributed in a  12 m by 12 m room, as
shown in~\autoref{fig:iotlab}.
The B-L072z-lrwan1 platform consists of an ARM Cortex-M0 CPU, which runs at 32 MHz, provides 192\,kB of ROM/20\,kB of RAM,
and contains a SX1276 LoRa transceiver.
The testbed contributes  a \textit{serial\_aggregator} tool that aggregates all
UART output of the deployment and adds a timestamp.
We add logging to our measurement firmware for packet schedule, transmission, reception, and \acrshort{mac} queue lengths and use this information to calculate transmission delay, \gls{prr}, and time on air.

\paragraph{Multisuperframe structure}
We configure \gls{dsme} to one superframe per multisuperframe, which results in
a multisuperframe duration of $T_{msf}$=7.68\,s. This configuration exposes 7 \gls{gts} over 16 channels,
which enables 112 unique time-frequency slots. We use a beacon interval
of two multisuperframes which results in a beacon period of 15.36\,s.

\paragraph{Network topology}
A variable number of source devices transmits data to three sink devices using
direct communication (gateway-less).
This mapping accommodates solely \gls{gts}
transmissions on the proposed multisuperframe configuration. During bootstrap,
a random sink is assigned to each source device. We use static allocation for \gls{gts}.
With that, we imitate \gls{gts} allocation during device association with the \gls{dsme} Association Request command (see~\autoref{sec:back_dsme}).
We deploy an extra device that operates as the \gls{pan} coordinator, that establish the superframe structure by transmitting \textit{enhanced beacons}.
Although any sink or source device may operate as the \gls{pan} coordinator, we opt
for this approach to simplify the deployment.

\paragraph{\gls{mac} configurations} \label{sec:mac_config}
If not mentioned otherwise, we configure the \gls{csma} backoff parameters to \texttt{macMinBE}=7, \texttt{macMaxCsmaBackoff}=5, \texttt{macMaxBE}=8 (see~\autoref{sec:superframe_structure}),
which are close to the maximum values, in order to cope with long time on air.
\autoref{sec:csma_ca} further compares these values to 802.15.4 default values. In agreement with
the 802.15.4 standard, we set the maximum number of retransmissions to \texttt{macMaxFrameRetries}=4.
We utilize the channel hopping mode for \gls{gts} transmissions.

\subsection{Data transmission in \gls{cap} and \gls{cfp}}\label{sec:data_transmission}

\begin{figure}
    \tikzsetnextfilename{data_transmission}

\begin{tikzpicture}
\definecolor{color0}{rgb}{0.905882352941176,0.16078431372549,0.541176470588235}
\definecolor{color1}{rgb}{0.850980392156863,0.372549019607843,0.00784313725490196}
\definecolor{color2}{rgb}{0.458823529411765,0.43921568627451,0.701960784313725}
\definecolor{color3}{rgb}{0.105882352941176,0.619607843137255,0.466666666666667}
\definecolor{color4}{rgb}{0.4,0.650980392156863,0.117647058823529}
\begin{groupplot}[
group style={
  group size=3 by 2,
  horizontal sep=0.25cm,
  vertical sep=0.35cm,
},
height=0.3\textwidth,
width=0.35\textwidth,
xmax=99,
ymax=1.03,
axis on top,
yticklabel pos=left,
xmin=0, ymin=0,
legend cell align={left},
legend style={
  fill opacity=0.8,
  draw opacity=1,
  text opacity=1,
  at={(0.97,0.03)},
  nodes={scale=0.8, transform shape},
  anchor=south east,
  draw=white!80!black,
},
tick align=outside,
tick pos=left,
x grid style={white!69.0196078431373!black},
xtick style={color=black},
y grid style={white!69.0196078431373!black},
ytick style={color=black},
cycle list name=exotic,
]
\nextgroupplot[title={\textbf{TX interval=20\,s}},xticklabels={},xlabel={}]
\addplot+ [mark repeat=20] table [x index=14, y index=15] {data/tx_cap.csv};
\addlegendentry{N=5}
\addplot+ [mark repeat=20] table [x index=2, y index=3] {data/tx_cap.csv};
\addlegendentry{N=10}
\addplot+ [mark repeat=20] table [x index=8, y index=9] {data/tx_cap.csv};
\addlegendentry{N=15}
\nextgroupplot[title={\textbf{TX interval=10\,s}},xticklabels={}, yticklabels={},xlabel={}]
\addplot+ [mark repeat=20] table [x index=12, y index=13] {data/tx_cap.csv};
\addplot+ [mark repeat=20] table [x index=0, y index=1] {data/tx_cap.csv};
\addplot+ [mark repeat=20] table [x index=6, y index=7] {data/tx_cap.csv};
\nextgroupplot[title={\textbf{TX interval=5\,s}},xticklabels={}, yticklabels={},xlabel={}]
\addplot+ [mark repeat=20] table [x index=16, y index=17] {data/tx_cap.csv};
\addplot+ [mark repeat=20] table [x index=4, y index=5] {data/tx_cap.csv};
\addplot+ [mark repeat=20] table [x index=10, y index=11] {data/tx_cap.csv};
\nextgroupplot[]
\addplot+ [mark repeat=20] table [x index=14, y index=15] {data/cfp.csv};
\addplot+ [mark repeat=20] table [x index=2, y index=3] {data/cfp.csv};
\addplot+ [mark repeat=20] table [x index=8, y index=9] {data/cfp.csv};
\nextgroupplot[yticklabels={}]
\addplot+ [mark repeat=20] table [x index=12, y index=13] {data/cfp.csv};
\addplot+ [mark repeat=20] table [x index=0, y index=1] {data/cfp.csv};
\addplot+ [mark repeat=20] table [x index=6, y index=7] {data/cfp.csv};
\nextgroupplot[yticklabels={}]
\addplot+ [mark repeat=20] table [x index=16, y index=17] {data/cfp.csv};
\addplot+ [mark repeat=20] table [x index=4, y index=5] {data/cfp.csv};
\addplot+ [mark repeat=20] table [x index=10, y index=11] {data/cfp.csv};
\end{groupplot}
\node[anchor=base,rotate=-90,yshift=0.275cm] at (group c3r1.east) {\textbf{\acrshort{cap}}};
\node[anchor=base,rotate=-90,yshift=0.275cm] at (group c3r2.east) {\textbf{\acrshort{cfp}}};
\node[anchor=base,rotate=90,yshift=1cm] at (group c1r1.south west) {CDF};
\node[anchor=base,rotate=0,yshift=-1cm] at (group c2r2.south) {Transmission delay [s]};
\end{tikzpicture}
     \caption{Transmission delay for unconfirmed transmissions during the \gls{cap} (\gls{csma}) and \gls{cfp} (\gls{gts}) with a varying number (N) of source devices and transmission intervals.}
    \label{fig:data_transmission}
\end{figure}
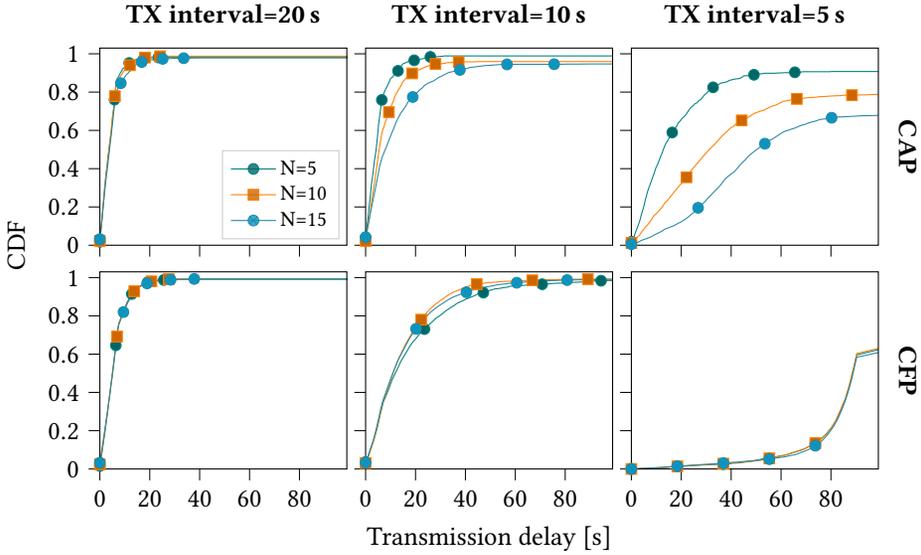

\autoref{fig:data_transmission} shows the distribution of transmission
delay for unconfirmed transmissions during \gls{cap} (\gls{csma}) and \gls{cfp} (\gls{gts}), for different
network sizes and transmission interval. The intersections between each curve
and the right axis reflect the \gls{prr}

\paragraph{\gls{csma} transmission}
The transmission delay increases with the network size and
lower transmission intervals. In both cases, the on-air traffic increases, which leads to enhanced wireless interference.
As a result, the \gls{cca} procedure faces more often a busy channel which increments the number of \gls{csma} backoff periods per transmission.
This effect increases the delay between packet schedule and the
actual transmission and causes a higher transmission delay.
Similarly, the \gls{prr} decreases with the network size and lower transmission intervals.
Multiple \gls{cca} failures enhance the probability of exceeding the maximum number of \gls{cca}
retries.
The \acrshort{mac} drops the packet in such case, hence, it decreases the \gls{prr}.
Note that due to inaccuracies of the \gls{cca} procedure, the \acrshort{mac} transmits a fraction of packets
even when the channel is busy which decreases the \gls{prr} marginally. We further analyze this
effect in~\autoref{sec:csma_ca}.
Higher transmission delays increase the \gls{cap} queue stress, since packets have to be buffered until they are actually sent.
Hence,  a fraction of packet losses occur due to \gls{cap} queue overflows. The stressed scenario (TX interval=5\,s) reflects this
situation, in which an increasing number of senders decrease the reception ratio.

\paragraph{\gls{gts} transmission}
The transmission delay increases with a lower transmission
interval during \gls{cfp}, but it does not vary with the network size. In contrast to the \gls{csma} scenario,
each source device transmits data during a dedicated time slot which repeats every
$T_{msf} \approx $7.68\,s.
In the advent of packet queuing for a particular sink device,
the last queued packet delays until the \acrshort{mac} transmitted all preceding frames. As a consequence,
a lower transmission interval (TX interval=5\,s) increases the transmission delay, by increasing the
average queue occupation -- introducing \acrshort{mac} queue stress. This situation
explains a larger transmission delay in \gls{csma} than in \gls{gts}, most notable
in scenarios with TX interval=5\,s/10\,s, even with
high backoff exponent configuration for \gls{csma}.
Note that the network size only affects the number of allocated time slots during one multisuperframe. Consequently,
the network size does not affect the transmission delay as long as a sufficient number of
\gls{gts} in the multisuperframe structure exist.

\paragraph{Effect of retransmissions}
\begin{figure}
    \tikzsetnextfilename{data_transmission_ackreq}

\begin{tikzpicture}
\definecolor{color0}{rgb}{0.905882352941176,0.16078431372549,0.541176470588235}
\definecolor{color1}{rgb}{0.850980392156863,0.372549019607843,0.00784313725490196}
\definecolor{color2}{rgb}{0.458823529411765,0.43921568627451,0.701960784313725}
\definecolor{color3}{rgb}{0.105882352941176,0.619607843137255,0.466666666666667}
\definecolor{color4}{rgb}{0.3,0.650980392156863,0.117647058823529}
\definecolor{color5}{rgb}{0.4,0.650980392156863,0.117647058823529}
\definecolor{color6}{rgb}{0.6,0.650980392156863,0.117647058823529}
\begin{groupplot}[
group style={
  group size=3 by 2,
  horizontal sep=0.25cm,
  vertical sep=0.35cm,
},
height=0.3\textwidth,
width=0.35\textwidth,
xmax=99,
ymax=1.03,
axis on top,
yticklabel pos=left,
xmin=0, ymin=0,
legend cell align={left},
legend style={
  fill opacity=0.8,
  draw opacity=1,
  text opacity=1,
  at={(0.97,0.03)},
  nodes={scale=0.8, transform shape},
  anchor=south east,
  draw=white!80!black,
},
tick align=outside,
tick pos=left,
x grid style={white!69.0196078431373!black},
xtick style={color=black},
y grid style={white!69.0196078431373!black},
ytick style={color=black},
cycle list name=exotic,
]
\nextgroupplot[title={\textbf{TX interval=20\,s}},xlabel={},xticklabels={}]
\addplot+[mark repeat=20] table [x index=2, y index=3] {data/tx_cap_ackreq.csv};
\addlegendentry{Confirmed}
\addplot+ [mark repeat=20] table [x index=2, y index=3] {data/tx_cap.csv};
\addlegendentry{Unconfirmed}
\nextgroupplot[title={\textbf{TX interval=10\,s}},xlabel={},xticklabels={},yticklabels={}]
\addplot+ [mark repeat=20] table [x index=0, y index=1] {data/tx_cap_ackreq.csv};
\addplot+ [mark repeat=20] table [x index=0, y index=1] {data/tx_cap.csv};
\nextgroupplot[title={\textbf{TX interval=5\,s}},xlabel={},xticklabels={},yticklabels={}]
\addplot+ [mark repeat=20] table [x index=4, y index=5] {data/tx_cap_ackreq.csv};
\addplot+ [mark repeat=20] table [x index=4, y index=5] {data/tx_cap.csv};
\nextgroupplot[]
\addplot+ [mark repeat=20] table [x index=8, y index=9] {data/tx_cfp_ackreq.csv};
\addplot+ [mark repeat=20] table [x index=2, y index=3] {data/cfp.csv};
\nextgroupplot[yticklabels={}]
\addplot+ [mark repeat=20] table [x index=6, y index=7] {data/tx_cfp_ackreq.csv};
\addplot+ [mark repeat=20] table [x index=0, y index=1] {data/cfp.csv};
\nextgroupplot[yticklabels={}]
\addplot+ [mark repeat=20] table [x index=4, y index=5] {data/tx_cfp_ackreq.csv};
\addplot+ [mark repeat=20] table [x index=4, y index=5] {data/cfp.csv};
\end{groupplot}
\node[anchor=base,rotate=-90,yshift=0.275cm] at (group c3r1.east) {\textbf{\acrshort{cap}}};
\node[anchor=base,rotate=-90,yshift=0.275cm] at (group c3r2.east) {\textbf{\acrshort{cfp}}};
\node[anchor=base,rotate=90,yshift=1cm] at (group c1r1.south west) {CDF};
\node[anchor=base,rotate=0,yshift=-1cm] at (group c2r2.south) {Transmission delay [s]};
\end{tikzpicture}
     \caption{Comparison of transmission delays for confirmed and unconfirmed transmissions during \gls{cap} (\gls{csma}) and \gls{cfp} (\gls{gts}), for ten source devices and varying transmission intervals.}
    \label{fig:data_transmission_ackreq}
\end{figure}
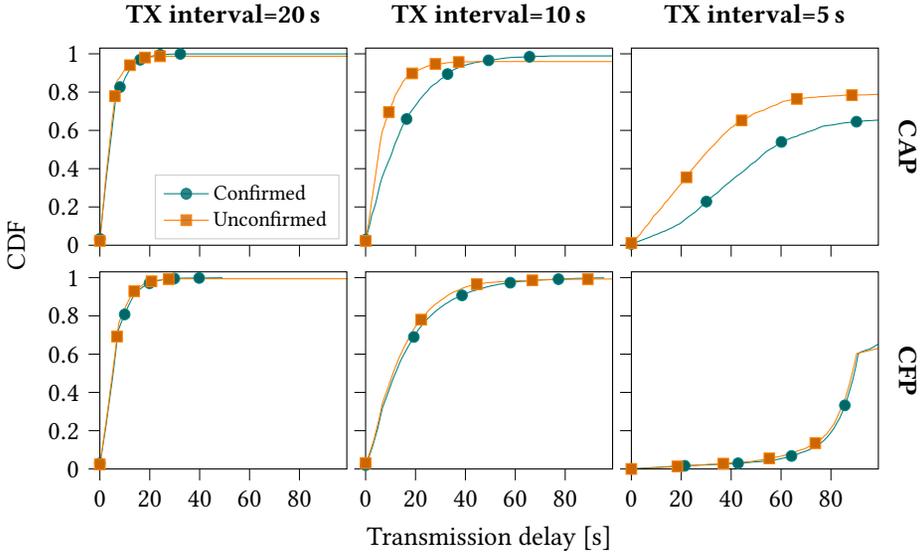
\autoref{fig:data_transmission_ackreq} compares the delay and \gls{prr}
for confirmed and unconfirmed transmissions during \gls{cap} and \gls{cfp}.
Confirmed frames under transmission stay in the \acrshort{mac} queue until
the reception of a valid ACK frame. In case of packet loss, the \acrshort{mac} retransmits
a pending frame until reception of the ACK frame or running out of retransmission
attempts.
Following our preceding measurements, we vary transmission intervals in a network of ten source devices.
Confirmed transmissions during \gls{cap}
reveal a higher transmission delay in relaxed scenarios (TX interval=10\,s/20\,s), \ie 90\% of confirmed packets finish within 40\,s, whereas the same amount of unconfirmed packets finish in less than 20\,s.
Therefore, the reception ratio increases from 95\% to 100\% with confirmed traffic.
In the stressed scenario (TX interval=5\,s), the transmission delay of the confirmed scenario
increases as well, while the \gls{prr} decreases in comparison to the unconfirmed scenario. Two causes are worth stressing: 
\one retransmissions increase the on-air traffic, which leads to collisions and a high number of \gls{cca} failures;
\two frames in retransmit occupy the \acrshort{mac} queue for
a longer time and are dropped occasionally due to \gls{cap} queue overflow.

In the \gls{cfp}, confirmed packets improve the \gls{prr} by only $\approx$ 0.5\% to achieve 100\% success.
Similar to the \gls{csma} scenario, frame retransmissions increase the probability
of packet reception, however, since \gls{gts} transmissions are exclusive,
retransmits are barely required.
Therefore, the contribution of frame retransmissions to \acrshort{mac} queue stress is
negligible.
Only a few retransmitted frames slightly increase the transmission delay.
This effect is notable in the scenario with TX interval=10\,s.
In contrast to the stressed \gls{csma} scenario, the queue load in the stressed \gls{gts} scenario (TX interval=5\,s)
with confirmed transmissions is similar to unconfirmed transmissions. Hence, the \gls{prr} does not decrease any further.

\subsection{Effect of \gls{cad} under different \gls{csma} configurations}\label{sec:csma_ca}

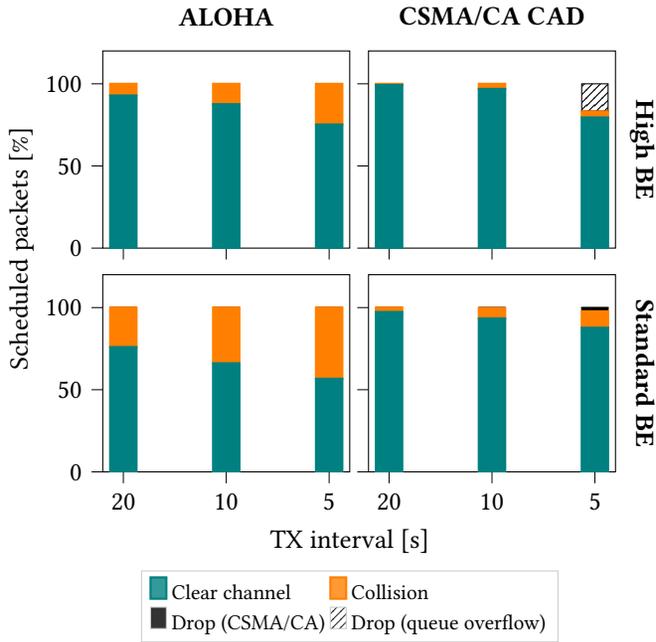
\begin{figure}
    \tikzsetnextfilename{cadfx}
\begin{tikzpicture}
\definecolor{color0}{rgb}{0.905882352941176,0.16078431372549,0.541176470588235}
\definecolor{color1}{rgb}{0.850980392156863,0.372549019607843,0.00784313725490196}
\definecolor{color2}{rgb}{0.458823529411765,0.43921568627451,0.701960784313725}
\definecolor{color3}{rgb}{0.105882352941176,0.619607843137255,0.466666666666667}
\definecolor{color4}{rgb}{0.4,0.650980392156863,0.117647058823529}
\begin{groupplot}[
group style={
  group size=2 by 2,
  horizontal sep=0.25cm,
  vertical sep=0.35cm,
},
height=0.3\textwidth,
width=0.35\textwidth,
xtick={5,10,20},
ymin=0,
legend cell align={left},
legend columns=2,
legend style={
  fill opacity=0.8,
  draw opacity=1,
  text opacity=1,
  at={(1,-0.5)},
  nodes={scale=0.8, transform shape},
  anchor=north,
  draw=white!80!black,
},
tick align=outside,
tick pos=left,
x grid style={white!69.0196078431373!black},
xtick style={color=black},
y grid style={white!69.0196078431373!black},
ytick style={color=black},
xminorticks=true,
ybar stacked,
/pgf/bar width=10pt,
xtick=data,
ymax=120,
x dir=reverse,
symbolic x coords={5,10,20},
cycle list name=exotic_bars,
]
\nextgroupplot[xticklabels={},xlabel={},title={\textbf{ALOHA}}]
\addplot+ [] table [y index=7] {data/capfx.csv};
\addplot+ [] table [y index=2] {data/capfx.csv};
\addplot [fill=black] table [y index=9] {data/capfx.csv};
\addplot [pattern=north east lines] table [y index=11] {data/capfx.csv};
\nextgroupplot[xticklabels={},xlabel={},yticklabels={}, title={\textbf{\acrshort{csma} \acrshort{cad}}}]
\addplot+ [] table [y index=6] {data/capfx.csv};
\addplot+ [] table [y index=1] {data/capfx.csv};
\addplot [fill=black] table [y index=8] {data/capfx.csv};
\addplot [pattern=north east lines] table [y index=10] {data/capfx.csv};
\nextgroupplot[]
\addplot+ [] table [y index=7] {data/capfx_standard.csv};
\addlegendentry{Clear channel}
\addplot+ [] table [y index=2] {data/capfx_standard.csv};
\addlegendentry{Collision}
\addplot [fill=black] table [y index=9] {data/capfx_standard.csv};
\addlegendentry{Drop (\acrshort{csma})}
\addplot [pattern=north east lines] table [y index=11] {data/capfx_standard.csv};
\addlegendentry{Drop (queue overflow)}
\nextgroupplot[yticklabels={}]
\addplot+ [] table [y index=6] {data/capfx_standard.csv};
\addplot+ [] table [y index=1] {data/capfx_standard.csv};
\addplot [fill=black] table [y index=8] {data/capfx_standard.csv};
\addplot [pattern=north east lines] table [y index=10] {data/capfx_standard.csv};
\end{groupplot}
\node[anchor=base,rotate=-90,yshift=0.275cm] at (group c2r1.east) {\textbf{High BE}};
\node[anchor=base,rotate=-90,yshift=0.275cm] at (group c2r2.east) {\textbf{Standard BE}};
\node[anchor=base,rotate=90,yshift=1cm] at (group c1r1.south west) {Scheduled packets [\%]};
\node[anchor=base,rotate=0,yshift=-1cm] at (group c1r2.south east) {TX interval [s]};
\end{tikzpicture}
     \caption{Proportion of \gls{cap} packets that face a clear channel on transmission, collide, or are dropped by \gls{csma}. Results are separated into high BE and standard BE \gls{csma} configurations with and without \gls{cad}, for ten source devices and varying transmission intervals.}
    \label{fig:cadfx}
\end{figure}

\paragraph{Effect on collisions}
We evaluate the collision avoidance capabilities of the \gls{cad} feature for transmissions during the \gls{cap} and compare to the ALOHA protocol, \ie randomized delay before transmissions. Thereby, we utilize two timing parameters sets for \gls{csma} with \gls{cap} and the initial ALOHA backoff.
\begin{enumerate}
\item \textbf{High BE}: our default choice (compare~\autoref{sec:methodology}) with \texttt{macMinBE}=7, \texttt{macMaxCsmaBackoff}=5, \texttt{macMaxBE}=8.
\item \textbf{Standard BE}: 802.15.4e standard values (for radios that operate in the 2.4GHz band) with \texttt{macMinBE}=3, \texttt{macMaxCsmaBackoff}=4, \texttt{macMaxBE}=5.
\end{enumerate}
\autoref{fig:cadfx} displays the fraction of packets that face a clear channel, collide, are dropped due to maximum number of \gls{csma} reattempts or queue overflow. Naturally, the latter options do not occur using ALOHA.
In the ALOHA scenario (\autoref{fig:cadfx} left), the results show that collisions increase with a lower transmission interval.
Due to higher traffic on air, the chances of packet collision increase
up to 38\% with ten nodes and TX interval=5\,s in the standard BE scenario (\autoref{fig:cadfx}, bottom left).
Note that scenarios with
standard BE \gls{csma} settings show higher collision rates than scenarios with high BE
settings (\autoref{fig:cadfx} left bottom vs top). The higher backoff exponent  increases the initial TX delay, hence, the average transmission interval, and thereby reduces the probability of collisions.
The number of transmitted packets stays constant (100\%) in all ALOHA scenarios, regardless of interference or busy channel.
This is because the \acrshort{mac} always assumes a free channel at the end of the
backoff period and unconditionally transmits the frame.

The number of collisions in the \gls{csma} \gls{cad} scenario (\autoref{fig:cadfx} right), is smaller than in the
ALOHA scenario and increases at a lower rate with a lower transmission interval. In contrast to ALOHA, the number of transmitted packets decreases with a lower transmission interval.
This is the positive effect of channel sensing indicated by \gls{cca} failures, which avoids sending during ongoing transmissions on the channel.
Packets that are delayed due to \gls{cca} remain queued and the \acrshort{mac} drops a pending frame if \gls{csma} runs out of retries. This decreases the number of transmitted frames in a stressful scenario (\gls{csma} standard BE, TX interval=5\,s).
It is worth noting that \gls{cad} is affected by inaccuracies; 
it detects a clear channel if two nodes start \gls{csma} about simultaneously. As a result,
a fraction of packets collides despite \gls{cca}.
Similar to the ALOHA scenario, high BE \gls{csma} settings trigger fewer collisions than standard BE settings.
A reduction in transmission rate due to higher backoff delays relaxes the channel, though,
a negative side effect is additional \gls{cap} queue load which increases packet losses due to occasional queue overflows (\gls{csma} high BE, TX interval=5\,s).

\begin{figure}
    \tikzsetnextfilename{capfx_ct}

\begin{tikzpicture}
\definecolor{color0}{rgb}{0.905882352941176,0.16078431372549,0.541176470588235}
\definecolor{color1}{rgb}{0.850980392156863,0.372549019607843,0.00784313725490196}
\definecolor{color2}{rgb}{0.458823529411765,0.43921568627451,0.701960784313725}
\definecolor{color3}{rgb}{0.105882352941176,0.619607843137255,0.466666666666667}
\definecolor{color4}{rgb}{0.4,0.650980392156863,0.117647058823529}
\begin{groupplot}[
group style={
  group size=3 by 2,
  horizontal sep=0.25cm,
  vertical sep=0.35cm,
},
height=0.3\textwidth,
width=0.35\textwidth,
axis on top,
xmin=2, ymin=0,
xmax = 80, ymax=1,
ytick = {0,0.25,0.5,0.75,1},
legend cell align={left},
legend style={
  fill opacity=0.8,
  draw opacity=1,
  text opacity=1,
  at={(0.97,0.05)},
  nodes={scale=0.8, transform shape},
  anchor=south east,
  draw=white!80!black,
},
tick align=outside,
tick pos=left,
x grid style={white!69.0196078431373!black},
xtick style={color=black},
y grid style={white!69.0196078431373!black},
ytick style={color=black},
cycle list name=exotic,
]
\nextgroupplot[title={\textbf{TX interval=20\,s}},xmin=0,xmax=29,xlabel={},xticklabels={}]
    \addplot+ [mark repeat=20
        ] table [x index=4, y index=5] {data/capfx_ct.csv};
\addlegendentry{High BE}
\addplot+ [mark repeat=20
] table [x index=2, y index=3] {data/tx_cap_standard_nocad.csv};
\addlegendentry{Standard BE}
\nextgroupplot[xlabel={}, xticklabels={},yticklabels={}, title={\textbf{TX interval=10\,s}},xmin=0,xmax=42]
\addplot+ [mark repeat=20
] table [x index=0, y index=1] {data/capfx_ct.csv};
\addplot+ [mark repeat=20
] table [x index=0, y index=1] {data/tx_cap_standard_nocad.csv};
\nextgroupplot[yticklabels={},xticklabels={},title={\textbf{TX interval=5\,s}},xmin=0,xlabel={}]
\addplot+ [mark repeat=20
] table [x index=8, y index=9] {data/capfx_ct.csv};
\addplot+ [mark repeat=20
] table [x index=4, y index=5] {data/tx_cap_standard_nocad.csv};
\nextgroupplot[xmin=0,xmax=29]
\addplot+ [mark repeat=20] table [x index=6, y index=7] {data/capfx_ct.csv};
\addplot+ [mark repeat=20] table [x index=2, y index=3] {data/tx_cap_standard.csv};
\nextgroupplot[yticklabels={},xmin=0,xmax=42]
\addplot+ [mark repeat=20] table [x index=2, y index=3] {data/capfx_ct.csv};
\addplot+ [mark repeat=20] table [x index=0, y index=1] {data/tx_cap_standard.csv};
\nextgroupplot[yticklabels={},xmin=0]
\addplot+ [mark repeat=20] table [x index=10, y index=11] {data/capfx_ct.csv};
\addplot+ [mark repeat=20] table [x index=4, y index=5] {data/tx_cap_standard.csv};
\end{groupplot}
\node[anchor=base,rotate=-90,yshift=0.275cm] at (group c3r1.east) {\textbf{ALOHA}};
\node[anchor=base,rotate=-90,yshift=0.275cm] at (group c3r2.east) {\textbf{\acrshort{csma} \acrshort{cad}}};
\node[anchor=base,rotate=90,yshift=1cm] at (group c1r1.south west) {CDF};
\node[anchor=base,rotate=0,yshift=-1cm] at (group c2r2.south) {Transmission delay [s]};
\end{tikzpicture}
     \caption{Comparison of transmission delays between \gls{cap} transmissions with \gls{csma} \gls{cad} and ALOHA for ten source devices, with varying transmission intervals
    and \gls{csma} backoff exponent (BE) configurations.}
    \label{fig:capfx_ct}
\end{figure}
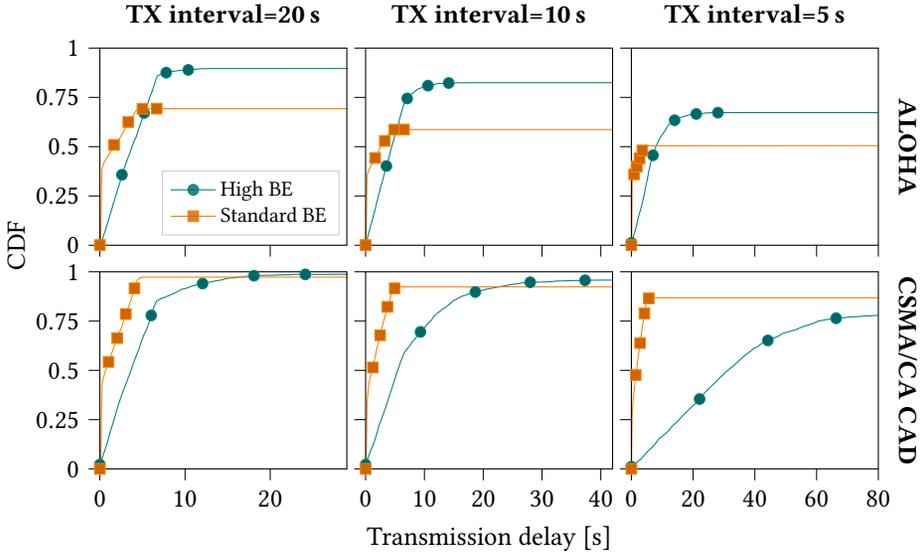

\paragraph{Effect on transmission delay}
The results in~\autoref{fig:capfx_ct} show that the transmission delay in the \gls{csma}
scenario increases with decreasing TX intervals.
Increased channel access failures with \gls{csma}  delay the
transmissions (until \gls{cca} reports clear channel). Hence, the average transmission
delay increases.
Increasing TX intervals with ALOHA do not affect the transmission delay, since sending is independent of the channel state.
ALOHA therefore suffers from wireless interference.
In all cases, \gls{csma} leads to a higher \gls{prr} than ALOHA transmission. Although
\gls{csma} with \gls{cad} reduces the proportion of transmitted packets, the number of non-transmitted
packets (which avoided a collision) is smaller than collisions upfront. As a result, the \gls{prr} increases.
Scenarios with high BE \gls{csma} settings reveal higher packet reception ratios,
as a result of reduced collisions and higher transmission delay as a result of
higher backoff delay.

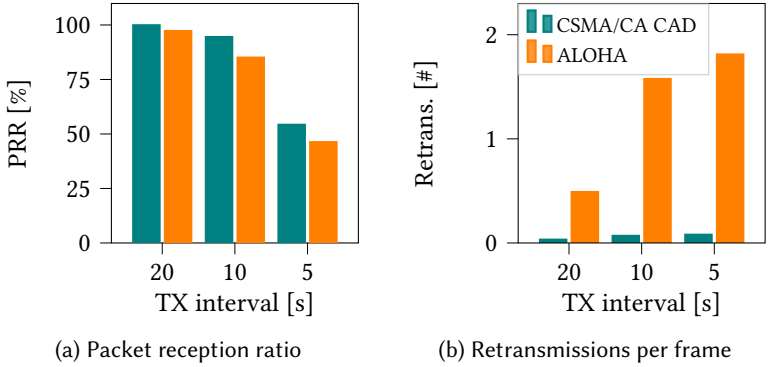
\begin{figure}
	\subfloat[Packet reception ratio]{\tikzsetnextfilename{capfx_tx_prr}

\begin{tikzpicture}
\definecolor{color0}{rgb}{0.905882352941176,0.16078431372549,0.541176470588235}
\definecolor{color1}{rgb}{0.850980392156863,0.372549019607843,0.00784313725490196}
\definecolor{color2}{rgb}{0.458823529411765,0.43921568627451,0.701960784313725}
\definecolor{color3}{rgb}{0.105882352941176,0.619607843137255,0.466666666666667}
\definecolor{color4}{rgb}{0.4,0.650980392156863,0.117647058823529}
\begin{axis}[
height=0.34\textwidth,
width=0.35\textwidth,
legend cell align={left},
legend style={
  fill opacity=0.8,
  draw opacity=1,
  text opacity=1,
  at={(0.9,1)},
  nodes={scale=0.8, transform shape},
  anchor=south east,
  draw=white!80!black,
},
xmin=-0.7, ymin=0,
xmax=2.7,
/pgf/bar width=10pt,
tick align=outside,
tick pos=left,
x grid style={white!69.0196078431373!black},
xtick style={color=black},
y grid style={white!69.0196078431373!black},
ytick style={color=black},
ybar, 
xlabel={TX interval [s]},
xtick={0,1,2},
xticklabels={20,10,5},
ylabel={\acrshort{prr} [\%]},
ytick={0,0.25,0.5,0.75,1},
yticklabels={0,25,50,75,100},
cycle list name=exotic_bars,
]

\addplot+ [] table [x index=0, y index=10] {data/capfx_tx.csv};
\addplot+ [] table [x index=0, y index=10] {data/capfx_tx_no_cad.csv};
\end{axis}
\end{tikzpicture}
 }
    \hspace{1em}
    \subfloat[Retransmissions per frame]{\tikzsetnextfilename{capfx_tx_retrans}

\begin{tikzpicture}
\definecolor{color0}{rgb}{0.905882352941176,0.16078431372549,0.541176470588235}
\definecolor{color1}{rgb}{0.850980392156863,0.372549019607843,0.00784313725490196}
\definecolor{color2}{rgb}{0.458823529411765,0.43921568627451,0.701960784313725}
\definecolor{color3}{rgb}{0.105882352941176,0.619607843137255,0.466666666666667}
\definecolor{color4}{rgb}{0.4,0.650980392156863,0.117647058823529}
\begin{axis}[
height=0.34\textwidth,
width=0.35\textwidth,
axis on top,
xmin=-0.7, ymin=0,
xmax=2.7,
legend cell align={left},
legend style={
  fill opacity=0.8,
  draw opacity=1,
  text opacity=1,
  at={(0,1)},
  nodes={scale=0.8, transform shape},
  anchor=north west,
  draw=white!80!black,
},
tick align=outside,
tick pos=left,
x grid style={white!69.0196078431373!black},
xtick style={color=black},
y grid style={white!69.0196078431373!black},
ytick style={color=black},
/pgf/bar width=10pt,
ybar,
xlabel={TX interval [s]},
xtick={0,1,2},
xticklabels={20,10,5},
ymax=2.3,
cycle list name=exotic_bars,
ylabel={Retrans. [\#]}
]
\addplot+ [] table [x index=0, y index=11] {data/capfx_tx.csv};
\addlegendentry{\acrshort{csma} \acrshort{cad}}
\addplot+ [] table [x index=0, y index=11] {data/capfx_tx_no_cad.csv};
\addlegendentry{ALOHA}
\end{axis}
\end{tikzpicture}
 \label{fig:capfx_tx_retrans}}
    \caption{Packet reception ratio (left) and average retransmissions per frame (right) for fifteen source devices and varying transmission intervals.}
    \label{fig:capfx_tx_prr}
\end{figure}

\paragraph{Effect on retransmissions}\label{sec:eff_cad_retrans}
 \autoref{fig:capfx_tx_prr} analyzes the effect of using \gls{cad} when enabling confirmable traffic and retransmissions.
We compare the \gls{prr} and the average number of retransmissions per packet (Figure~\autoref{fig:capfx_tx_retrans}) for both the \gls{csma} with \gls{cad} and ALOHA scenarios.
The \gls{prr} decreases with higher TX intervals and \gls{csma} \gls{cad} measurements outperform ALOHA.
The effect is most notable in the stressed scenario (TX interval=5\,s) where the difference amounts to $\approx$ 8\%.
Retransmissions remain rare with \gls{csma} which indicates that losses are mainly caused by avoided transmissions in stressed cases.
Nevertheless, \gls{csma} outperforms ALOHA in terms of reception ratio.
In contrast, nodes retransmit every packet up to 1.5x (on average) using ALOHA, without improving packet reception.
This useless amount of retransmissions demonstrates the advantage of \gls{csma} with \gls{cad}.

\subsection{Time on air and duty cycle compliance}\label{sec:time_on_air}

\begin{figure}
    \subfloat[Source device]{\tikzsetnextfilename{duty_cycle_sensor}

\begin{tikzpicture}
\definecolor{color0}{rgb}{0.905882352941176,0.16078431372549,0.541176470588235}
\definecolor{color1}{rgb}{0.850980392156863,0.372549019607843,0.00784313725490196}
\definecolor{color2}{rgb}{0.458823529411765,0.43921568627451,0.701960784313725}
\definecolor{color3}{rgb}{0.105882352941176,0.619607843137255,0.466666666666667}
\definecolor{color4}{rgb}{0.4,0.650980392156863,0.117647058823529}
\begin{groupplot}[
group style={
  group size=2 by 2,
  horizontal sep=0.5cm,
  vertical sep=0.35cm,
},
height=0.27\textwidth,
width=0.27\textwidth,
xmin=-0.5, ymin=0,
xmax=1.5,ymax=80,
legend cell align={left},
legend style={
  fill opacity=0.8,
  draw opacity=1,
  text opacity=1,
  at={(0.9,0.65)},
  nodes={scale=0.8, transform shape},
  anchor=south east,
  draw=white!80!black,
},
tick align=outside,
tick pos=left,
x grid style={white!69.0196078431373!black},
xtick style={color=black},
y grid style={white!69.0196078431373!black},
ytick style={color=black},
xtick={0,1},
xticklabels={N=5,N=15},
ytick={0,20,40,60,80},
]

\nextgroupplot[title={\textbf{TX interval=10\,s}},xticklabels={},xlabel={}]
\addplot [thick, color1, only marks,mark size=1] table [x index=16, y index=10] {data/duty_cycle.csv};
\addplot [thick, color1, only marks,mark size=1] table [x index=17, y index=2] {data/duty_cycle.csv};

\nextgroupplot[title={\textbf{TX interval=5\,s}},yticklabels={}, xticklabels={},xlabel={}]
\addplot [thick, color1, only marks,mark size=1] table [x index=16, y index=14] {data/duty_cycle.csv};
\addplot [thick, color1, only marks,mark size=1] table [x index=17, y index=6] {data/duty_cycle.csv};

\nextgroupplot[]
\addplot [thick, color1, only marks,mark size=1] table [x index=16, y index=8] {data/duty_cycle.csv};
\addplot [thick, color1, only marks,mark size=1] table [x index=17, y index=0] {data/duty_cycle.csv};
\path [draw=black, draw opacity=0.3, semithick, dash pattern=on 5.55pt off 2.4pt]
(axis cs:-1,36)
--(axis cs:4,36) node [pos=0.24, text opacity=0.5, anchor=south] {1\%};

\nextgroupplot[yticklabels={}]
\addplot [thick, color1, only marks,mark size=1] table [x index=16, y index=12] {data/duty_cycle.csv};
\addplot [thick, color1, only marks,mark size=1] table [x index=17, y index=4] {data/duty_cycle.csv};
\path [draw=black, draw opacity=0.3, semithick, dash pattern=on 5.55pt off 2.4pt]
(axis cs:-1,36)
--(axis cs:4,36) node [pos=0.24, text opacity=0.5, anchor=south] {1\%};

\end{groupplot}
\node[anchor=base,rotate=90,yshift=1cm] at (group c1r1.south west) {Time on air [s]};
\end{tikzpicture}
 \label{fig:duty_cycle_sensor}}
    \hspace{1em}
    \subfloat[Sink device]{\tikzsetnextfilename{duty_cycle_actuator}

\begin{tikzpicture}
\definecolor{color0}{rgb}{0.905882352941176,0.16078431372549,0.541176470588235}
\definecolor{color1}{rgb}{0.850980392156863,0.372549019607843,0.00784313725490196}
\definecolor{color2}{rgb}{0.458823529411765,0.43921568627451,0.701960784313725}
\definecolor{color3}{rgb}{0.105882352941176,0.619607843137255,0.466666666666667}
\definecolor{color4}{rgb}{0.4,0.650980392156863,0.117647058823529}
\begin{groupplot}[
group style={
  group size=2 by 2,
  horizontal sep=0.5cm,
  vertical sep=0.35cm,
},
height=0.27\textwidth,
width=0.27\textwidth,
xmin=-0.5, ymin=0,
xmax=1.5,ymax=80,
legend cell align={left},
legend style={
  fill opacity=0.8,
  draw opacity=1,
  text opacity=1,
  at={(0.9,0.65)},
  nodes={scale=0.8, transform shape},
  anchor=south east,
  draw=white!80!black,
},
tick align=outside,
tick pos=left,
x grid style={white!69.0196078431373!black},
xtick style={color=black},
y grid style={white!69.0196078431373!black},
ytick style={color=black},
xtick={0,1},
xticklabels={N=5,N=15},
ytick={0,20,40,60,80},
]

\nextgroupplot[title={\textbf{TX interval=10\,s}},xlabel={},xticklabels={}]
\addplot [thick, color1, only marks,mark size=1] table [x index=16, y index=11] {data/duty_cycle.csv};
\addplot [thick, color1, only marks,mark size=1] table [x index=17, y index=3] {data/duty_cycle.csv};

\nextgroupplot[yticklabels={},title={\textbf{TX interval=5\,s}},xlabel={},xticklabels={}]
\addplot [thick, color1, only marks,mark size=1] table [x index=16, y index=15] {data/duty_cycle.csv};
\addplot [thick, color1, only marks,mark size=1] table [x index=17, y index=7] {data/duty_cycle.csv};

\nextgroupplot[]
\addplot [thick, color1, only marks,mark size=1] table [x index=16, y index=9] {data/duty_cycle.csv};
\addplot [thick, color1, only marks,mark size=1] table [x index=17, y index=1] {data/duty_cycle.csv};
\path [draw=black, draw opacity=0.3, semithick, dash pattern=on 5.55pt off 2.4pt]
(axis cs:-1,36)
--(axis cs:4,36) node [pos=0.24, text opacity=0.5, anchor=south] {1\%};

\nextgroupplot[yticklabels={}]
\addplot [thick, color1, only marks,mark size=1] table [x index=16, y index=13] {data/duty_cycle.csv};
\addplot [thick, color1, only marks,mark size=1] table [x index=17, y index=5] {data/duty_cycle.csv};
\path [draw=black, draw opacity=0.3, semithick, dash pattern=on 5.55pt off 2.4pt]
(axis cs:-1,36)
--(axis cs:4,36) node [pos=0.24, text opacity=0.5, anchor=south] {1\%};

\end{groupplot}
\node[anchor=base,rotate=-90,yshift=0.275cm] at (group c2r1.east) {\textbf{\acrshort{cap}}};
\node[anchor=base,rotate=-90,yshift=0.275cm] at (group c2r2.east) {\textbf{\acrshort{cfp}}};
\end{tikzpicture}
 \label{fig:duty_cycle_actuator}}
    \caption{Time on air of data frames on source devices (left) and ACK frames on sink devices (right) compared to 1\% band limitations (gray line). Frames are sent in \gls{cap} (\gls{csma}) or \gls{cfp} (\gls{gts}) and we vary the number of source devices (N) and transmission intervals.}
    \label{fig:duty_cycle}
\end{figure}
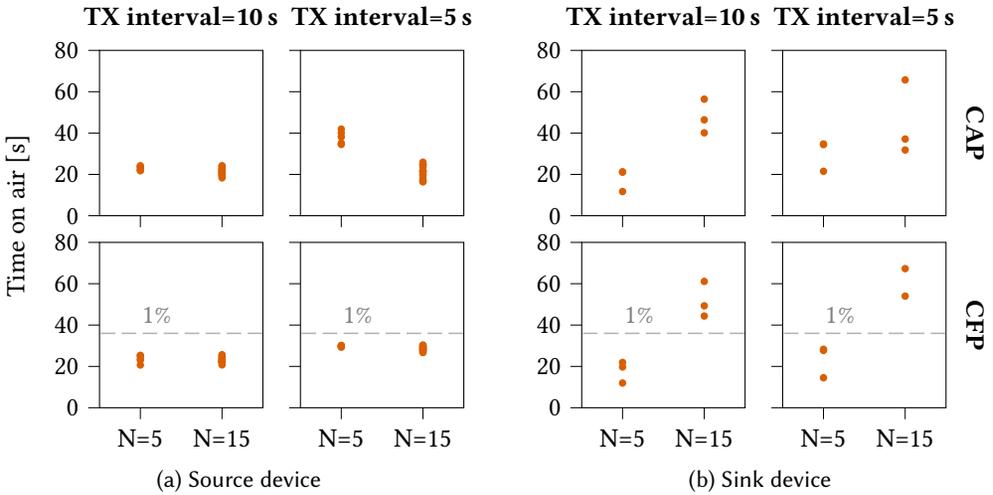

Transmission time with LoRa radios is limited by band regulations. We analyze the on-air time with respect to duty-cycle compliance. As depicted in~\autoref{sec:methodology}, we set up topologies in which a variable number of source devices sends data to three sink devices. The assignment is uniformly random. Sink devices, in turn, reply with an ACK to every incoming packet.
Note that a data frame contains 27 Bytes of data, whereas the ACK frame contains only 5 Bytes. Due to the LoRa \acrshort{phy} frame overhead, however, the ACK packet takes $\approx$ 31ms on air, which is around half of the data frame ($\approx$ 67ms).
We set up mostly stressed TX intervals to foster duty cycle violations and present our results of the time on air per node in~\autoref{fig:duty_cycle}. Thereby, the dashed gray line indicates the maximum on air time to comply with a 1\% band occupation during \gls{cfp}. For the \gls{cap} we utilize a 10\% of the band (see~\autoref{sec:mappings}) which is not visible on this scale.

\paragraph{\gls{csma} transmission}
The time on air increases with a lower transmission interval as a result of a higher transmission rate on the \acrshort{mac} (Figure~\autoref{fig:duty_cycle_sensor} top left to right).
For the N=5 case, the time almost doubles from $\approx$ 20\,s to 40\,s with half the TX interval. Note that we apply a 10\% band to the \gls{cap}, hence, on-air times of up to 360\,s per node do comply with duty cycle restrictions.

Our results further show that the average time on air of source devices decreases in larger networks (N increases). The effect is most notable in the scenario with TX interval=5\,s.
In that scenario, \gls{cca} fails more often due to higher on-air
traffic, which persuades the \acrshort{mac} to drop a fraction of data frames, either due to exceeded \gls{csma} attempts or overflowed \gls{cap} queue (see~\autoref{sec:csma_ca}).
The case with N=15 source devices does not vary  on air time with varying TX intervals due to these \acrshort{mac} drops.

In the case of sink devices, the time on air increases with a lower transmission interval, similarly to source devices. In  contrast, though, the time on air also increases with the network size.
ACK packets are sent in response to every incoming source device frame and do not utilize \gls{csma}.
Due to our topology choice, a sink device has to return multiple ACK packets to satisfy all its assigned source devices. Consequently, a higher number of source devices leads to a higher ACK frame transmission rate per sink device, which increases the time on air up to 60\,s for \,scenarios with N=15 source devices. This is still in line with 10\% restrictions.
Note that the random source-sink assignment leads to a different number of source devices per sink on each scenario, which introduces variations between time on air measurements across sink devices.

Sink devices send multiple ACK packets back by back, contrasting a `simultaneous' channel access of source devices which introduced \acrshort{mac} dropping.
Overall, increasing transmission rates have a less severe impact on node duty cycles than increasing number of nodes that try to access the medium during the same time period.

\paragraph{\gls{gts} transmission}
The average time on air of source devices increases with a lower TX interval, which is in agreement with \gls{csma} transmissions (Figure~\autoref{fig:duty_cycle_sensor} bottom left to right).
Our \gls{cfp} assignment with one \gls{gts} per source-sink link, however, limits the effective TX interval to $T_{msf}$=7.68\,s in our multisuperframe configuration as described in~\autoref{sec:methodology}.
Note that this configuration preserves duty cycle compliance natively. Sending a 67ms long packet every 7.68\,s results in 31.4\,s active send time per hour, which is below the 1\% regulation mark of 36\,s per device and hour. Conversely, we intentionally chose a very stressful measurement setup with TX interval=5\,s.

Similar to the relaxed \gls{csma} scenario, the average time on air of sink devices increases with a lower TX interval (Figure~\autoref{fig:duty_cycle_actuator} bottom left to right) due to an increased frame rate.
Increasing the network size, in contrast to \gls{csma}, further increases the on-air times of sink devices.
Observe that the time on air exceeds 1\% of duty cycle in scenarios with N=15. The reasons for this are threefold.
\one Due to our topology choice, each sink device has to confirm five source packets on average in the N=15 scenario. This amplification burdens the link budget of a single sink device. Hence, we deliberately violate the duty cycle regulations by our experiment setup.
\two Sink devices only send ACK frames back by back and without \gls{csma}. Hence, sink devices transmit 100\% of the scheduled ACK frames.
\three \gls{gts} transmissions utilize guaranteed resources, which increases the reception ratio and decreases losses in comparison to \gls{csma} transmissions. As a result, the number of transmitted ACK frames is in line with the number of transmitted data frames, regardless of the network size.

\subsection{Coexistence with LoRaWAN}\label{sec:coexistence}
\begin{figure}
    \tikzsetnextfilename{coexistency}

\begin{tikzpicture}
\definecolor{color0}{rgb}{0.905882352941176,0.16078431372549,0.541176470588235}
\definecolor{color1}{rgb}{0.850980392156863,0.372549019607843,0.00784313725490196}
\definecolor{color2}{rgb}{0.458823529411765,0.43921568627451,0.701960784313725}
\definecolor{color3}{rgb}{0.105882352941176,0.619607843137255,0.466666666666667}
\definecolor{color4}{rgb}{0.4,0.650980392156863,0.117647058823529}
\begin{groupplot}[
group style={
  group size=1 by 1,
  horizontal sep=0.25cm,
  vertical sep=0.25cm,
},
height=0.34\textwidth,
width=0.75\textwidth,
legend cell align={left},
legend style={
  fill opacity=0.8,
  draw opacity=1,
  text opacity=1,
  at={(0.9,1)},
  nodes={scale=0.8, transform shape},
  anchor=north east,
  draw=white!80!black,
},
/pgf/bar width=10pt,
tick align=outside,
tick pos=left,
x grid style={white!69.0196078431373!black},
xtick style={color=black},
y grid style={white!69.0196078431373!black},
ytick style={color=black},
ybar, 
xlabel={Scenario},
ylabel={\acrshort{prr} [\%]},
cycle list name=exotic_bars,
ymin = 90,
xtick=data,
]

\nextgroupplot[xticklabels={}]
\addplot+ [] table [x index=0, y index=1] {data/dsme.csv};
\addlegendentry{Baseline}
\addplot+ [] table [x index=0, y index=2] {data/dsme.csv};
\addlegendentry{With cross-traffic}
\end{groupplot}
\end{tikzpicture}
     \caption{Comparison of \gls{prr} for ten \gls{dsme}-LoRa source devices with and without
    cross traffic from a LoRaWAN network with ten class A devices.
    All devices transmit unconfirmed frames with 16 bytes payload and uniformly distributed interarrival times between 7 and 13\,s.
    \gls{dsme}-LoRa devices transmit during \gls{cfp} (\gls{gts}).}
    \label{fig:coexistence}
\end{figure}
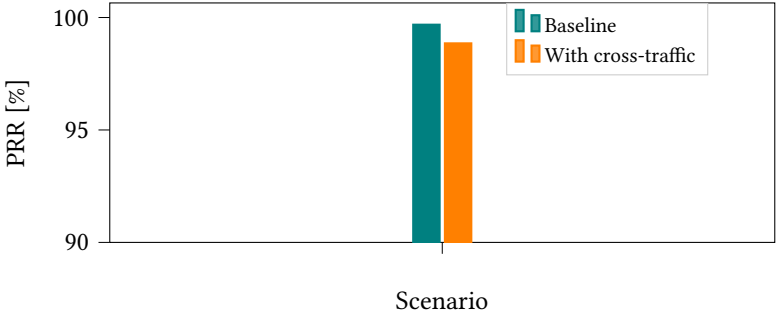

The proposed \gls{dsme}-LoRa \gls{phy} channels overlap with channels of LoRaWAN
networks. Therefore,
we are interested in the effects of LoRaWAN cross-traffic in \gls{dsme}-LoRa networks.
For this analysis we focus only on cross-traffic between \gls{gts} transmissions and LoRaWAN
traffic for two reasons: \one the common
\gls{cap} channel does not overlap with standard LoRaWAN uplink channels. \two LoRaWAN
downlink traffic is typically transmitted using
a higher spreading factor and thereby do not collide with \gls{dsme}-LoRa packets.

For the evaluation, we deploy \gls{dsme}-LoRa and LoRaWAN networks simultaneously and
measure the \gls{prr} of the \gls{dsme}-LoRa network. We compare these values against
the same \gls{dsme} network without cross-traffic.

For the LoRaWAN network, we set up ten nodes with class A transmissions and DR5 (spreading factor 7, bandwidth 125\,kHz).
The deployment uses a single 8-channel TTN~\cite{ttn} LoRaWAN gateway, available in the testbed.
The \gls{dsme}-LoRa network consist of ten source devices with the topology and
configuration in agreement with~\autoref{sec:methodology}.
All devices transmit 16 bytes payload using unconfirmed transmissions and uniformly distributed interarrival times between
7 and 13\,s.

\autoref{fig:coexistence} shows the \gls{prr} of the isolated \gls{dsme}-LoRa
network (baseline) and the network with LoRaWAN cross-traffic. 
With \gls{dsme} cross-traffic, the \gls{prr} reduces $\approx 0.7$\%. From all available \gls{dsme}-LoRa channels, only seven overlap
with TTN LoRaWAN channels, which means 56.25\% of transmissions are collision free.
A fraction of the remaining packets is transmitted concurrently with \gls{dsme}-LoRa
transmissions, which reflect the \gls{prr} reduction.
The LoRaWAN traffic does not collide with \gls{dsme} beacons and therefore, device
desynchronization as a result of collision between LoRaWAN packets and \gls{dsme}-LoRa
beacons is negligible.
Even though LoRaWAN traffic degrades \gls{prr} as a result of concurrent transmissions
on shared channels, the cross-traffic does not prevent normal operation of the \gls{dsme}-LoRa
network. We conclude that \gls{dsme}-LoRa traffic is compatible with standard LoRaWAN uplink traffic.

\subsection{Effect of interference in common channel}\label{sec:common_channel}
\autoref{sec:coexistence} confirmed that \gls{dsme}-LoRa networks tolerate channel
interference in \gls{gts} channels. However, the results do not reflect tolerance
to noise in the common channel used for \gls{cap} and beacon transmissions. 
While common LoRaWAN deployments in the EU868 region do not transmit in the 10\%
band (\gls{cap} channel) using the same \gls{phy} settings as \gls{dsme}-LoRa,
the LoRaWAN network server does not prevent the configuration of a downlink channel
using spreading factor 7, which may cause LoRaWAN frames to collide with \gls{dsme}-LoRa frames.
Since synchronization to the \gls{dsme} superframe structure relies
on beacons, we evaluate whether \gls{dsme}-LoRa can operate under noise in the common channel.
We focus only on \gls{gts} transmission, because the effect of noise during \gls{csma} transmissions
has already been analyzed in~\autoref{sec:data_transmission}.

We generate a harsh environment by deploying five jammer devices that send 127 bytes
payload data in the common channel with uniformly distributed interarrival times between 1 s
and 6 s, next to the \gls{dsme}-LoRa network with ten source devices. The interarrival
time of packets and the \gls{mac} configurations of the \gls{dsme}-LoRa network are
identical to the network in~\autoref{sec:coexistence}.

\begin{figure}
    \tikzsetnextfilename{nb_noise}

\begin{tikzpicture}
\definecolor{color0}{rgb}{0.905882352941176,0.16078431372549,0.541176470588235}
\definecolor{color1}{rgb}{0.850980392156863,0.372549019607843,0.00784313725490196}
\definecolor{color2}{rgb}{0.458823529411765,0.43921568627451,0.701960784313725}
\definecolor{color3}{rgb}{0.105882352941176,0.619607843137255,0.466666666666667}
\definecolor{color4}{rgb}{0.4,0.650980392156863,0.117647058823529}
\begin{axis}[
height=0.3\textwidth,
width=0.75\textwidth,
axis on top,
yticklabel pos=left,
xmin=0, ymin=0,
legend cell align={left},
legend style={
  fill opacity=0.8,
  draw opacity=1,
  text opacity=1,
  at={(0.0,0.03)},
  nodes={scale=0.8, transform shape},
  anchor=south west,
  draw=white!80!black,
},
tick align=outside,
tick pos=left,
x grid style={white!69.0196078431373!black},
xtick style={color=black},
y grid style={white!69.0196078431373!black},
ytick style={color=black},
cycle list name=exotic,
ylabel={PRR [\%]},
xlabel={Time [s]},
xmax=400,
ytick={0,20,40,60,80,100},
]
\addplot+ [mark repeat=20] table [x index=0, y index=1] {data/nb_1.csv};
\addlegendentry{R1}
\addplot+ [mark repeat=20] table [x index=0, y index=1] {data/nb_2.csv};
\addlegendentry{R2}
\addplot+ [mark repeat=20] table [x index=0, y index=1] {data/nb_3.csv};
\addlegendentry{R3}
\end{axis}
\end{tikzpicture}
     \caption{Evolution of \gls{prr} for three replicas of a \gls{dsme}-LoRa deployment with ten source devices, unconfirmed \gls{gts} transmissions
    and TX interval=10\,s, under heavy interference on the common channel.}
    \label{fig:nb_noise}
\end{figure}
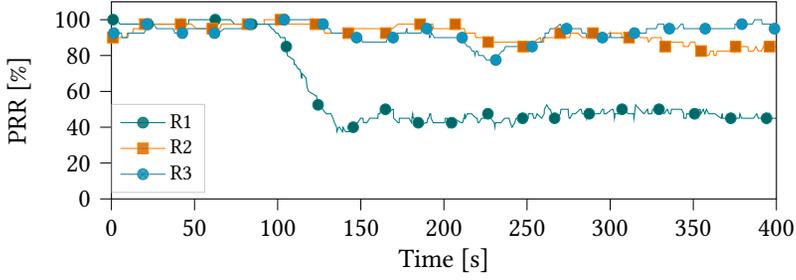

\autoref{fig:nb_noise} shows the moving average \gls{prr} over time for three replicas
(R1, R2 and R3).
The \gls{prr} of all replicas oscillates around 92\% before T=100 s.
After T=100 s the \gls{prr} in R3 reduces to $\approx 50$\% and does not recover.
Similarly, around T=270 s  the \gls{prr} in R2 drops to $\approx 89$\%.
To understand these results, observe that the common channel is included as one of the transmission channels
in \gls{cfp}. Therefore, \gls{gts} transmissions in the common channel
are likely to collide with traffic from the jammers. Assuming $\frac{1}{16}$
of \gls{gts} vulnerable transmissions  (\ie one channel gets jammed), around 94\% of transmission are collision
free. This reflects the lower average \gls{prr} before T=100 s.
The \gls{prr} drops in R2 and R3 are caused by the desynchronization of two sink devices and
a source device, respectively, as a result of beacon loss. When the \gls{mac} misses
a number of consecutive beacons (4 by default), the device disassociates and ignores all transmission request
and \gls{gts} reception slots. Therefore, incoming and outgoing packets are simply
discarded.

To summarize, interference on a single channel reduces the efficiency of \gls{gts}
transmissions, but does not prevent normal operation. Interference on the
common channel, however, increases beacon loss, which desynchronizes devices from coordinators.
While raising the threshold of consecutive missed beacons can delay desynchronization
on sporadic interference, it cannot solve the problem. Additionally, long time on air
and long range of LoRa frames make wireless attacks plausible, in which an attacker
blocks beacon reception through channel jamming. We discuss potential solutions
to this problem in~\autoref{sec:xtraffic}.

\subsection{Energy consumption}\label{sec:energy_consumption}

\newcommand{\energycsma}{S1\xspace}
\newcommand{\energygtstx}{S2\xspace}
\newcommand{\energygtsrx}{S3\xspace}

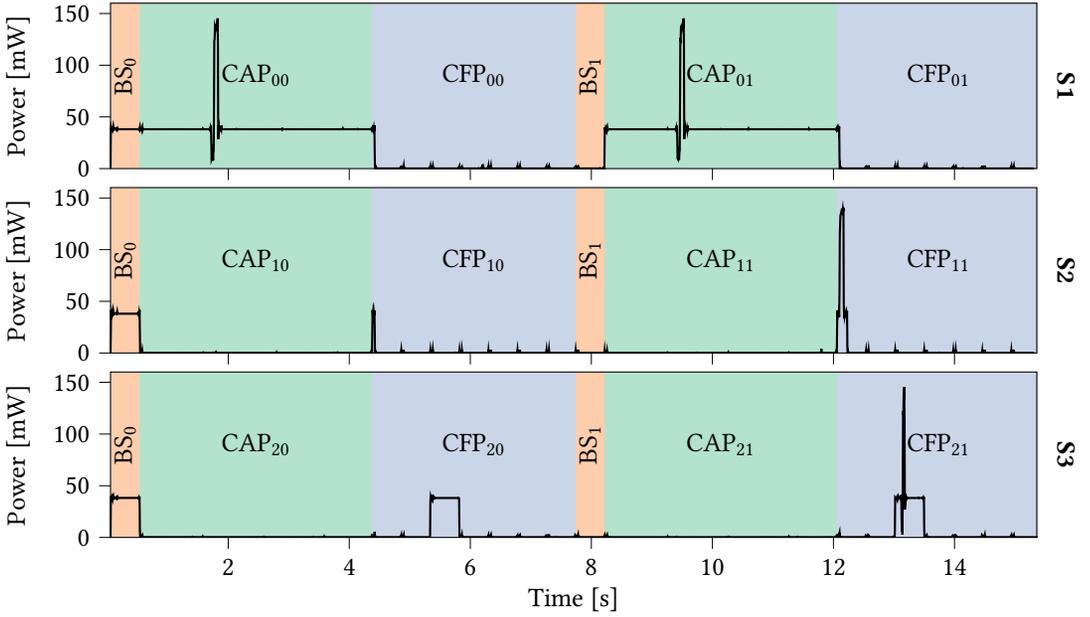
\begin{figure}
    \tikzsetnextfilename{energy}

\begin{tikzpicture}
\definecolor{beacon}{RGB}{253,205,172}
\definecolor{cap}{RGB}{179,226,205}
\definecolor{cfp}{RGB}{203,213,232}
\begin{groupplot}[
group style={
  group size=1 by 3,
  horizontal sep=0.25cm,
  vertical sep=0.25cm,
},
height=0.27\textwidth,
width=\textwidth,
axis on top,
ymax=160,
legend cell align={left},
legend style={
  fill opacity=0.8,
  draw opacity=1,
  text opacity=1,
  at={(0.97,0)},
  nodes={scale=0.8, transform shape},
  anchor=south east,
  draw=white!80!black,
},
tick align=outside,
tick pos=left,
x grid style={white!69.0196078431373!black},
xtick style={color=black},
y grid style={white!69.0196078431373!black},
ytick style={color=black},
xmin=0,ymin=0,
xlabel={Time [s]},
ylabel={Power [mW]},
xmax=15.36,
xmin=0.055,
]
\nextgroupplot[xlabel={},xticklabels={}]
\foreach \s in {0,1} {
\pgfmathsetmacro{\bs}{0.06+\s*7.68}
\pgfmathsetmacro{\caps}{\bs+0.48}
\pgfmathsetmacro{\cfp}{\caps+8*0.48}
\pgfmathsetmacro{\nbs}{\cfp+7*0.48}
\edef\temp{\noexpand
    \path [fill=beacon] (axis cs: \bs,0) rectangle (axis cs:\caps,180) node [pos=0.5,rotate=90] {\acrshort{bs}$_\s$};
}
\temp
\edef\temp{\noexpand
    \path [fill=cap] (axis cs: \caps,0) rectangle (axis cs:\cfp,180) node [pos=0.5] {\acrshort{cap}$_{0\s}$};
}
\temp
\edef\temp{\noexpand
\path [fill=cfp] (axis cs: \cfp,0) rectangle (axis cs:\nbs,180) node [pos=0.5] {\acrshort{cfp}$_{0\s}$};
}
\temp
}
\addplot [thick, black] table [x index=0, y index=1] {data/en_detail_cap.csv};
\nextgroupplot[]
\foreach \s in {0,1} {
\pgfmathsetmacro{\bs}{0.06+\s*7.68}
\pgfmathsetmacro{\caps}{\bs+0.48}
\pgfmathsetmacro{\cfp}{\caps+8*0.48}
\pgfmathsetmacro{\nbs}{\cfp+7*0.48}
\edef\temp{\noexpand
\path [fill=beacon] (axis cs: \bs,0) rectangle (axis cs:\caps,180) node [pos=0.5,rotate=90] {\acrshort{bs}$_\s$};
}
\temp
\edef\temp{\noexpand
\path [fill=cap] (axis cs: \caps,0) rectangle (axis cs:\cfp,180) node [pos=0.5] {\acrshort{cap}$_{1\s}$};
}
\temp
\edef\temp{\noexpand
\path [fill=cfp] (axis cs: \cfp,0) rectangle (axis cs:\nbs,180) node [pos=0.5] {\acrshort{cfp}$_{1\s}$};
}
\temp
}
\addplot [thick, black] table [x index=0, y index=1] {data/en_detail_cfp.csv};
\nextgroupplot[]
\foreach \s in {0,1} {
\pgfmathsetmacro{\bs}{0.06+\s*7.68}
\pgfmathsetmacro{\caps}{\bs+0.48}
\pgfmathsetmacro{\cfp}{\caps+8*0.48}
\pgfmathsetmacro{\nbs}{\cfp+7*0.48}
\edef\temp{\noexpand
\path [fill=beacon] (axis cs: \bs,0) rectangle (axis cs:\caps,180) node [pos=0.5,rotate=90] {\acrshort{bs}$_\s$};
}
\temp
\edef\temp{\noexpand
\path [fill=cap] (axis cs: \caps,0) rectangle (axis cs:\cfp,180) node [pos=0.5] {\acrshort{cap}$_{2\s}$};
}
\temp
\edef\temp{\noexpand
\path [fill=cfp] (axis cs: \cfp,0) rectangle (axis cs:\nbs,180) node [pos=0.5] {\acrshort{cfp}$_{2\s}$};
}
\temp
}
\addplot [thick, black] table [col sep=comma, x index=0, y index=1] {data/en_gts_bi.csv};
\end{groupplot}
\node[anchor=base,rotate=-90,yshift=0.275cm] at (group c1r1.east) {\textbf{S1}};
\node[anchor=base,rotate=-90,yshift=0.275cm] at (group c1r2.east) {\textbf{S2}};
\node[anchor=base,rotate=-90,yshift=0.275cm] at (group c1r3.east) {\textbf{S3}};
\end{tikzpicture}
     \caption{Power consumption during one beacon interval with two multisuperframes separated into \gls{csma} transmissions (top), \gls{gts} transmissions with transceiver off during \gls{cap} (middle) and \gls{gts} receptions with transceiver off during \gls{cap} (bottom).}
    \label{fig:energy}
\end{figure}
\begin{figure}
    \tikzsetnextfilename{energy_tx_details}

\begin{tikzpicture}
\definecolor{color0}{rgb}{0.905882352941176,0.16078431372549,0.541176470588235}
\definecolor{color1}{rgb}{0.850980392156863,0.372549019607843,0.00784313725490196}
\definecolor{color2}{rgb}{0.458823529411765,0.43921568627451,0.701960784313725}
\definecolor{color3}{rgb}{0.105882352941176,0.619607843137255,0.466666666666667}
\definecolor{color4}{rgb}{0.4,0.650980392156863,0.117647058823529}
\definecolor{beacon}{RGB}{253,205,172}
\definecolor{cap}{RGB}{179,226,205}
\definecolor{cfp}{RGB}{203,213,232}
\begin{groupplot}[
group style={
  group size=2 by 2,
  horizontal sep=0.25cm,
  vertical sep=0.35cm,
},
height=0.27\textwidth,
width=0.5\textwidth,
axis on top,
xmax = 175,
ymax=180,
xmin=0,
ymin=0,
legend cell align={left},
legend style={
  fill opacity=0.8,
  draw opacity=1,
  text opacity=1,
  at={(0.97,0)},
  nodes={scale=0.8, transform shape},
  anchor=south east,
  draw=white!80!black,
},
tick align=outside,
tick pos=left,
x grid style={white!69.0196078431373!black},
xtick style={color=black},
y grid style={white!69.0196078431373!black},
ytick style={color=black},
xlabel={Time [ms]},
ylabel={Power [mW]}
]
\nextgroupplot[xticklabels={},xlabel={},title={\textbf{\acrshort{csma}}}]
\path [fill=cap] (axis cs: 0,0) rectangle (axis cs:180,180);
\addplot [thick, black] table [x index=0, y index=1] {data/csma_tx.csv};
\path [draw=gray, thick,dashed] (axis cs:46,0) -- (axis cs:46,180);
\path [draw=gray, thick,dashed] (axis cs:117.4,0) -- (axis cs:117.4,180);
\node [] at (axis cs:23,160) {\acrshort{cca}};
\node [] at (axis cs:81.7,160) {TX};
\node [] at (axis cs:146.2,160) {ACK RX};
\nextgroupplot[yticklabels={},ylabel={},xticklabels={},xlabel={},title={\textbf{\acrshort{gts}}}]
\path [fill=cfp] (axis cs: 0,0) rectangle (axis cs:180,180);
\addplot [thick, black] table [x index=0, y index=1] {data/en_cfp_tx.csv};
\path [draw=gray, thick,dashed] (axis cs:71.3,0) -- (axis cs:71.3,180);
\path [draw=gray, thick,dashed] (axis cs:129,0) -- (axis cs:129,180);
\node [] at (axis cs:35.65,160) {TX};
\node [] at (axis cs:100.15,160) {ACK RX};
\nextgroupplot[]
\path [fill=cap] (axis cs: 0,0) rectangle (axis cs:180,180);
\addplot [thick, black] table [col sep=comma,x index=0, y index=1] {data/en_csma_rx.csv};
\path [draw=gray, thick,dashed] (axis cs:46,0) -- (axis cs:46,180);
\path [draw=gray, thick,dashed] (axis cs:117.4,0) -- (axis cs:117.4,180);
\node [] at (axis cs:81.70,160) {RX};
\node [] at (axis cs:146.2,160) {ACK TX};
\nextgroupplot[yticklabels={},ylabel={}]
\path [fill=cfp] (axis cs: 0,0) rectangle (axis cs:180,180);
\addplot [thick, black] table [col sep=comma,x index=0, y index=1] {data/en_gts_rx.csv};
\path [draw=gray, thick,dashed] (axis cs:71.3,0) -- (axis cs:71.3,180);
\path [draw=gray, thick,dashed] (axis cs:129,0) -- (axis cs:129,180);
\node [] at (axis cs:35.65,160) {RX};
\node [] at (axis cs:100.15,160) {ACK TX};
\end{groupplot}
\node[anchor=base,rotate=-90,yshift=0.275cm] at (group c2r1.east) {\textbf{TX}};
\node[anchor=base,rotate=-90,yshift=0.275cm] at (group c2r2.east) {\textbf{RX}};
\end{tikzpicture}
     \caption{Power consumption for transmission and reception of one packet with \gls{csma} (left) and in a guaranteed time slot (right).}
    \label{fig:energy_details}
\end{figure}
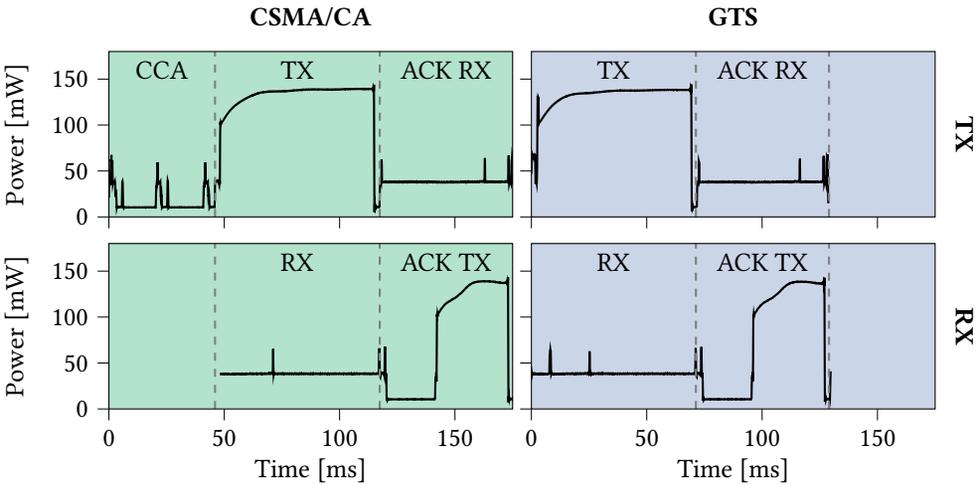

\begin{table}
\centering
\addtolength{\tabcolsep}{3pt} \caption{Energy consumption for each superframe period. Passive (top) components relate the maintenance of the superframe structure under varying transmissions options. Active (bottom) components relate to the actual data transmission. \textit{State} relates to the following components. idle: device is ready for operation w/o ongoing transmissions, off: device is not operable to save energy, TX/RX: transmission/reception of frames.}
    \label{tab:slot-power}
    \begin{tabular}{cclrr}
        \toprule
        & \textbf{Period} & \textbf{State} & \textbf{Additional Description} & \textbf{Energy [mJ]}\\
        \midrule
        \multirow{7}{*}{\rotatebox[origin=c]{90}{\textbf{Passive}}}
        &  \multirow{2}{*}{\gls{bs}}& RX & Beacon synchronization & 18.28\\
        &                     & RX off& Inactive beacon slot & 0.08\\
        \cmidrule(lr){2-5}
        & \multirow{2}{*}{\gls{cap}}& RX idle & \textit{macRxOnWhenIdle=1} & 146.12\\
        &                           & RX off& \textit{macRxOnWhenIdle=0} & 0.50\\
        \cmidrule(lr){2-5}
        & \multirow{3}{*}{\gls{cfp}} & TX idle & Single \gls{gts} allocation & 2.20\\
        &                            & RX idle & Single \gls{gts} allocation & 19.31\\
        &                            & RX off & No \gls{gts} allocation & 0.69\\
        \midrule
        \multirow{4}{*}{\rotatebox[origin=c]{90}{\textbf{Active}}}
        & \multirow{2}{*}{\gls{cap}} & TX & \multirow{2}{*}{Single \gls{csma} TX} & 12.04\\
        &  & RX &  & 6.44\\
        \cmidrule(lr){2-5}
        & \multirow{2}{*}{\gls{cfp}}  & TX & \multirow{2}{*}{Single \gls{gts} TX} & 11.27 \\
        &   & RX &   & 6.44\\
        \bottomrule
    \end{tabular}
\end{table}

We evaluate the power consumption on the target board using a digital multimeter (Keithley DMM7510 7 1/2). Therefore, we sample the current consumption at 100kHz and provide the board with an externally stabilized voltage supply.
Our analyses are separated into passive and active consumption. Passive consumption includes the maintenance of the superframe structure without data transmission. Active consumption, in contrast, includes the transmission and reception of data and ACK frames in different superframe periods. Hence, the total consumption of a node consists of both passive and active components.
\autoref{fig:energy} represents the passive power consumption over time during one beacon interval and three traffic options. It is noteworthy, though, that we exemplary include active TX/RX spikes in the plot, for presentative reasons.

\begin{enumerate}
\item \label{itm:energycsma} \textbf{\energycsma} enables transmission during \gls{cap}. This requires both sender and receiver to enable the transceiver during that period (\autoref{fig:energy} top).
\item \label{itm:energygtstx} \textbf{\energygtstx} disables the \gls{cap} to save power and represents the case for sending data during one \gls{gts} (\autoref{fig:energy} middle).
\item \label{itm:energygtsrx} \textbf{\energygtsrx} is similar to (\ref{itm:energygtstx}), however, it displays data reception during one \gls{gts} (\autoref{fig:energy} bottom).
\end{enumerate}

\autoref{fig:energy_details} represents active power consumption for the sender and receiver of a frame with \gls{csma} (used in the \gls{cap}) as well as without channel sensing (in the \gls{cfp}).
\autoref{tab:slot-power} integrates the power over dedicated intervals and presents the energy consumption for passive (top part) and active (bottom part) actions.
In the reminder of this section, we will first analyze passive and active components separately. We then evaluate the total energy consumption and present our results in~\autoref{tab:power}.
In all measurement configurations we set the transmission interval to TX interval=20\,s and the payload size
to 16 bytes.

\paragraph{Passive consumption (\autoref{tab:slot-power} top)}
During \gls{bs}\textsubscript{0} the \acrshort{mac} turns the transceiver on for the duration of the beacon slot (0.48\,s), in order to receive the beacon from its coordinator. The energy for beacon synchronization is 18.28\,mJ for CPU processing, listening and receiving (RX).
During \gls{bs}\textsubscript{1} the \acrshort{mac} repeats the superframe structure and begins with a new inactive beacon slot, reserved for beacon collision avoidance (see~\autoref{sec:back_dsme}). The \acrshort{mac} keeps the transceiver off (RX off) during that time and the energy consumption reduces to 0.08\,mJ.

The \acrshort{mac} switches to the \gls{cap} after a beacon slot. In the  \energycsma scenario (\autoref{fig:energy} top), the transceiver stays idle listening during \gls{cap}\textsubscript{00}  for 3.84\,s (RX idle, \textit{macRxOnWhenIdle=1}), which consumes 146.12\,mJ. This high consumption shows the need for battery powered devices to turn the transceiver off during \gls{cap}.

The \energygtstx scenario (\autoref{fig:energy} middle) reflects that the transceiver is turned off (RX off, \textit{macRxOnWhenIdle=0}) during \gls{cap}\textsubscript{10}, as it reduces the consumption during \gls{cap} to  0.5\,mJ, only for maintenance purposes (\ie timers, interrupts, \etc). This makes the node, however, unavailable for packet reception during that period.
The \gls{cfp} follows the \gls{cap} (T=4.32\,s) and the \gls{mac} switches to slot mode. Without a \gls{gts} allocation, the transceiver stays off (RX off) and a system wake-up for internal housekeeping requires 0.69\,mJ, which is similarly low as the sleep mode of the \gls{cap}.
In the presence of an allocated \gls{gts} TX slot in the \gls{cfp}, the \acrshort{mac} turns the transceiver on (TX idle) before the \gls{gts} in order to prepare the next transmission. An empty transmission queue triggers the immediate shut down of the transceiver, to save energy. This situation reflects the power peak in \gls{cfp}\textsubscript{10} (T=4.32), which consumes no more than 2.20\,mJ and can be mitigated by slot deallocation. An actual \gls{gts} transmission is visible in \gls{cfp}\textsubscript{11}.

Scenario \energygtsrx (\autoref{fig:energy} bottom) presents the corresponding consumption in \gls{cfp}\textsubscript{20} to receive during one \gls{gts}. Here, the \gls{mac} enables the transceiver during one full \gls{gts} duration (RX idle), which requires 19.31\,mJ and provides a frugal alternative to the \gls{cap} receiver. An actual \gls{gts} reception on top of the baseline is displayed in \gls{cfp}\textsubscript{21}.

\paragraph{Active consumption (\autoref{tab:slot-power} bottom)}
At the beginning of \gls{csma} transmission (\autoref{fig:energy_details}, top left), the
\acrshort{mac} waits for the duration of the backoff period and performs three consecutive
\gls{cca} measures which consumes 0.73\,mJ.
On clear channel, the transceiver loads the frame and performs the frame transmission (TX), which consumes 9.07\,mJ, followed by ACK frame reception at 2.23\,mJ. In total, this makes $\approx$ 12.04\,mJ for \gls{csma} transmission during \gls{cap}.

The \gls{csma} receiver (\autoref{fig:energy_details}, bottom left) consumes 2.75\,mJ for the bare frame, however, awaiting the turnaround time and sending the ACK back consumes additional 3.69\,mJ. Hence, pure receiving (RX) requires 6.44\,mJ, which  appears low compared to transmission. It however requires an active \gls{cap}, which consumes > 20 times more energy (see~\autoref{tab:slot-power}).

On transmission during \gls{gts} (\autoref{fig:energy_details}, top right), the \gls{mac} loads the frame into the transceiver buffer without a preceding \gls{cca}, immediately transmits, and  receives an ACK. This total consumption of 11.27\,mJ outperforms the \gls{csma} sender slightly.
Note that the device turns off the transceiver if the \gls{mac} queue is empty, which further reduces the passive \gls{cfp} consumption occasionally.

Similar to the reception during \gls{csma}, the reception during \gls{gts} (\autoref{fig:energy_details}, bottom right) turns the receiver on for the duration of the frame, delays, and transmits the ACK frame, which leads to the same consumption. In contrast, however, \gls{gts} receivers can turn the transceiver  off during \gls{cap}.

\begin{table}
\centering
    \caption{Passive and active energy consumption per beacon interval [mJ], for the \gls{csma} scenario \energycsma and both \gls{gts} scenarios \energygtstx \& \energygtsrx.}
    \label{tab:power}
    \begin{tabular}{lrrrrrr}
        \toprule
& \multicolumn{2}{c}{\textbf{\energycsma}} & \multicolumn{2}{c}{\textbf{\energygtstx}} & \multicolumn{2}{c}{\textbf{\energygtsrx}}\\
        \cmidrule(lr){2-3}
        \cmidrule(lr){4-5}
        \cmidrule(lr){6-7}
        \textbf{Period} & Energy [mJ] & Prop. [\%]& Energy [mJ] & Prop. [\%]& Energy [mJ] & Prop. [\%]\\
        \midrule
        \gls{bs}         & 18.36  &5.68  & 18.36 &45.15  & 18.36 &28.74\\
        \gls{cap}&&&&&&\\
        \hspace{3mm} passive & 292.23 &90.45 & 0.99 &2.43    & 0.99 &1.55\\
        \hspace{3mm} active  & 11.12  &3.44  & 0.00 &0.00    & 0.00 &0.00\\
        \gls{cfp}&&&&&&\\
        \hspace{3mm} passive & 1.38   &0.43  & 4.40 &10.82   & 38.62 &60.45\\
        \hspace{3mm} active  & 0.00   &0.00  & 16.91 &41.59  & 5.92 &9.27\\
        \midrule
        \textbf{Total } & \textbf{323.09} & \textbf{100} & \textbf{40.66} & \textbf{100} & \textbf{63.89} & \textbf{100}\\
        \bottomrule
    \end{tabular}
\end{table}

\paragraph{Total energy consumption}
\autoref{tab:power} presents the total energy consumption and proportions during \gls{bs}, \gls{cap}, and \gls{cfp}, for the three scenarios in~\autoref{fig:energy}. We normalize the consumption to one beacon interval and present average values from ten measurements. Results are separated into passive and active operations in alignment with the preceding micro analysis.
All three scenarios unsurprisingly consume the same amount of energy (18.36\,mJ) for maintaining the beacon slot.
In the \energycsma scenarios, over 90\% of the consumed energy accounts to passive \gls{cap} consumption---for keeping the radio on---whereas only $\approx$ 3.5 \% is used for sending. Since the \gls{cfp} is not actively used, the overhead remains small ($<$ 0.5\%).
We also conducted measurements at the receiver side in this scenario, but the variation remains negligible.
As the results do not contribute to additional insights, we excluded these experiments.
The overall consumption for scenarios \energygtstx and \energygtsrx behaves similar.
The \gls{cap} remains unused to save energy. Thus, only the passive component consumes 1\,mJ ($<$2.5\%). The \gls{gts} scenarios save about 290\,mJ over \energycsma.
The main proportion is spent in the \gls{cfp}.
Active sending in a \gls{gts} ($\approx$ 17\,mJ) is more expensive than receiving ($\approx$ 6\,mJ). Conversely, turning on the transceiver for a \gls{gts} duration ($\approx$ 39\,mJ) consumes 4 times more energy than scheduling a transmission slot without sending ($\approx$ 11\,mJ).
In total, this leads to a 50\% higher consumption of the \gls{gts} receiver scenario \energygtsrx.
Comparing \energycsma to \energygtstx and \energygtsrx, a total consumption of 323.09\,mJ reveals a notable overhead to the \gls{gts} alternatives, which require 5-8 times less energy. Stress during \gls{cap} (see~\autoref{sec:csma_ca}) can further worsen the energy excess for \gls{csma} reattempts and retransmits.
Conversely, \energycsma scenarios \one are required to associate slots in \energygtstx and \energygtsrx in a real-word deployment and \two enable sporadic transmissions without association overhead.
In practice, however, mixed scenarios can build a compromise to facilitate moderate consumption in flexible deployments.

 \section{Analytical stochastic model}\label{sec:analytics}

An important measure for the feasibility of our solution is its  performance under continuous network load. To evaluate this, we introduce an analytical stochastic model, which allows to calculate the stationary probability
distributions of the \acrshort{mac} queue length at an arbitrary time
and  for transmissions during \gls{cfp} (\gls{gts}). 
Our symbols and nomenclature are summarized in~\autoref{tab:variables}.

\begin{table}
\centering
\caption{List of variables used in our model for \gls{dsme}-LoRa.}
\label{tab:variables}
\begin{tabular}{ll}
    \toprule
    \textbf{Variable} & \textbf{Description} \\
    \midrule
    $L(t)$ & Packets in queue at time $t$ \\
    $T_{n}$ & Time at the end of the slot $n$\\
    $L_{n}$ & Queue length at $T_{n}$, equivalent to $L(T_{n})$\\
    $S$ &  Elapsed time since the end of the last slot ($0 \leq S<T_{msf}$)\\
    $N(t,s)$ & Number of packets scheduled between $t$ and $t+s$\\
    $\lambda$ & Transmission schedule rate \\
    $T_{msf}$ &  Duration of a multisuperframe \\
    $\rho$ & System utilization, defined as $\lambda \cdot T_{msf}$ \\
    $D$ & Number of neighbouring devices\\
    $L_{d}(t)$ & Number of queued packets with destination to neighbour $d \in [0,D)$\\
    & by definition, $L(t) = \sum_{d=0}^{D-1} L_{d}(t)$\\
    \bottomrule
\end{tabular}
\end{table}

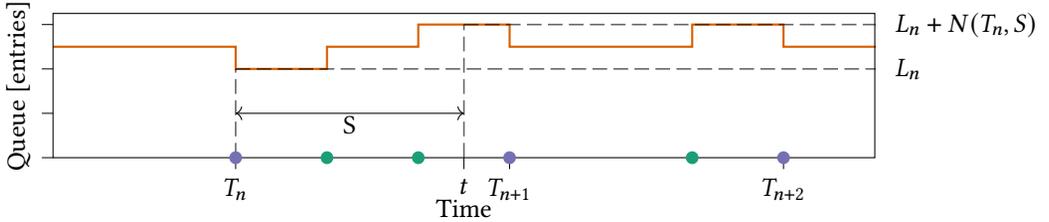
\begin{figure}
    \tikzsetnextfilename{lt_deduction}

\begin{tikzpicture}
\definecolor{color0}{rgb}{0.905882352941176,0.16078431372549,0.541176470588235}
\definecolor{color1}{rgb}{0.850980392156863,0.372549019607843,0.00784313725490196}
\definecolor{color2}{rgb}{0.458823529411765,0.43921568627451,0.701960784313725}
\definecolor{color3}{rgb}{0.105882352941176,0.619607843137255,0.466666666666667}
\definecolor{color4}{rgb}{0.4,0.650980392156863,0.117647058823529}
  \begin{groupplot}[
    group style={
      group size=1 by 1,
      horizontal sep=0.25cm,
      vertical sep=0.25cm,
    },
    height=0.25\textwidth,
    width=0.9\textwidth,
yticklabels={,,$L_{n}$, $L_{n} + N({T_{n},S})$},
    ytick={0,2,4,6},
    xtick={0,3,5.5,6,9,12},
      xticklabels={,$T_{n}$,$t$,$T_{n+1}$,$T_{n+2}$,$T_{n+3}$},
    xmax=10,
    ymax=6.5,
    axis on top,
    yticklabel pos=right,
xmin=1, ymin=0,
    legend cell align={left},
    legend style={
      fill opacity=0.8,
      draw opacity=1,
      text opacity=1,
      at={(0.9,0.03)},
      nodes={scale=0.8, transform shape},
      anchor=south east,
      draw=white!80!black,
    },
tick align=outside,
tick pos=left,
x grid style={white!69.0196078431373!black},
xtick style={color=black},
y grid style={white!69.0196078431373!black},
ytick style={color=black},
]
\nextgroupplot[]
\addplot [thick, only marks, color2,mark=*]
table {0 0
    3 0
    6 0
    9 0
    12 0
};
\addplot [thick, const plot, color1]
table {0 5
    3 4
    4 5
    5 6
    6 5
    8 6
    9 5
    12 4
};
\addplot [thick, only marks, color3, mark=*]
table {4 0
    5 0
    8 0
};
\path [draw=black, draw opacity=0.6, semithick, dash pattern=on 5.55pt off 2.4pt]
(axis cs:5.5,0)
--(axis cs:5.5,6);
\path [draw=black, draw opacity=0.6, semithick, dash pattern=on 5.55pt off 2.4pt]
(axis cs:3,0)
--(axis cs:3,4);
\path [draw=black, draw opacity=0.6, semithick, dash pattern=on 5.55pt off 2.4pt]
(axis cs:3,4)
--(axis cs:12,4);
\path [draw=black, draw opacity=0.6, semithick, dash pattern=on 5.55pt off 2.4pt]
(axis cs:5.5,6)
--(axis cs:12,6);
\draw[<->,draw=black] (axis cs:3,2) -- (axis cs:5.5,2);
\draw (axis cs:4.25,1.5) node[
  scale=1,
  fill=white,
  draw=none,
  line width=0.4pt,
  inner sep=0pt,
  text=black,
  rotate=0.0,
  align=center
]{S};
\end{groupplot}
\node[anchor=base,rotate=90,yshift=0.375cm] at (group c1r1.west) {Queue [entries]};
\node[anchor=base,yshift=-0.775cm] at (group c1r1.south) {Time};
\end{tikzpicture}
 \caption{Qualitative evolution of the \acrshort{mac} queue  over time: occupation grows on packet scheduling events (green) and  reduces at slot occurrences (purple).}
    \label{fig:lt_deduction}
\end{figure}

The temporal evolution of the \gls{mac} queue at a \gls{dsme}-LoRa device is visualized in~\autoref{fig:lt_deduction}. Packets arrive randomly over time and are added to the queue. At the end of an arbitrary slot $T_n$, packets are transmitted and removed from the queue. Correspondingly, the number of queue entries at the end of slot $n$ is $L_n$, while $N(T_n,S)$ packets are added in the time span $S$ after $T_n$. In a homogeneous process,  $N(T_n,S)$ is proportional to $S$.

\subsection{A Markov queuing process}

We model our \gls{dsme}-LoRa transmission system as a simple Markov queuing process,  For this we make the following simplifying assumptions.
\begin{enumerate}
    \item packets arrive independently 
    \item exponentially distributed interarrival times between scheduled packets with $\rho < 1$ ($\lambda < \frac{1}{T_{msf}}$)
    \item \acrshort{mac} queue has unlimited capacity
\item unacknowledged transmissions and 100\% packet reception ratio
    \item transmission time on air is neglected.
\end{enumerate}

We start by considering transfer to only one neighbour ($D=1$). Thereafter, we extend the model under moderate conditions to scenarios with $D > 1$.
We further analyze this situation in~\autoref{sec:multiple_gts}.

\begin{figure}
    \tikzsetnextfilename{markov}

\begin{tikzpicture}[->, >=stealth', auto, semithick, node distance=3cm]
\tikzstyle{every state}=[fill=white,draw=black,thick,text=black,scale=1]
\node[state]    (A)                     {$L_0$};
\node[state]    (B)[right of=A]   {$L_1$};
\node[state]    (C)[right of=B]   {$L_2$};
\node[state]    (D)[right of=C]   {$L_3$};
\path
(A) edge[loop above]    node{$k_0 + k_1$}      (A)
    edge[bend right = 40]    node{$k_2$}           (B)
    edge[bend right = 40]    node{$k_3$}     (C)
    edge[bend right = 40]    node{$k_4$}      (D)
(B) edge[loop above]    node{$k_1$}     (B)
    edge[bend right = 40]    node{$k_2$}     (C)
    edge[bend right = 40]    node{$k_3$}      (D)
    edge[bend right = 40]    node{$k_0$}           (A)
(C) edge[loop above]    node{$k_1$}     (C)
    edge[bend right = 40]    node{$k_2$}     (D)
    edge[bend right = 40]    node{$k_0$}           (B)
(D) edge[loop above]    node{$k_1$}     (D)
    edge[bend right = 40]    node{$k_0$}     (C)
    ;
\node[right=.1cm of D] {$\dots$};
\end{tikzpicture}
 	\caption{Embedded Markov chain: The queue occupation $L_i$ remains constant between slots, if only one packet arrives ($k_1$). It increases by $i-1$ for $k_i$ packet arrivals, and decreases for $k_0$.}
     \label{fig:markov}
\end{figure}
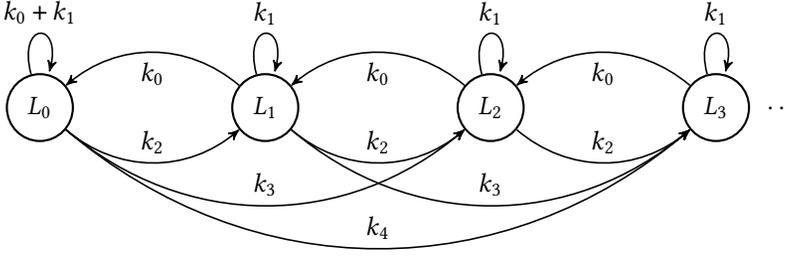

Our Markov queuing model is shown in~\autoref{fig:markov}. The state of the queue $L_i$ is reduced by one at the end of every time slot. Packets arrive randomly during any time interval $s$  in the queue of the system and follow a Poisson process with parameter $\lambda s$. For a complete multisuperframe time, let us denote \[k_{i} = P\{N(T_n, T_{msf}) = i\} = \frac{\rho^i \cdot  e^{-\rho}}{i!}\] 
the probability of $i$ packets arriving during one multisuperframe ($\rho = \lambda \cdot T_{msf}$ denotes the arrival intensity, \ie system utilization). 
Then the transition arcs of the Markov matrix $P$ are defined by 
\begin{numcases}{P[i,j] =}
    k_0 & \text{if} j-i= -1 \label{eq:m1}\\
    k_0 + k_1 & \text{if} i=0, j=0 \label{eq:m2}\\
    k_{j-i+1} & \text{if} j $\geq$ i \label{eq:m3}\\
    0 & otherwise
\end{numcases}
The probability of a queue reduction corresponds to no packet arrival  (\autoref{eq:m1}), of a growth by (j-i) to {j-i+1} packet arrivals (\autoref{eq:m3}), and a constant initial condition to either none or one packet arriving (\autoref{eq:m2}).

\subsection{Queue length}\label{sec:queue_length}

For calculating the actual queue occupation, we note that the number of queued packets at an arbitrary time is $L(t) = L_{n} + N(t-S,S)$,
as seen in~\autoref{fig:lt_deduction}.

Consequently, the distribution of queue length is given as

\begin{eqnarray}
	P\{L(t) = i\} &=& P\{L_{n} + N(t-S,S) = i\} \nonumber\\
	&=& \sum_{j=0}^{i} P\{L_{n}=i-j,N(t-S,S)=j\}\nonumber\\
	&=& \sum_{j=0}^{i} P\{L_{n}=i-j\} \cdot P\{N(t-S,S)=j\}\label{eq_L_general}
\end{eqnarray}

We first derive a result for $P\{N(t-S,S)=i\}$. Note that $P\{N(t-S,S) = i \mid S=s\}$ is a Poissonian with parameter $(\lambda \cdot s)$ and $S$ is uniform in $(0,T_{msf})$. Therefore we can calculate $P\{N(t-S,S) = j\}$ via the law of total probability:

\begin{eqnarray}\label{eq_ns}
	P\{N(t-S,S) = j\}&=&\int_{0}^{\infty} P\{N(t-S,S) = j | S=s\} \cdot P\{S=s\} ds \nonumber\\
    &=&\int_{0}^{T_{msf}} \frac{(\lambda s)^j e^{-\lambda s}}{j!} \cdot \frac{1}{T_{msf}} ds
	=\frac{1}{\rho} \cdot \Gamma(j+1,\rho)\label{eq_ns}
\end{eqnarray}

where $\Gamma(j,x) = \frac{\int_{0}^{x} t^{j-1} e^{-t}}{\int_{0}^{\infty} t^{j-1} e^{-t}}$ is the regularized lower incomplete gamma function.

For the calculation of $P(L_{n} = i)$, observe that $L_{i}, \, \forall i \in [0,\infty)$ is a Markov chain (\autoref{fig:markov}), for which we search 
the stationary distribution.
We also observe that the Markov chain is ergodic (positive recurrent and aperiodic). Thus,
the stationary distribution $\pi_{i}=\lim_{n \to \infty} P(L_{n} = i)$ exists
and complies with $P^T \vec{\pi} = \vec{\pi}$.
The calculation of a closed-form analytical solution for $\pi_{i}$ is not trivial. 
We describe a detailed numeric procedure to calculate the vector in~\autoref{sec:pi_calculation}.

Combining the stationary distribution of the Markov chain and~\autoref{eq_ns}
into~\autoref{eq_L_general} leads to the distribution of queue length:

$$
P\{L(t) = i\} = \sum_{j=0}^{i} \pi_{i-j} \frac{\Gamma(j+1,\rho)}{\rho}
$$

It is possible to calculate the average queue length directly. Observe
that $E(L(t)) = E(L_n) + E(N(S))$, where $E(L_n)=\sum_{i=0}^{\infty} i \pi_i$ and
$E(N(S)) = \int_{0}^{T_{msf}} \rho \frac{1}{T_{msf}} dt =
\frac{\rho}{2}$.

Therefore
\begin{equation}
E(L(t)) = \sum_{i=0}^{\infty} i \pi_i + \frac{\rho}{2}.
\end{equation}

\subsection{Transmission delay}\label{sec:delay}
We now calculate the distribution of transmission delay in multiples of $T_{msf}$ from
the distribution of queue length:
$$
P\{W \leq n T_{msf}\} = P\{L(t) \leq n - 1\}
$$

As an example, the fraction of packets with transmission delay less than one multisuperframe
is $P\{L(t) \leq 0\} = \pi_0\frac{1-e^{-\rho}}{\rho}$

Little's Law~\cite{l-pqf-61} $L = \lambda   W$ calculates the
average number of queued items (L) using the arrival rate ($\lambda$) and average waiting
time W. We use the result to calculate the average transmission delay directly

\begin{eqnarray}
	W &=& \frac{1}{\lambda} E(L(t))
	=\frac{1}{\lambda} \left(\sum_{i=0}^{\infty} i \pi_i + \frac{\rho}{2}\right)
\end{eqnarray}

\subsection{Allocation of multiple \gls{gts}}\label{sec:multiple_gts}
So far the model assumes only one neighbour device ($D = 1$) and the
allocation of only one slot. The model, however, is still valid for $D > 1$ if
each target device allocates only one slot per multisuperframe. In such case, the \acrshort{mac}
utilizes the queue as multiple independent FIFO sub-queues ($L_{d}$).
As a result, the model is valid for each sub-queue  and the distribution of the
total queue length is:

$$
P\{L(t) = i\} = P\left\{\sum_{d=0}^{D-1} L_{d}(t) = i\right\}
$$

Note that the average queue length is the sum of all average sub-queue length
and the average transmission delay the fraction of the average queue length and
the total schedule rate (Little equation).

The proposed formulas, however, cease to hold if the \acrshort{mac} allocates more than one
slot to the same neighbour. Nevertheless,  $D = 1$ sets the worst case 
scenario for transmission delay and queue length over these scenarios.

\subsection{Validation of the model}

We validate the model accuracy for the distribution of queue length and transmission
delay. We chose the \gls{gts} transmission scenario with the highest rate of transmitted packets,
namely with 15 source devices, in order minimize the effect of the transient queue. We
do not include the scenario with TX interval=5\,s,
because it is shorter than the multisuperframe duration (7.68\,s). In such case
the model does not converge.
For the calculation of theoretical results we clipped the Markov matrix to 100 elements.

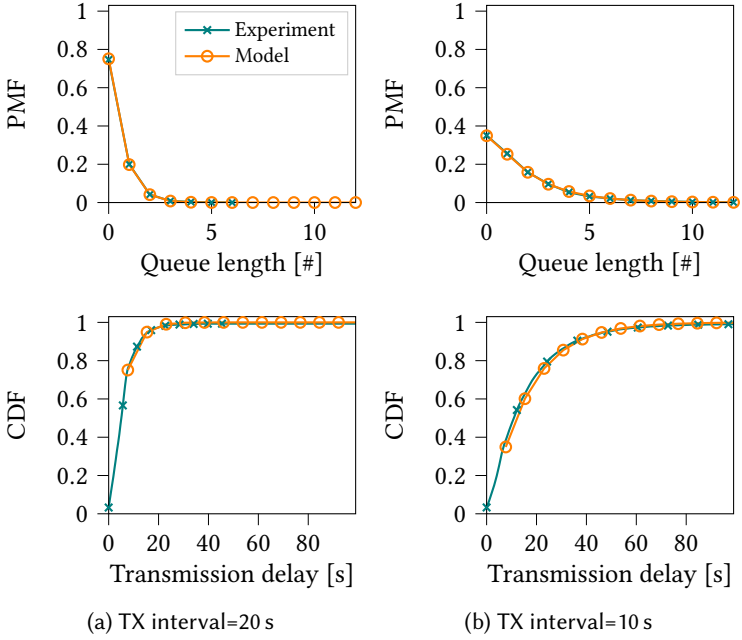
\begin{figure}
    \subfloat[TX interval=20\,s]{\tikzsetnextfilename{model_validation_ql}

\begin{tikzpicture}
\definecolor{color0}{rgb}{0.905882352941176,0.16078431372549,0.541176470588235}
\definecolor{color1}{rgb}{0.850980392156863,0.372549019607843,0.00784313725490196}
\definecolor{color2}{rgb}{0.458823529411765,0.43921568627451,0.701960784313725}
\definecolor{color3}{rgb}{0.105882352941176,0.619607843137255,0.466666666666667}
\definecolor{color4}{rgb}{0.4,0.650980392156863,0.117647058823529}
\begin{groupplot}[
group style={
  group size=1 by 2,
  horizontal sep=0.25cm,
  vertical sep=1.50cm,
},
height=0.3\textwidth,
width=0.35\textwidth,
ymax=1.03,
axis on top,
yticklabel pos=left,
xmin=0, ymin=0,
legend cell align={left},
legend style={
  fill opacity=0.8,
  draw opacity=1,
  text opacity=1,
  at={(0.97,0.97)},
  nodes={scale=0.8, transform shape},
  anchor=north east,
  draw=white!80!black,
},
tick align=outside,
tick pos=left,
x grid style={white!69.0196078431373!black},
xtick style={color=black},
y grid style={white!69.0196078431373!black},
ytick style={color=black},
cycle list name=exotic,
]
\nextgroupplot[ylabel={PMF},xlabel={Queue length [\#]},xmax=12]
\addplot+ [thick, mark=x] table [x index=0, y index=1] {data/exp_ql_20.csv};
\addlegendentry{Experiment}
\addplot+ [thick, mark=o] table [x index=2, y index=3] {data/model_ql.csv};
\addlegendentry{Model}
\nextgroupplot[ylabel={CDF},xmax=99, xlabel={Transmission delay [s]}]
\addplot+ [thick, mark=x,mark repeat=12] table [x index=8, y index=9] {data/cfp.csv};
\addplot+ [thick, mark=o] table [x index=2, y index=3] {data/model.csv};

\end{groupplot}
\end{tikzpicture}
 \label{fig:model_validation_ql}}
    \subfloat[TX interval=10\,s]{\tikzsetnextfilename{model_validation_ct}

\begin{tikzpicture}
\definecolor{color0}{rgb}{0.905882352941176,0.16078431372549,0.541176470588235}
\definecolor{color1}{rgb}{0.850980392156863,0.372549019607843,0.00784313725490196}
\definecolor{color2}{rgb}{0.458823529411765,0.43921568627451,0.701960784313725}
\definecolor{color3}{rgb}{0.105882352941176,0.619607843137255,0.466666666666667}
\definecolor{color4}{rgb}{0.4,0.650980392156863,0.117647058823529}
\begin{groupplot}[
group style={
  group size=1 by 2,
  horizontal sep=0.25cm,
  vertical sep=1.50cm,
},
height=0.3\textwidth,
width=0.35\textwidth,
ymax=1.03,
axis on top,
yticklabel pos=left,
xmin=0, ymin=0,
legend cell align={left},
legend style={
  fill opacity=0.8,
  draw opacity=1,
  text opacity=1,
  at={(0.97,0.97)},
  nodes={scale=0.8, transform shape},
  anchor=north east,
  draw=white!80!black,
},
tick align=outside,
tick pos=left,
x grid style={white!69.0196078431373!black},
xtick style={color=black},
y grid style={white!69.0196078431373!black},
ytick style={color=black},
cycle list name=exotic,
]
\nextgroupplot[ylabel={PMF},xlabel={Queue length [\#]},xmax=12]
\addplot+ [thick, mark=x] table [x index=0, y index=1] {data/exp_ql_10.csv};
\addplot+ [thick, mark=o] table [x index=0, y index=1] {data/model_ql.csv};
\nextgroupplot[ylabel={CDF}, xmax=99, xlabel={Transmission delay [s]}]
\addplot+ [thick, mark=x,mark repeat=12] table [x index=6, y index=7] {data/cfp.csv};
\addplot+ [thick, mark=o] table [x index=0, y index=1] {data/model.csv};
\end{groupplot}
\end{tikzpicture}
 \label{fig:model_validation_ct}}
    \caption{Validation of the analytical stochastic model with experimental results. Comparison of distributions of the queue length and transmission delay for varying transmission intervals.}
    \label{fig:model_validation}
\end{figure}

In~\autoref{fig:model_validation} we validate our model by comparing to experimental results (see~\autoref{sec:eval_hw}).
Figure~\autoref{fig:model_validation_ql} compares the probability mass function of the queue
length at packet schedule between the results of the experiments and the
model. The model predicts the distribution of queue length with more
than 99.99\% of accuracy. In the relaxed scenario the probability of more than five elements in the queue
is $6.17\cdot10^{-5}$, which is consistent with the observation that the queue does not
exceed this value.
Similarly, the model predicts the transmission delay with an accuracy of 99.99\%, as seen
in Figure~\autoref{fig:model_validation_ct}.

The small variations between the experiment results and the model are
due to the effect of the transient queue and a small fraction of packet losses. The former effect mitigates either with a bigger network size or with a longer experiment run.

 \section{Simulation Study: Assessment of Large Scale Ensembles}\label{sec:simulation}

\begin{figure}
    \tikzsetnextfilename{sim_arch}

\begin{tikzpicture}[node distance=0.5cm and 0.3cm]
\definecolor{opendsme}{RGB}{255,255,179}
\definecolor{flora}{RGB}{141,211,199}
\definecolor{inet}{RGB}{190,186,218}
\definecolor{own}{RGB}{251,128,114} \tikzstyle{process} = [rectangle, minimum width=5cm, minimum height=0.4cm, text centered, draw=black, fill=orange!30]
\tikzstyle{own} = [rectangle, minimum width=5cm, minimum height=0.7cm, text centered, draw=black, fill=orange!20]
\tikzstyle{big} = [rectangle, minimum width=5cm, minimum height=1.5cm, text centered, draw=black, fill=orange!30]
\tikzstyle{medium} = [rectangle, minimum width=5cm, minimum height=.8cm, text centered, draw=black, fill=orange!30]
\tikzstyle{small} = [rectangle, minimum width=2.2cm, minimum height=.4cm, text centered, draw=black, fill=orange!30]
\tikzstyle{dsme_lora} = [rectangle, minimum width=4cm, minimum height=0.4cm, text centered, draw=black, fill=orange!30]
\tikzstyle{bullet} = [rectangle, minimum width=0.4cm, minimum height=0.4cm, draw=black,anchor=north west, xshift=0.1cm]
\tikzstyle{line} = [thick,-,>=stealth]
\tikzstyle{arrow} = [thick,->,>=stealth]

\node (prrtrafgen) [medium,fill=opendsme] {};
\node (app) at (prrtrafgen.north west) [anchor=north west, xshift=0.1cm, yshift=-0.1cm] {Application};
\node at (prrtrafgen) [small,fill=inet,xshift=1cm] {Traffic generator};

\node (nexthop) [process,fill=inet,below=of prrtrafgen.south, anchor=north] {L3 address resolver};

\node (inetdsme) [big,fill=opendsme,below=of nexthop.south, anchor=north] {};
\node  at (inetdsme.north west) [anchor=north west,yshift=-0.1cm,xshift=0.1cm] {inet-dsme};
\node at (inetdsme) [small,fill=opendsme,xshift=1cm,yshift=0.3cm]  {DSME MAC};
\node at (inetdsme.south) [dsme_lora, fill=own, anchor=south] {\textbf{\acrshort{dsme}-LoRa}};

\node (radio) [process,fill=flora,below=of inetdsme, anchor=north] {LoRa Radio};

\newcommand\omnetdistx{10pt}
\newcommand\omnetdisty{10pt}
\draw[dashed, very thick] ([xshift=-\omnetdistx, yshift=\omnetdisty]prrtrafgen.north west) 
-- ([xshift=\omnetdistx, yshift=\omnetdisty]prrtrafgen.north east)
-- ([xshift=\omnetdistx, yshift=-\omnetdisty]radio.south east) 
-- ([xshift=-\omnetdistx, yshift=-\omnetdisty]radio.south west)
-- ([xshift=-\omnetdistx, yshift=\omnetdisty]prrtrafgen.north west);

\node (legend) [rectangle, minimum width=4.0cm, minimum height=2.5cm, right=of inetdsme, xshift=.5cm, yshift=.5cm] {};
\node (b_opendsme) at (legend.west) [bullet,fill=opendsme, yshift=1cm] {};
\node (b_flora) at (legend.west) [bullet,fill=flora, yshift=0.5cm] {};
\node (b_own) at (legend.west) [bullet,fill=own, yshift=0cm] {};
\node (b_inet) at (legend.west) [bullet,fill=inet, yshift=-1cm] {};
\node (b_omnet) at (legend.west) [bullet,dashed,very thick, yshift=-1.5cm] {};

\node [right=of b_opendsme] {\textit{openDSME}};
\node [right=of b_flora] {\textit{FLoRa}};
\node [right=of b_inet] {\textit{INET}};
\node [right=of b_omnet] {\textit{OMNeT++}};
\node [right=of b_own] {Own extension};

\draw [<->,>={[scale=1]Latex}] (prrtrafgen) -- (nexthop);
\draw [<->,>={[scale=1]Latex}] (nexthop) -- (inetdsme);
\draw [<->,>={[scale=1]Latex}] (inetdsme) -- (radio);

\end{tikzpicture}
     \caption{\gls{dsme}-LoRa simulation environment and our contribution.}
    \label{fig:sim_arch}
\end{figure}
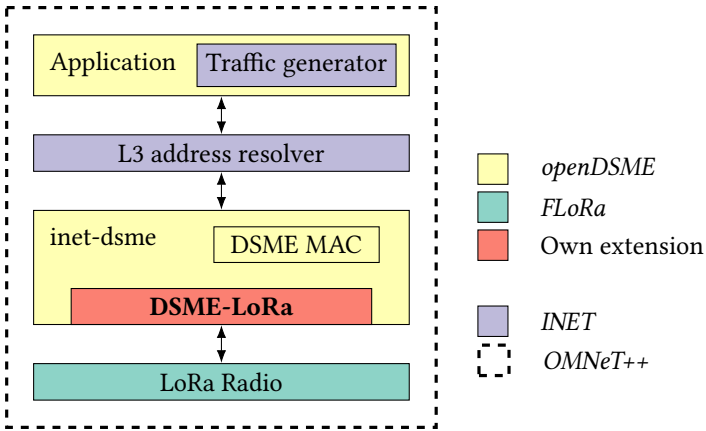

We proceed to evaluate the performance of \gls{dsme}-LoRa for a larger networks using
the \textit{INET \cite{inet-framework-21} /OMNeT++ \cite{v-odess-01}} based on our simulation environment~\cite{aksw-dfml-21}.
The simulator utilizes the radio module of \textit{FLoRa}~\cite{spd-aclnd-18} and the OMNeT++ adaptation of \textit{openDSME}~\cite{kkt-rwmnd-18}, namely \textit{inet-dsme}, for the \acrshort{mac}
implementation. We extend \textit{inet-dsme} to enable \gls{dsme} communication over the LoRa radio, as shown in~\autoref{fig:sim_arch}.
We reuse the traffic generator application of \textit{inet-dsme}, namely \textit{PRRTrafGen}, which bases on the \textit{IpvxTrafGen} traffic generator module of INET.
We utilize the \textit{nextHop} module of INET to resolve L3 address from packets into the destination \acrshort{mac} address.

\subsection{Validation of the simulation environment}\label{sec:simulation_vali}

\begin{figure}
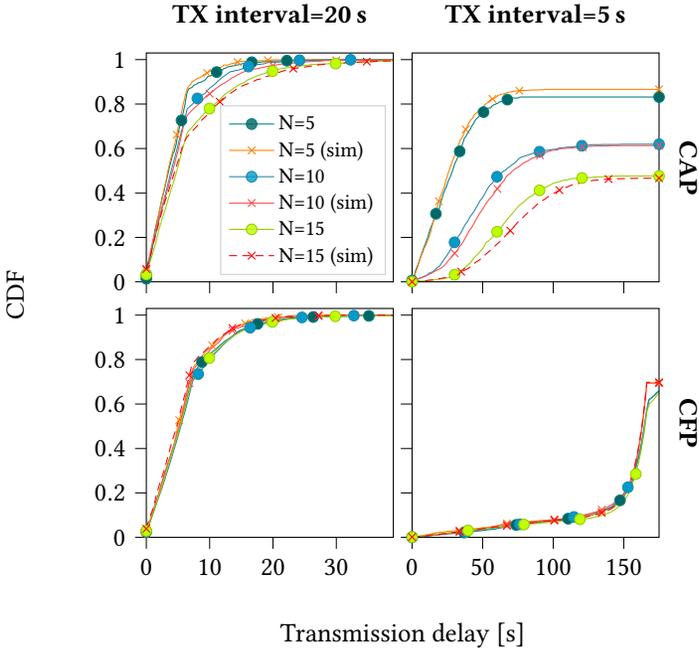

    \tikzsetnextfilename{sim_validation}


     \caption{Comparison of transmission delays from simulations and experiments for confirmed transmissions during \gls{cap} (\gls{csma}) and \gls{cfp} (\gls{gts}). We vary the number (N) of source devices and the transmission interval.}
    \label{fig:sim_validation}
\end{figure}

To validate our simulator, we first compare simulation results with real-world measurements on hardware (see~\autoref{sec:eval_hw}).
In particular, \autoref{fig:sim_validation} compares simulation results and our experiments conducted in~\autoref{sec:data_transmission}.
In \gls{csma} transmissions, a fraction of collided
packets are successfully decoded (capture effect). The fraction of packets
varies between the simulation and the experiment, which lead to a different number
of retransmissions and dropped frames by the \gls{mac}.
In \gls{gts} transmissions, the collision free transmission renders high reception ratio for
\gls{mac} transmissions in the simulation. In the experiment, in practice, a tiny fraction of transmitted frames is lost, which increase the number of retransmissions. 
Overall, the results of the simulator converge with the experiments results. Differing
behavior between the physical channel and the simulation channel model explain small variations.

\subsection{Large scale peer to peer communication}\label{sec:simulation_large}

\begin{figure}
    \tikzsetnextfilename{sim_100_300}

\begin{tikzpicture}
\definecolor{color0}{rgb}{0.905882352941176,0.16078431372549,0.541176470588235}
\definecolor{color1}{rgb}{0.850980392156863,0.372549019607843,0.00784313725490196}
\definecolor{color2}{rgb}{0.458823529411765,0.43921568627451,0.701960784313725}
\definecolor{color3}{rgb}{0.105882352941176,0.619607843137255,0.466666666666667}
\definecolor{color4}{rgb}{0.4,0.650980392156863,0.117647058823529}
\begin{groupplot}[
group style={
  group size=2 by 2,
  horizontal sep=0.25cm,
  vertical sep=0.35cm,
},
height=0.3\textwidth,
width=0.35\textwidth,
ymax=1.03,
axis on top,
yticklabel pos=left,
xmin=0, ymin=0,
legend cell align={left},
legend style={
  fill opacity=0.8,
  draw opacity=1,
  text opacity=1,
  at={(0.97,0.03)},
  nodes={scale=0.8, transform shape},
  anchor=south east,
  draw=white!80!black,
},
tick align=outside,
tick pos=left,
x grid style={white!69.0196078431373!black},
xtick style={color=black},
y grid style={white!69.0196078431373!black},
ytick style={color=black},
cycle list name=exotic,
]
\nextgroupplot[title={\textbf{\acrshort{cap}}},xticklabels={},xlabel={},xmin=0,xmax=9.5]
\addplot+ [mark repeat=12, mark size=2] table [x index=2, y index=3] {data/sim_100_and_300.csv};
\addplot+ [mark repeat=12, mark size=2] table [x index=0, y index=1] {data/sim_100_and_300.csv};
\nextgroupplot[title={\textbf{\acrshort{cfp}}},xlabel={},xticklabels={},xmax=220,width=0.65\textwidth,yticklabels={}]
\addplot+ [mark repeat=12, mark size=2] table [x index=8, y index=9] {data/sim_100_and_300.csv};
\addlegendentry{N=100}
\addplot+ [mark repeat=12, mark size=2] table [x index=6, y index=7] {data/sim_100_and_300.csv};
\addlegendentry{N=300}
\nextgroupplot[xmin=0,xmax=9.5]
\addplot+ [mark repeat=12, mark size=2] table [x index=28, y index=29] {data/sim_100_and_300.csv};
\addplot+ [mark repeat=12, mark size=2] table [x index=38, y index=39] {data/sim_100_and_300.csv};
\nextgroupplot[yticklabels={},xmax=220,width=0.65\textwidth]
\addplot+ [mark repeat=12, mark size=2] table [x index=44, y index=45] {data/sim_100_and_300.csv};
\addplot+ [mark repeat=12, mark size=2] table [x index=10, y index=11] {data/sim_100_and_300.csv};
\end{groupplot}
\node[anchor=base,rotate=-90,yshift=0.275cm] at (group c2r1.east) {\textbf{TX interval=80\,s}};
\node[anchor=base,rotate=-90,yshift=0.275cm] at (group c2r2.east) {\textbf{TX interval=40\,s}};
\node[anchor=base,rotate=90,yshift=1cm] at (group c1r1.south west) {CDF};
\path (group c1r2.south west) -- (group c2r2.south east) node[midway,sloped,below=1cm,anchor=north] {Transmission delay [s]};
\end{tikzpicture}
     \caption{Comparison of transmission delays for relaxed and stressed scenarios, during \gls{cap} (\gls{csma}) and \gls{cfp} (\gls{gts}), for confirmed transmissions and a varying number (N) of nodes.}
    \label{fig:sim_100_300}
\end{figure}
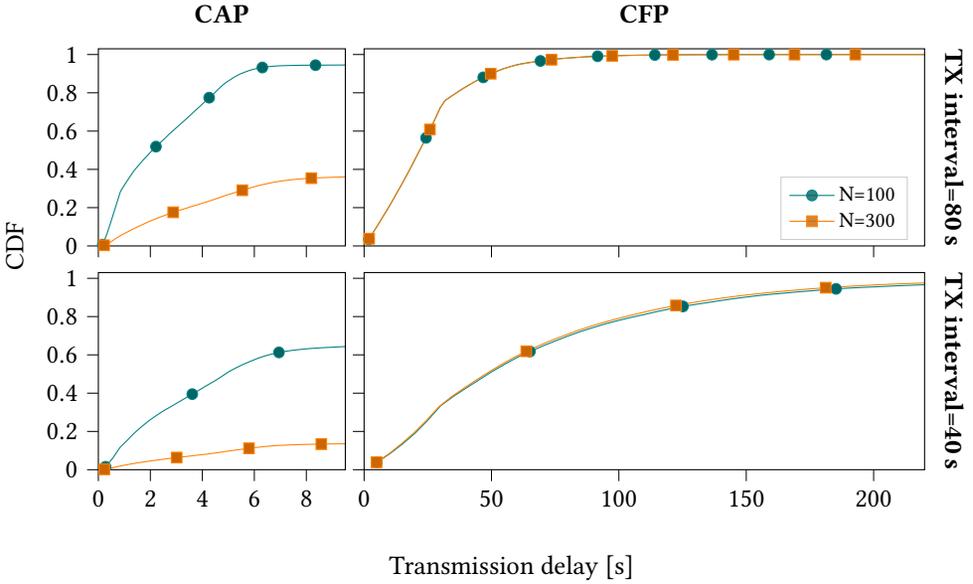

We evaluate the transmission delay and packet reception ratio of confirmed \gls{csma} and \gls{gts} transmissions,
for varying network sizes (N=100 and N=300) and varying transmission interval
(\autoref{fig:sim_100_300}). In agreement with~\autoref{sec:mac_config}, we use high backoff exponent settings to minimize packet collision. To accommodate one slot for every source device during \gls{cfp}, we
configure the multisuperframe order to 5, which renders 28 \gls{gts} and
a multisuperframe duration of 30.72\,s (\autoref{tab:msf_duration}).
\autoref{fig:sim_100_300} presents our results.

\paragraph{\gls{csma} transmission}. 
Our results show that the small network (N=100) renders a 95\% packet reception ratio
in the relaxed scenario (\autoref{fig:sim_100_300}, top left). The high on-air
traffic due to the network size increases packet collisions and \gls{cca}
failure rates, as analyzed in~\autoref{sec:eval_hw}. As a result, a fraction of
packet is lost. With half of the
transmission interval (\autoref{fig:sim_100_300}, bottom left), the small network
reduces the packet reception ratio to $\approx$ 60\%, as a result of the higher
on-air traffic.

Scenarios with big networks render an even higher on-air time, which reflects
$\approx$ 38\% packet reception ratio in the relaxed scenario
(\autoref{fig:sim_100_300}, top left, N=300) and $\approx$ 14\% in the stressed
scenario (\autoref{fig:sim_100_300}, bottom left).
This concludes that \gls{csma} is not reliable for large scale deployments.

Observer that the transmission delay of the majority frames do not exceed 10\,s even
in the stressed scenario (\autoref{fig:sim_100_300}, bottom left, N=300).
This value reflects the worst case transmission (maximum \gls{csma} retries and
maximum frame retransmissions).
The delay in the worst case is lower than the TX interval in both \gls{cap} scenarios.
Therefore, the stress in the \gls{cap} queue is low, hence the transmission delay.

\paragraph{\gls{gts} transmission}. 
In the relaxed scenario (\autoref{fig:sim_100_300}, top right), the transmission delay hits
the maximum value at $\approx$ 120\,s in both network sizes, as a result of the delay of queued packets. As per~\autoref{sec:eval_hw}, the transmission delay does not vary with the network sizes, because all
devices have equal \gls{gts} resources (one slot per multisuperframe).
In contrast, in the stressed scenario (\autoref{fig:sim_100_300}, bottom right) the transmission delay, as a result of the higher \gls{mac} queue stress, hits the maximum at $\approx$ 500\,s (not shown in the subfigure). Similar to the relaxed scenario, the transmission delay does not vary with the network size.

The packet reception ratio hits $\approx$ 100\% in all \gls{gts} scenarios, as a result of
the slot allocation.
The results reflect the robustness of \gls{gts} transmissions over \gls{csma},
which make it suitable for large scale scenarios. We further analyze this in~\autoref{sec:design_tx}.

\subsection{Impact of the multisuperframe duration}\label{sec:impact_msf}

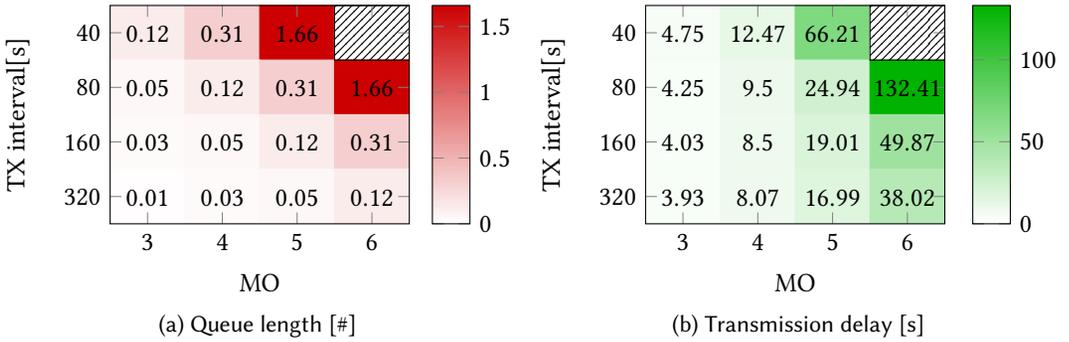
\begin{figure}
    \subfloat[{Queue length [\#]}]{\tikzsetnextfilename{msf_heatmap}

\begin{tikzpicture}
\begin{axis}[enlargelimits=false,
    colormap={col_mod}{
rgb(0cm)=(1,1,1); rgb(1cm)=(0.8,0,0)
  },
  colormap name=col_mod,
  width=0.4\textwidth,
  height=0.32\textwidth,
  colorbar,
    yticklabels={0,40,80,160,320},
    xticklabels={2,3,4,5,6},
  ylabel={TX interval[s]},
  xlabel={MO},
  point meta min=0,
  point meta max=1.66,
  nodes near coords={\pgfmathfloatifflags{\pgfplotspointmeta}{0}{}{\pgfmathprintnumber{\pgfplotspointmeta}}},
  nodes near coords style={
      /pgf/number format/fixed,
  },
  every node near coord/.append style={xshift=0pt,yshift=-7pt, black},
]

\addplot[
matrix plot,
mesh/cols=4,
mesh/rows=4,
point meta=explicit]
table[meta=C]{
  x y C
  0 0 0.12
  1 0 0.31
  2 0 1.66
  3 0 0
  0 1 0.05
  1 1 0.12
  2 1 0.31
  3 1 1.66
  0 2 0.03
  1 2 0.05
  2 2 0.12
  3 2 0.31
  0 3 0.01
  1 3 0.03
  2 3 0.05
  3 3 0.12
};
\draw [fill=white,pattern=north east lines] (axis cs: 2.5,0.5) rectangle (axis cs: 3.5,-0.5);
\end{axis}
\end{tikzpicture}
 \label{fig:heatmap_ql}}
    \subfloat[{Transmission delay [s]}]{\tikzsetnextfilename{msf_heatmap_ct}

\begin{tikzpicture}
\begin{axis}[enlargelimits=false,
    colormap={col_mod}{
rgb(0cm)=(1,1,1); rgb(1cm)=(0,0.7,0)
  },
  colormap name=col_mod,
  width=0.4\textwidth,
  height=0.32\textwidth,
  colorbar,
    yticklabels={0,40,80,160,320},
    xticklabels={2,3,4,5,6},
  ylabel={TX interval[s]},
  xlabel={MO},
  point meta min=0,
  point meta max=133,
  nodes near coords={\pgfmathfloatifflags{\pgfplotspointmeta}{0}{}{\pgfmathprintnumber{\pgfplotspointmeta}}},
  nodes near coords style={
      /pgf/number format/fixed,
  },
  every node near coord/.append style={xshift=0pt,yshift=-7pt, black},
]

\addplot[
matrix plot,
mesh/cols=4,
mesh/rows=4,
point meta=explicit]
table[meta=C]{
  x y C
  0 0 4.75
  1 0 12.47
  2 0 66.21
  3 0 0
  0 1 4.25
  1 1 9.50
  2 1 24.94
  3 1 132.41
  0 2 4.03
  1 2 8.50
  2 2 19.01
  3 2 49.87
  0 3 3.93
  1 3 8.07
  2 3 16.99
  3 3 38.02
};
\draw [fill=white,pattern=north east lines] (axis cs: 2.5,0.5) rectangle (axis cs: 3.5,-0.5);
\end{axis}
\end{tikzpicture}
 \label{fig:heatmap_ct}}
    \caption{Queue length \ref{fig:heatmap_ql} and transmission delay \ref{fig:heatmap_ct} for varying transmission
    intervals and multisuperframe orders. Invalid configurations are marked with hatches.}
    \label{fig:heatmap}
\end{figure}

We analyze the effect of the multisuperframe duration on the average queue length and
transmission delay on transmissions during \gls{cfp}.
\autoref{fig:heatmap} shows the results of the analytical stochastic model (\autoref{sec:analytics}) for queue length
(left) and transmission delay (right) of unconfirmed transmissions during \gls{cfp}, for
different multisuperframe configurations and transmission intervals.
The results show that the average queue length increases with
the multisuperframe order. An increment of one multisuperframe order duplicates
the number of superframes per multisuperframe, ergo the multisuperframe duration.
With a fixed transmission interval this situation increases the system utilization
($\rho = \lambda \cdot T_{msf}$),
which reflects the higher queue length. The queue lengths in the diagonals are
equal, as a result of equal system utilization.

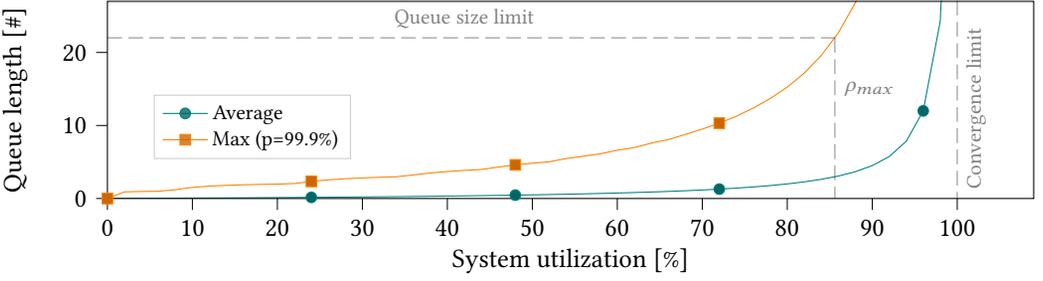
\begin{figure}
    \tikzsetnextfilename{ql_rho}

\begin{tikzpicture}
\definecolor{color0}{rgb}{0.905882352941176,0.16078431372549,0.541176470588235}
\definecolor{color1}{rgb}{0.850980392156863,0.372549019607843,0.00784313725490196}
\definecolor{color2}{rgb}{0.458823529411765,0.43921568627451,0.701960784313725}
\definecolor{color3}{rgb}{0.105882352941176,0.619607843137255,0.466666666666667}
\definecolor{color4}{rgb}{0.4,0.650980392156863,0.117647058823529}
\begin{groupplot}[
group style={
  group size=1 by 1,
  horizontal sep=0.25cm,
  vertical sep=0.25cm,
},
height=0.3\textwidth,
width=\textwidth,
xmax=1.09,
ymax=27,
axis on top,
xmin=0, ymin=0,
legend cell align={left},
legend style={
  fill opacity=0.8,
  draw opacity=1,
  text opacity=1,
  at={(0.05,0.2)},
  nodes={scale=0.8, transform shape},
  anchor=south west,
  draw=white!80!black,
},
tick align=outside,
tick pos=left,
x grid style={white!69.0196078431373!black},
xtick style={color=black},
y grid style={white!69.0196078431373!black},
ytick style={color=black},
xlabel={System utilization [\%]},
ylabel={Queue length [\#]},
cycle list name=exotic,
xtick={0,0.1,0.2,0.3,0.4,0.5,0.6,0.7,0.8,0.9,1},
xticklabels={0,10,20,30,40,50,60,70,80,90,100}
]
\nextgroupplot[]
\addplot+ [mark repeat=12, mark size=2] table [col sep=comma,x index=1, y index=3] {data/ql_rho.csv};
\addlegendentry{Average}
\path [draw=black, draw opacity=0.3, semithick, dash pattern=on 5.55pt off 2.4pt]
(axis cs:1,30)
--(axis cs:1,0);
\addplot+ [mark repeat=12, mark size=2] table [col sep=comma,x index=1, y index=2] {data/ql_rho.csv};
\addlegendentry{Max (p=99.9\%)}
    \node at (axis cs:1,0) [text opacity=0.5, rotate=90,anchor = north west] {\footnotesize{Convergence limit}};
\path [draw=black, draw opacity=0.3, semithick, dash pattern=on 5.55pt off 2.4pt]
(axis cs:0.856,22)
--(axis cs:0.856,0);
\path [draw=black, draw opacity=0.3, semithick, dash pattern=on 5.55pt off 2.4pt]
(axis cs:0,22)
--(axis cs:0.856,22);
    \node at (axis cs:0.42,22) [text opacity=0.5, anchor = south] {\footnotesize{Queue size limit}};
\node at (axis cs:0.856,15) [text opacity=0.5, anchor = west] {\footnotesize{$\rho_{max}$}};
\end{groupplot}
\end{tikzpicture}

     \caption{Average and maximum queue length for varying system utilization cases ($\rho$).}
    \label{fig:ql_rho}
\end{figure}

Observe that variations in the queue length, as a result of variations in the
system utilization, are higher as the system utilization approaches 100\%, as depicted
in~\autoref{fig:ql_rho}. This reflects in the higher queue utilization in the upper
right corner (Figure~\autoref{fig:heatmap_ql}). Transmissions with TX interval=40\,s
and multisuperframe order of 6 ($T_{msf}$=61.44\,s) are out of the convergence
region, hence, they are not shown in the figure.

Figure~\autoref{fig:heatmap_ct} reflects that the transmission delay is the product
between the transmission interval and the average queue length, as seen in~\autoref{sec:delay}. The increase in queue length on varying multisuperframe
reflects in the increased transmission delay on higher multisuperframe orders.
Note that equal queue lengths reflect different transmission delays, as a result of
the longer multisuperframe duration.

We use the model to calculate the worst case scenario of queue length for
a given system utilization (\autoref{fig:ql_rho}). We define the worst case scenario as the maximum
queue length with a confidence of 99.9\%. The results show that the queue length in the worst case
scenario increases linearly until a system utilization of 60\%, from where the
queue length grows exponentially.
The queue length exceeds the maximum of \textit{openDSME} (22 frames)
at $\rho_{max}$=85.6\%. We use this value to calculate
the throughput ($\lambda_{max}$) for each multisuperframe order, given that
$\rho_{max} = \lambda_{max} \cdot T_{msf}$. We compare the throughput results from the
model against simulation environment results (\autoref{tab:throughput}).
The model show the maximum throughput (401.24 packets/hour) with \gls{mo}=3.
Observe that an increment in \gls{mo} halves the maximum throughput (\autoref{tab:throughput},
second column, top to bottom). This is required to maintain the maximum system
utilization (85.6\%), because an increment in \gls{mo} duplicates the
multisuperframe duration (\autoref{tab:msf_duration}),
The model shows a deviation of less than 0.02\% with respect to
the simulation.

\begin{table}
\centering
    \caption{Comparison of throughput ($\lambda_{max}$) using results from the analytical stochastic model and simulation, for \gls{gts} transmission with 85.6\% system utilization (maximum queue length), for single \gls{gts} allocation and varying multisuperframe orders.}
    \label{tab:throughput}
    \begin{tabular}{crrr}
        \toprule
\textbf{\acrlong{mo}} & \multicolumn{2}{c}{\textbf{Throughput [$\frac{packets}{hour}$]}} & \textbf{Error [\%]}\\
        \cmidrule(lr){2-3}
        & Model & Simulation & \\
        \midrule
        3 & 401.24  & 401.25   & $2.5\cdot10^{-3}$\\
        4 & 200.58  & 200.62  & $1.9\cdot10^{-2}$\\
        5 & 100.29 & 100.31 & $1.9\cdot10^{-2}$\\
        6 & 50.15 & 50.16 & $2.5\cdot10^{-3}$\\
        \bottomrule
    \end{tabular}
\end{table}

 \section{Design discussions}\label{sec:discussion}

Based on the evaluation results and the analytical stochastic model, we
can now discuss optimal transmission patterns for different scenarios and trade-offs
between different superframe configurations. We also compare design options to
comply with local regulations and improve the energy consumption.
Finally, we discuss \gls{dsme}-LoRa operation under cross-traffic.

\subsection{Analysis on data transmission}\label{sec:design_tx}

Our evaluation reveals different properties for \gls{csma} and \gls{gts} transmissions.
On the one hand,  \gls{csma} transmissions in low on-air traffic show
lower transmission delays than \gls{gts} transmissions and similar packet
reception ratio. In high on-air traffic scenarios, \gls{csma} failures and
packet collisions increase transmission delays and sharply reduces packet reception, 
which render \gls{csma} transmission unusable for periodic communication in
large scale networks.
Confirmed transmissions improve the packet reception ratio for \gls{csma}, but
increases transmission delay.

On the other hand, \gls{gts} transmissions admit 
$\approx$ 100\% packet reception ratios and transmission delays for transmission
intervals below the maximum system utilization, (85.6\% of the multisuperframe duration, see~\autoref{sec:impact_msf}).
The  delay of \gls{gts} transmission depends
only on the \gls{mac} queue length at the moment of packet scheduling. For applications
that require class-based service differentiation, priority level of \gls{dsme} transmissions
has potentials to reduce transmission delay of high priority messages.
The network size does not affect transmission delay nor packet reception ratio 
for devices in a network that allocate a \gls{gts}.
In contrast to \gls{csma}, unconfirmed transmissions in \gls{gts} perform similar to
confirmed transmissions.
Hence, we recommend to use confirmed transmission in \gls{gts} only for high
priority data.

Due to the deterministic behavior and very high packet reception ratio,
\gls{gts} transmission is a better alternative than \gls{csma} transmissions
for reliable large scale unicast communication. However, \gls{csma} transmissions are still important
for two reasons:
\one \gls{csma} support broadcast frame transmissions. This
makes \gls{csma} transmissions effective for application where a small group
of devices broadcasts data to multiple receivers, such as firmware update scenarios.
We leave the evaluation of broadcast transmissions for future work.
\two the \gls{cap} is used for transmission of \gls{mac} commands required for
association and slot allocation.

To optimize \gls{csma} transmissions, we recommend the usage of small \gls{csma} backoff exponent
settings for scenarios with low on-air traffic, aiming to reduce transmission delay. On the
other hand, we recommend a high backoff exponent to increase the packet reception ratio for
higher on-air traffic, in order to increase \gls{prr}.

In common LoRaWAN deployments, the addition of gateways increases \gls{prr} by
exploiting the capture effect (see~\autoref{sec:lora_modulation}). On frame collision,
a fraction of LoRaWAN gateways can still recover a frame if the power difference with the colliding frames is large enough. Due to the gateway-less
nature of \gls{dsme}-LoRa, it is not possible to increase \gls{prr} of unicast frames by adding more
receivers. However, the capture effect can improve delivery of broadcast
frames, in which a group of devices successfully decode the broadcast frame despite collision.
For example, devices at a close distance to a coordinator may still successfully decode beacons under
LoRa cross-traffic interference.

We show that \gls{cad} improves the performance of \gls{csma} transmissions by reducing
collisions, which effectively increases \gls{prr} and reduces
frame retransmissions (\autoref{sec:eff_cad_retrans}). The latter
not only reduce the time on air of devices, but reduces energy consumption.
We believe it is possible to reduce collisions even further by utilizing a more sophisticated
\gls{csma} mechanism such as LMAC~\cite{glgtl-lecmal-20}.

\subsection{Selection of multisuperframe configuration}

Our evaluation shows that smaller multisuperframe orders decrease the delay
of \gls{gts} transmissions.
However, this reduces the \gls{gts} resources (\autoref{tab:msf_duration}).
Although each \gls{gts} defines multiple unique frequency slots, a device can only
allocate one frequency slot per \gls{gts}. This limits the number of \gls{gts} links
per device to the number of \gls{gts} in the multisuperframe structure, which in turn 
favors cluster-tree and peer to peer topologies over star topologies.

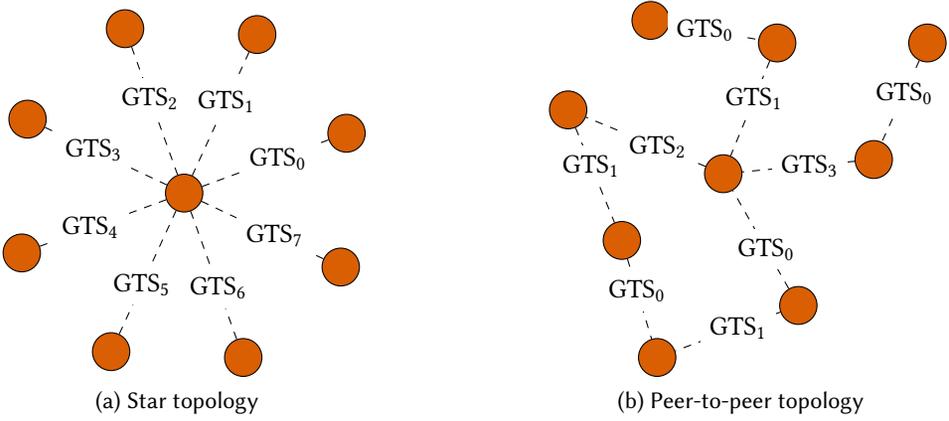
\begin{figure}
    \subfloat[Star topology]{\tikzsetnextfilename{gts_resources_start}

\definecolor{color1}{rgb}{0.850980392156863,0.372549019607843,0.00784313725490196}
\definecolor{color2}{rgb}{0.458823529411765,0.43921568627451,0.701960784313725}
\begin{tikzpicture}[>={Latex}]
  \tikzset{
    source/.style = {draw, fill=color1, circle, inner sep=5pt },
    sink/.style = {draw, fill=color2, circle, inner sep=5pt },
  }

  \node[source] (sink) at (0,0) {};
  \foreach \y in {0,...,7} {
      \node[source] (source_\y) at ({20+\y*45:2.3cm}) {};
      \draw [dashed] (source_\y) -- (sink) node[pos=0.4,fill=white] {GTS\textbf{$_{\y}$}};
  }

\end{tikzpicture}
 \label{fig:gts_topology_star}}
    \hspace{2cm}
    \subfloat[Peer-to-peer topology]{\tikzsetnextfilename{gts_resources_p2p}

\definecolor{color1}{rgb}{0.850980392156863,0.372549019607843,0.00784313725490196}
\definecolor{color2}{rgb}{0.458823529411765,0.43921568627451,0.701960784313725}
\begin{tikzpicture}[>={Latex}]
  \tikzset{
    source/.style = {draw, fill=color1, circle, inner sep=5pt },
    sink/.style = {draw, fill=color2, circle, inner sep=5pt },
  }

  \node[source] (node0) at (0cm,0cm) {};
  \node[source] (node1) at ({120:2cm}) {};
  \node[source] (node2) at ({160:2.5cm}) {};
  \node[source] (node3) at ({200:2cm}) {};
  \node[source] (node4) at ([shift=({140:2cm})] {140:2cm}) {};
  \node[source] (node5) at ([shift=({140:2cm})] {100:2.5cm}) {};
  \node[source] (node6) at ([shift=({140:2cm})] {60:2.5cm}) {};
  \node[source] (node7) at ([shift=({-50:2cm})]node6) {};
  \node[source] (node8) at ([shift=({0:2cm})]node6) {};

  \draw [dashed] (node0) -- (node1) node [pos=0.4,fill=white] {GTS$_0$};
  \draw [dashed] (node0) -- (node3) node [pos=0.4,fill=white] {GTS$_1$};
  \draw [dashed] (node2) -- (node3) node [pos=0.4,fill=white] {GTS$_0$};
  \draw [dashed] (node4) -- (node2) node [pos=0.4,fill=white] {GTS$_1$};
  \draw [dashed] (node1) -- (node4) node [pos=0.4,fill=white] {GTS$_2$};
  \draw [dashed] (node5) -- (node6) node [pos=0.4,fill=white] {GTS$_0$};
  \draw [dashed] (node6) -- (node1) node [pos=0.4,fill=white] {GTS$_1$};
  \draw [dashed] (node7) -- (node1) node [pos=0.4,fill=white] {GTS$_3$};
  \draw [dashed] (node8) -- (node7) node [pos=0.4,fill=white] {GTS$_0$};

\end{tikzpicture}
 \label{fig:gts_topology_p2p}}
    \caption{\gls{gts} assignment between neighbour devices for star topology (left)
    and peer-to-peer topology (right)}
    \label{fig:gts_topology}
\end{figure}

Consider the two example topologies from \autoref{fig:gts_topology} with 9 devices.
In the star topology (Figure~\autoref{fig:gts_topology_star}) each child device allocates
a transmission \gls{gts} with the coordinator. In the peer-to-peer topology each
device allocates a transmission \gls{gts} with another device of the network.
The star topology requires 8 \gls{gts}, one per child device, to establish all
links. Hence, the superframe structure requires at least two superframes per multisuperframe,
which sets the multisuperframe duration to at least 15.36\,s.

On the other hand, the peer-to-peer topology (Figure~\autoref{fig:gts_topology_p2p}) allows transmitting frames
in the same \gls{gts}, using different channels. As a result, only 4 \gls{gts} are
needed to schedule all transmissions. Therefore, a configuration with one superframe
per superframe suffices, which sets the multisuperframe duration to 7.68\,s.
Under the same data transmission rates, the peer-to-peer topology reflects
shorter transmission delays than the star topology, as a result of the shorter
multisuperframe duration.
In contrast to the star topology, the peer-to-peer topology does not use all available \gls{gts} resources, which allows to further extend the network.

For scenarios with more than two superframes per multisuperframe, the \gls{cap}
reduction mechanism offers a solution to extend the \gls{gts} resources, in which
the \gls{cap} of all superframes in a multisuperframe, excluding the first, is replaced
by 8 \gls{gts}. However, this reduces the \gls{cap} time of a multisuperframe, which
stresses \gls{csma} transmissions and thereby challenges dynamic \gls{gts} allocation.
We will analyze the impact of \gls{cap} reduction on slot allocation in
future work.

\subsection{Compliance with regional regulations}\label{sec:regulations}

\paragraph{Regions with duty cycle restrictions}
We show in \autoref{sec:time_on_air} that unconfirmed data transmission to
neighbouring nodes does not stress the time on air resources of the network in regions
with duty cycle restrictions, because
transmissions do not require an intermediate forwarder (in contrast to LoRaWAN).
This makes \gls{dsme}-LoRa suitable for scenarios, in which a
series of devices communicate directly with one or more sink devices (see~\autoref{sec:problem_statement}).
On the other hand, confirmed transmissions stress time on air resources of
sink devices (\autoref{sec:time_on_air}). It is therefore crucial to limit the transmission interval for scenarios, in which
a sink device receives packets from multiple source devices, to ensure
compliance with duty cycle restrictions.

\begin{figure}
    \tikzsetnextfilename{design_dc}

\begin{tikzpicture}
\definecolor{color0}{rgb}{0.905882352941176,0.16078431372549,0.541176470588235}
\definecolor{color1}{rgb}{0.850980392156863,0.372549019607843,0.00784313725490196}
\definecolor{color2}{rgb}{0.458823529411765,0.43921568627451,0.701960784313725}
\definecolor{color3}{rgb}{0.105882352941176,0.619607843137255,0.466666666666667}
\definecolor{color4}{rgb}{0.4,0.650980392156863,0.117647058823529}
\begin{groupplot}[
group style={
  group size=1 by 1,
  horizontal sep=0.25cm,
  vertical sep=0.25cm,
},
height=0.35\textwidth,
width=\textwidth,
axis on top,
xmin=0, ymin=0,
xmax=300,
legend cell align={left},
legend style={
  fill opacity=0.8,
  draw opacity=1,
  text opacity=1,
  at={(0.97,0.63)},
  nodes={scale=0.8, transform shape},
  anchor=south east,
  draw=white!80!black,
},
tick align=outside,
tick pos=left,
x grid style={white!69.0196078431373!black},
xtick style={color=black},
y grid style={white!69.0196078431373!black},
ytick style={color=black},
ytick={0,1,2,3,4},
yticklabels={$10^0$,$10^1$,$10^2$,$10^3$,$10^4$},
ylabel={Transmission rate [$\frac{packets}{hour}$]},
xlabel={Source devices per sink[\#]},
cycle list name=exotic,
]
\nextgroupplot[]
\addplot+ [thick, mark repeat=10, mark size=2] table [col sep=comma, x index=0, y index=2] {data/design_dc.csv};
\addlegendentry{10\% band}
\addplot+ [thick, mark repeat=10, mark size=2] table [col sep=comma, x index=0, y index=1] {data/design_dc.csv};
\addlegendentry{1\% band}
\end{groupplot}
\end{tikzpicture}
     \caption{Maximum transmission rate of a source device that transmits confirmed data (16 bytes payload) to
    a single sink (star topology), for varying number of source devices in the network and
    duty cycle}
    \label{fig:design_dc}
\end{figure}
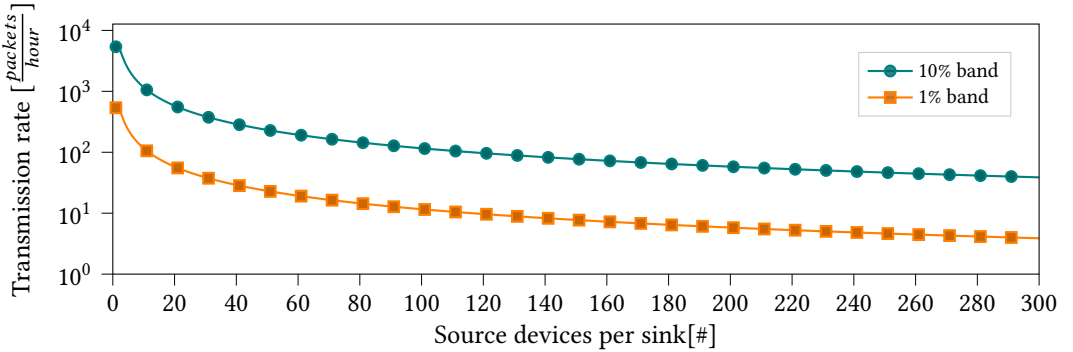

We show in \autoref{fig:design_dc} the theoretical limits of transmission rate
per source device as a function of source devices that transmit to a single
sink  (star topology). We assume that source devices do not retransmit
data (100\% packet reception ratio on first transmission).
For example, a star topology with 10 source devices allows transmission of
$\approx$ 100 packets per hour in the 1\% band and 1,000 packets per hour in
the 10\% band without exceeding the duty cycle restriction. On the other hand, a network with 115 source devices allows in the 1\% and 10\% bands transmission rates of 10 and 100 packets per hour, respectively.

\gls{csma} transmissions benefit from the 10\% band.
Nevertheless, the majority of channels in \gls{cfp} belong to the 1\% band,
which restricts data transmissions on source devices and ACK frames on sink devices.
We propose two potential solutions to overcome these limitations
\one use the group ACK feature (\autoref{sec:back_dsme}), which restricts
ACK transmission to one common ACK frame per multisuperframe.
Group ACKs do not contribute to better performance over regular ACK~\cite{mmdt-gawai-21}, but reduce the time on air utilization of sink devices by removing the
dependency to the number of source devices. \two distribute the \gls{dsme}-LoRa
channels in more bands. For example,
channels 11 \textendash{} 25 (see~\autoref{tab:frequencies}) can be arranged
into twelve channels in the \textit{g} band, two channels in the \textit{g1} band (868.0 \textendash{} 868.6 MHz) and
one channel in the \textit{g4} band (869.7 \textendash{} 870.0 MHz). If \gls{gts} transmissions
distribute evenly among channels. This allows for $\approx$ 20\% additional transmissions per device.

\paragraph{Regions with dwell time or channel hopping}\label{sec:analysis_chop}
Dwell time requirements (\eg in US902 \textendash{} 928 ) can be easily addressed by restricting
payload size, however, the channel hopping requirement (\eg in US902 \textendash{} 928 and CN779 \textendash{} 787) is incompatible with single channel
communication during \gls{cap}.
While on the one hand, limiting \gls{cap} transmissions in these regions is not an option, because
\gls{cap} is required for slot allocation and \gls{mac} control traffic,
enabling multichannel \gls{cap},  on the other hand, would require devices to listen on multiple channels
(\eg by using a LoRa concentrator). Although feasible, it increases
deployment costs.

We argue that \gls{fhss} transmissions (see~\autoref{sec:lora_modulation}) can enable transmissions 
during \gls{cap} for these regions.
In this regard, the channel number may dictate a unique \gls{fhss} sequence.
This addresses the problem of channel hopping and dwell time, since transmissions are spread
among different carrier frequencies. Thereby the transmission time per channel is reduced.
Still, it degrades \gls{cca} performance, because \gls{cad} can detect only one carrier frequency
at a time. Therefore, the \gls{cca} implementation requires a different strategy.
We will address this problem in future work.

\subsection{Energy considerations}\label{sec:improve}

On standard deployments, the passive consumption of \gls{cap} is 146.12\,mJ per superframe (19.02\,mW) and therefore not a good option for battery powered devices.
To overcome this problem, we proposed to turn off the \gls{cap} in battery powered
devices, as shown in \autoref{sec:energy_consumption}. Although this prevents frame
reception, the indirect transmission feature
(\autoref{sec:back_dsme}) provides a mechanism to communicate with a device
with the receiver off during \gls{cap}. On the other side, a device
can still turn on the transceiver during \gls{cap} to transmit data to other devices. This allows, for
example, to trigger \gls{gts} RX allocation from battery powered device.

\begin{figure}
    \tikzsetnextfilename{power_calc}

\begin{tikzpicture}
\definecolor{color0}{rgb}{0.905882352941176,0.16078431372549,0.541176470588235}
\definecolor{color1}{rgb}{0.850980392156863,0.372549019607843,0.00784313725490196}
\definecolor{color2}{rgb}{0.458823529411765,0.43921568627451,0.701960784313725}
\definecolor{color3}{rgb}{0.105882352941176,0.619607843137255,0.466666666666667}
\definecolor{color4}{rgb}{0.4,0.650980392156863,0.117647058823529}
\begin{axis}[
height=0.27\textwidth,
width=\textwidth,
legend cell align={left},
legend style={
  reverse legend,
  fill opacity=0.8,
  draw opacity=1,
  text opacity=1,
  at={(0.9,0.9)},
  nodes={scale=0.8, transform shape},
  anchor=north east,
  draw=white!80!black,
},
/pgf/bar width=10pt,
tick align=outside,
tick pos=left,
x grid style={white!69.0196078431373!black},
xtick style={color=black},
y grid style={white!69.0196078431373!black},
ytick style={color=black},
xlabel={Beacon order},
ylabel={Power [mW]},
ybar stacked,
symbolic x coords={3,4,5,6,7},
xtick=data,
cycle list name=exotic_bars,
]

\addplot+ [] coordinates {
    (3,2.392500)
    (4,1.196250)
    (5,0.598125)
    (6,0.299063)
    (7,0.149531)
};
\addlegendentry{Beacon slot}
\addplot+ [] coordinates {
    (3,0.344062)
    (4,0.344062)
    (5,0.344062)
    (6,0.344062)
    (7,0.344062)
};
\addlegendentry{\acrshort{cap} + \acrshort{cfp} period}
\end{axis}
\end{tikzpicture}
     \caption{Passive consumption of superframe structure with transceiver off during \gls{cap} (\textit{macRxOnWhenIdle=0}),
    for varying beacon order.}
    \label{fig:power_calc}
\end{figure}
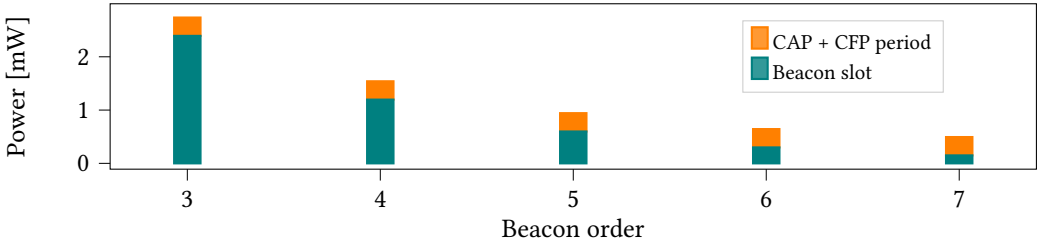

We show in \autoref{sec:energy_consumption} that the beacon
period has a high impact in the energy consumption. A way to improve this situation
is to configure a higher beacon order, which results on a higher beacon interval and
therefore reduces the passive consumption of the beacon slot.
We estimate the power consumption from the measurements in \autoref{sec:energy_consumption} and present the base consumption (superframe structure) for different beacon
order configurations in \autoref{fig:power_calc}.

Noteworthy, the beacon order has a high impact on the energy consumption. In the
scenario with \gls{bo}=3, the total passive consumption hits 2.73\,mW, in which
the beacon period consumes $\approx$ 87\%. On
the contrary in scenarios with \gls{bo}=7 the power consumption is 0.49\,mW, in which the beacon period consumes $\approx$ 30\% of the energy. Note that \gls{bo}=7
renders the beacon period to $\approx$ 122.88\,s -- in line with the duration of LoRaWAN
class B beacon period (128\,s).

Inherently, higher beacon intervals have two potential
problems:  The scanning procedure takes longer, which
increases the energy consumption during association, and the
devices synchronize to their neighbours less often, which
potentially leads to desynchronization due to clock drifts.
Adding more coordinators mitigates the longer association times, because the
frequency of beacons increases. The use of real time clocks, available in
common LoRa target platforms, mitigates desynchronization issues (\autoref{sec:implementation})

\begin{table}
\centering
    \caption{Comparison of average transmission delay, power, and lifetime for
    a \gls{dsme}-LoRa sender device with TX interval=15\,m and \gls{bo}=7, for varying multisuperframe order.}
    \label{tab:lifetime}
    \begin{tabular}{crrr}
        \toprule
        \textbf{\acrlong{mo}} & \textbf{Delay [s]} & \textbf{Power [mW]} & \textbf{Lifetime [y]} \\
        \midrule
        3 & 3.87 & 0.58 & 1.82 \\
        4 & 7.81 & 0.47 & 2.24 \\
        5 & 15.9 & 0.42 & 2.53 \\
        6 & 32.97 & 0.39 & 2.71 \\
        7 & 71.16 & 0.38 & 2.81 \\
        \bottomrule
    \end{tabular}
\end{table}

To illustrate the trade-off between transmission delay and energy consumption, we present
in \autoref{tab:lifetime} the energy consumption and lifetime of a \gls{dsme}-LoRa sender node for
different multisuperframe orders.
We assume exponentially distributed interarrival times with TX interval=15\,m.
The beacon interval is set to 122.28\,s (\gls{bo}=7) -- in line with LoRaWAN class B beacon interval -- 
and assume the device keeps the transceiver on for two beacon intervals to associate with a single coordinator.
We also estimate the voltage regulator efficiency to be 90\%.
For the lifetime estimation we assume the device operates on a battery of 2800\,mAh capacity -- in line
with common off the shelf AA alkaline batteries. We utilize the model (\autoref{sec:analytics}) to estimate transmission delay.

\gls{mo}=7 renders the lowest power consumption (0.38\,mW) and allows $\approx$
3 years of operation. However, it also depicts the highest average latency ($\approx$ 71\,s).
Note that the delay can increase up to the beacon duration (122.28\,s) if the packet
is scheduled right after the boundary of the \gls{gts}. If the use case does not
tolerate high delays, the device may opt for a lower multisuperframe order.
Please observe that \one the energy consumption decreases with an increase of multisuperframe order
and \two the energy consumption decreases at a lower rate on higher multisuperframe
order. For example, a reduction from \gls{mo}=7 to \gls{mo}=6 increases consumption only by 0.01\,mW,
while a reduction from \gls{mo}=4 to \gls{mo}=3 increases 0.11\,mW.
This effect occurs because a decrement in \gls{mo} duplicates the number of \gls{gts}
per beacon interval. Recall that a scheduled \gls{gts}-TX consumes 2.2\,mJ even if there is no transmission (see~\autoref{sec:energy_consumption}), which reflects the increase in energy
consumption.
A device with \gls{mo}=3 renders an average delay of 3.87\,s and lifetime of $\approx$ 2 years. If transmission delay is not critical, a device can extend the
lifetime up to ~1 year by setting \gls{mo}=7. Although \gls{mo} beyond 7
is possible, it increases the beacon interval, which
challenges device synchronization. In general, the energy footprint of \gls{dsme}-LoRA
for uplink oriented applications is higher than LoRaWAN, considering an equivalent LoRaWAN class A device can operate $>$ 10 years running on batteries.

Finally, we present potential optimizations for \textit{openDSME} and the integration
into RIOT, aiming to reduce power consumption: \one avoid turning on the
transceiver on a TX \gls{gts} if the \acrshort{mac} queue is empty, which
reduces the energy consumption by $\approx$ 1.5\,mJ per superframe ($\approx$
0.20\,mW) for the  allocation of one \gls{gts} TX slot. This is possible with a
minor change in the \gls{gts} management routines of \textit{openDSME}. \two turn off
the transceiver in the beacon slot right after the reception of the beacon,
which potentially reduces $\approx$ 16\,mJ in the receiving beacon slot. \three
use \gls{cad} to detect the preamble of a LoRa frame at the beginning of a \gls{gts}
RX and proceed to reception if \gls{cad} succeeds, instead of keeping the
transceiver idle listening for the duration of the slot (0.48\,s). For example. three
\gls{cad} attempts consume 0.73\,mJ, as analyzed in \autoref{sec:energy_consumption}.
Under this scenario, the passive consumption of the \gls{cfp} reduces by 18.58\,mJ per
superframe (2.41\,mW).

\subsection{Analysis of cross-traffic}\label{sec:xtraffic}

We show in \autoref{sec:coexistence} that \gls{dsme}-LoRa can coexist with
LoRaWAN in the EU868 region, because LoRaWAN traffic neither interferes  with \gls{dsme} \gls{mac} control
traffic nor with beacons. Therefore, concurrent \gls{dsme}-LoRa and LoRaWAN
communication only degrades \gls{prr}. As long as the common channel does not
overlap with LoRaWAN channels, the coexistence of \gls{dsme}-LoRa and LoRaWAN
in regions other than EU868 is feasible.

The channel hopping mode in \gls{gts} allows communication despite heavy
interference on a single channel, but degrades \gls{prr} as a result of a fraction
of packets being transmitted in the noisy channel. To overcome this problem,
a device may transmit confirmed messages. Thereby the \gls{mac} will perform the retransmission
on a channel with better quality. An alternative solution is to use the
channel adaptation mode of \gls{dsme}-LoRa (see~\autoref{sec:back_dsme}), in
which the source and target device agree on a different channel if the channel quality
is poor. Although \textit{openDSME} implements the channel adaptation mode, it
does not implement the required \gls{mac} command (\gls{dsme} Link Report) to request channel quality information. Therefore,
there is no way to infer channel quality and agree on a different channel.

Poor channel quality in the common channel challenges device synchronization (see~\autoref{sec:common_channel}), which prevents normal operation of the \gls{dsme}-LoRa network.
While this is also a problem for standard \gls{dsme}, the long time on air and long
range of LoRa packets represents a security thread for \gls{dsme}-LoRa networks. 
Attackers may desynchronize devices by jamming the channel during beacon transmissions.

To address this problem, coordinators may request children devices to switch to
a channel with better quality using the \acrshort{phy}-OP-SWITCH mechanism
(see~\autoref{sec:network_formation}). This requires the coordinator device to
estimate the channel quality, for example, by keeping track of the failed \gls{cca}
attempts during \gls{csma} transmissions. However, the \gls{mac} control frames required
by the \acrshort{phy}-OP-SWITCH mechanism are sent during \gls{cap}, which challenges packet
delivery under noisy conditions. Also, a device may detect good channel quality
during \gls{cap} even if an attacker jams only beacon frames.
An alternative solution is to transmit frames using the \gls{fhss}, as analyzed in 
\autoref{sec:analysis_chop}. Thereby packet transmissions can tolerate noise in a single
channel, by relying on forward error correction mechanisms on the LoRa \gls{phy}. To prevent
selective jam attacks, packets can be transmitted with a pseudorandom \gls{fhss} sequence shared by all devices. We will analyze this proposal in future work.
 \section{Related work}\label{sec:related}

\subsection{IEEE 802.15.4 standards \gls{tsch} and \gls{dsme}}\label{sec:related_802154}

The 802.15.4 \acrshort{mac} modes \gls{tsch} and \gls{dsme}~\cite{IEEE-802.15.4-16} have been analyzed~\cite{kskt-inspe-18}, modeled~\cite{jagfd-ecapi-16, cmml-paimm-20}, and simulated~\cite{jl-peidm-12, apmb-saidt-15,e-tssta-20}.
The results indicate that \gls{tsch} obtains lower latency and higher throughput for small networks ($<$ 30 nodes). \gls{dsme} outperforms \gls{tsch} for higher duty cycles and an increasing number of nodes.
Kauer~\etal~\cite{kkt-rwmnd-18} introduce \textit{openDSME}, an implementation that is available for \textit{OMNeT++} and as a portable C++ library. We utilize \textit{openDSME} in our work. The authors compare simulated performances to real-world measurements---which are on par---and further investigate group ACKs~\cite{mmdt-gawai-21} that do not contribute to better performance over direct ACKs.
Vallati~\etal~\cite{vbpa-infid-17} find inefficiencies in \gls{dsme} network formation and provide countermeasures, however, we move network formation to future work.
Improvements on the QoS of \gls{dsme} networks were proposed by Kurunathan~\cite{k-iqidn-21}.
Similar to the IETF standard solution \gls{6tisch}~\cite{RFC-8180} for IPv6 over \gls{tsch}, Kurunathan~\etal~ present RPL (Routing Protocol for Low power and Lossy Networks) over \gls{dsme}~\cite{kskt-srasd-20}.
The IETF further provides an applicability statement~\cite{RFC-8036} for RPL in metering use cases and proposes \gls{dsme} as a \acrshort{mac}.
The IEEE 802.15 working group, in contrast, introduces ''Low-Energy Critical Infrastructure Monitoring'' in the \textit{w}-amendment \cite{IEEE-802.15.4-20w}, which adds long-range radios that operate in the sub-GHz band.
These networks are primary defined to operate in star topologies, which supports our topology choice of a single-hop \gls{dsme} network.

\subsection{Analysis of LoRaWAN}\label{sec:sec:related_lorawan}
Existing LoRaWAN~\cite{lorawan-spec-11} networks are susceptible to collisions~\cite{f-cplal-17,ofjg-cccal-19} as well as energy depletion ~\cite{mfpmm-ealev-19}.
Liando~\etal~\cite{lgtl-kufle-19} provide real-world measurements of LoRa and LoRaWAN and explore the impact of transmission parameters of the chirp spread spectrum modulation. They thereby identify optimization potentials for the medium access layer.
Slabicki~\etal~\cite{spd-aclnd-18} contribute \textit{FLoRa}, a LoRa simulator for \textit{OMNeT++}, and improve the adaptive data rate (ADR) mechanism of LoRaWAN. We utilize \textit{FLoRa} in our simulations.
Rizzi~\etal~\cite{rffsg-uliwn-17} and Leonardi ~\etal~\cite{lbbp-calma-20} show that slight modifications of the LoRaWAN \acrshort{mac} already improve performance metrics of class~A deployments, which are centered around the concept of uplink packets from an end node.
LoRaWAN, however, poses a severe challenge on downlink traffic due to band limitations~\cite{ETSI-en3002001-211, shsp-iedct-18} in the sub-GHz band, and contention with unpredictable uplink packets~\cite{mpp-edtpl-18}.
Vincenzo~\etal~\cite{vht-idsl-19} propose countermeasures to that problem, by adding multiple gateways and a gateway selection mechanism. This decreases losses but adds deployment cost.

LoRaWAN class~B (see~\autoref{sec:intro}), though barely deployed, provides periodic downlink slots (unlike class~A\&C) and multicast capabilities~\cite{lorawan-mcast-spec-1} thorough these slots.
Elbsnir~\etal~\cite{ekbd-elced-20} confirm that class~B decreases downlink latency and loss over class~A. Ron~\etal~\cite{rllcl-paodt-20} derive an optimal class~B configuration to trade waiting time with energy consumption, and Pasetti~\etal~\cite{psfrd-eucbl-20} design a single-gateway class~B LoRaWAN network for 312 LoRa nodes. Unfortunately, a practical evaluation is missing.
Operating in class~B, however, suffers from scalability issues~\cite{fbf-eslgc-18,sak-lcbmc-20}.
Despite, class~B still burdens the gateway duty cycle and requires an infrastructure network, hence, it is not an option for long-range node to node communication.

\subsection{New protocols for LoRa}\label{sec:related_newlora}
The IETF standardized a compression~\cite{RFC-9011} scheme for LoRa networks. Similarly, Pere{\v{s}}{\'{i}}ni~\etal~\cite{pk-meict-17} introduce a slim packet format with a new LoRa link layer, to reduce effective payload.
Gonzalez~\etal~\cite{gbv-slpld-18} motivate the development of a new LoRa \acrshort{mac} and present LoRa \acrshort{phy} configurations to define logical channels, which assists frequency- and time division multiple access protocols. We apply these considerations in our work.

Cotrim~\etal~\cite{ck-lmnrc-20} provide a classification for multi-hop LoRaWAN  networks.
Enabling multi-hop with long-range radios is common desire~\cite{stbbs-ermcb-17,tbs-eticl-17,aa-mllbm-19,bgbts-teelm-19}.
New designs of time-slotted LoRa protocols~\cite{zf-tlndc-21} have been analyzed with simulations~\cite{lbpb-ilnma-18,yte-talsa-19} and practical deployments~\cite{zakp-ttlii-20}.
Haubro~\etal~\cite{hoof-tlrri-20} present an adaptation of the 802.15.4 \gls{tsch} mode~\cite{IEEE-802.15.4-16} for LoRa. Their real-world measurements show the applicability of 802.15.4 \acrshort{mac} layers for long-range communication, however, the experiment deployment consists of only three nodes and limited traffic. The analysis of duty-cycle compliance remains open.
In contrast, we focus on LoRa and the 802.15.4 \gls{dsme} mode in our work and aim to fill in the gap of a large-scale deployment which further includes duty-cycle analyses.

Several \gls{mac} and \gls{phy} approaches have been analyzed to overcome the problem of concurrent LoRa communication.
Xu~\etal~\cite{xlyhd-sahsas-20} propose S-MAC, an adaptive scheduling mechanism for \gls{lpwan} that exploits the fact many \gls{lpwan} applications transmit period uplink data. Devices with the same spreading factor and known transmission interval are grouped and assigned
a unique carrier frequency to minimize intergroup frame collisions. The approach brings a 4 \texttimes{} throughput improvement for periodic uplink communication, but does not address downlink limitations of LoRaWAN (see~\autoref{sec:problem_statement}).
The authors of~\cite{p-iecca-18,knwm-pecsl-20} present experimental results for contention based media access with LoRa. Kennedy~\etal~\cite{knwm-pecsl-20} explore \gls{csma} with \gls{cad}.
Results show that listen before talk performs better than ALOHA in dense deployments, which motivated our efforts of using \gls{csma} with \gls{cad} in the contention access period of a \gls{dsme} frame (see~\autoref{sec:csma_ca}). Gamage~\etal~\cite{glgtl-lecmal-20} propose LMAC, an improved \gls{csma} protocol, and evaluate on a testbed the design of three advancing versions of the protocol. Results indicate that the approach brings 2.2\texttimes{} goodput improvement and 2.4\texttimes{} reduction of energy
consumption. We motivate the LMAC approach for future work to reduce collisions during \gls{cap} transmissions (see~\autoref{sec:csma_ca}). There have been multiple proposals to resolve LoRa frames collisions at the physical layer (\cite{spcbk-cicdm-21,xhzg-pslctr-21,xzg-fpdlt-19}). Evaluations of those mechanisms on software defined radio show a clear improvement of throughput and overall network capacity, but add hardware complexity and extra deployment cost in comparison to common off the shelf LoRa nodes.

Little work analyzes alternative communication pattern over LoRa.
Lee~\etal~\cite{lk-mlisu-18} propose gateway driven requests. This approach follows a request-response pattern and indicates performance benefits over producer driven ALOHA. Similarly, the authors of~\cite{lndd-endna-20,dhnzl-tlaci-21,dlhc-bilsp-21} deploy information-centric networking over LoRa radios, which is a data request driven protocol. Their work showed, however, the need for a proper LoRa media access layer.

 \section{Conclusions and outlook}\label{sec:conclusions}
In this work, we exposed the problems of LoRaWAN for node to node communication
and motivated the usage of IEEE 802.15.4 \gls{dsme} over LoRa, which opens LoRa to general networking. We summarized the
\gls{dsme} mappings for the EU868 region with system integration into the operating system RIOT, and presented 
a comprehensive evaluation of \gls{dsme}-LoRa on an IoT testbed. The results revealed that
\gls{csma} transmissions during the contention access period provide a good trade-off
between transmission delay and packet reception ratio for networks with low traffic
and a few nodes. On the contrary, \gls{gts} transmissions show about 100\% packet
reception ratio and predictable transmission delays for networks with higher network
size and higher traffic. We could show that under the limits of available \gls{gts} resources,
these performance metrics do not degrade with the network size. The results also confirmed 
that coexistence between LoRaWAN and \gls{dsme}-LoRa is possible. Nevertheless, noise
in the common channel affects normal operation of the network due to beacon loss.

Our findings confirmed that the Channel Activity Detection feature of
LoRa radios is a powerful clear channel detection mechanism for \gls{csma}, and effectively
reduces the number of retransmissions $\approx$ 15 times in scenarios with moderate
traffic. We evaluated the effect of \gls{csma} backoff exponent settings and
could show that higher values mitigate frame collisions during \gls{cap}. The evaluation
evidenced that direct communication between devices facilitates compliance with regional
duty cycle regulations. We also confirmed that with optimal \gls{mac} configurations, \gls{dsme}-LoRa
offers a passive consumption of less than 1\,mW.
Based on a novel analytical stochastic model we calculated average queue length in the \acrshort{mac} for slotted transmission, from which we estimated the transmission delays.
Validation of the model with data from the experiments using real IoT hardware showed 
an accuracy of 99.99\%.
We also evaluated \gls{dsme}-LoRa for larger network sizes using a well-known simulation environment
and confirmed our experimental findings. We evaluated the effect of
the \acrshort{mac} configuration and utilized the model to optimize throughput
for each configuration.
From the evaluation results we built an overview of transmission patterns and configurations
aiming to provide a good trade-off between transmission delay, time on air, and energy consumption,
 which led to proposing changes in the \acrshort{mac} implementation for improving energy consumption.

There are three future directions of this research. First, recent IETF concepts of the \gls{6tisch} and IPv6
over \acrshort{lpwan} working groups should be adopted while taking advantage of built-in features of \gls{dsme} to enable IPv6 over \gls{dsme}-LoRa. Second, studying dynamic slot allocation between
\gls{dsme}-LoRa nodes can foster deployment experience for real world scenarios. Third, the study of suitable network layers on top of \gls{dsme}-LoRa and its performance under massive industrial deployment~\cite{gklpf-inpmm-21} shall open a new direction of LoRa-centric research.

\paragraph{Acknowledgment} This work was supported in part by the
German Federal Ministry for Education and Research (BMBF)
within the project \textit{PIVOT: Privacy-Integrated design and
Validation in the constrained IoT.}

\paragraph{Availability of software and reproducibility}
We strongly support reproducible research (\cite{acmrep,swgsc-terrc-17}) and utilize open source software
and open testbed platforms. All of our work is intended for public release.
The code of the software components (implementation of \gls{dsme}-LoRa on RIOT, simulation environment),
the implementation of the analytical stochastic model, documentation, data sets and related tools
are available on GitHub at \url{https://github.com/inetrg/tosn-dsmelora22}.

\bibliographystyle{ACM-Reference-Format}

\appendix
\clearpage
\section{Numerical calculation of stationary Markov distribution}\label{sec:pi_calculation}

In this appendix, we want to explain how to evaluate  the station Markov distribution. For this, 
let us define $\vec{x}$ as any eigenvector of $P$ to the value of 1. Note that:
\begin{equation}\label{eq:pi_calc}
\vec{\pi} = \frac{\vec{x}}{\sum_{i=0}^{\infty} x_i}.
\end{equation}

The equation system that describes the eigenvectors is:

\begin{eqnarray*}\label{eq_sys_eigen}
    (k_0 + k_1)x_0 +k_0x_1 &=& x_0 \\
    k_2x_0 +k_1x_1 + k_0x_2 &=& x_1 \\
    k_3x_0 +k_2x_1 + k_1x_2 + k_0x_3 &=& x_2 \\
...\\
    k_0x_{n+1} + \sum_{i=0}^{n} k_{n+1-i}x_i &=& x_{n} \\
\end{eqnarray*}

Which resolves to:
\begin{eqnarray*}
    x_1 &=& x_0\frac{(1-k_0-k_1)}{k_0}\\
    x_i &=& x_0\frac{(1-k_1)x_{i-1} - \sum_{j=0}^{i-2} k_{i-j} x_j}{k_0} ,\forall i \in [2,\infty]\\
\end{eqnarray*}

We set $x_0 = 1$ and calculate ${x_1,x_2,x_3...x_N}$ with $N$ a big enough number.
We then obtain $\vec{\pi}$ using \autoref{eq:pi_calc}.
This method is not practical because it requires the calculation of all vector members.
Therefore, we propose to approximate $\pi_0$ with a polynomial function.

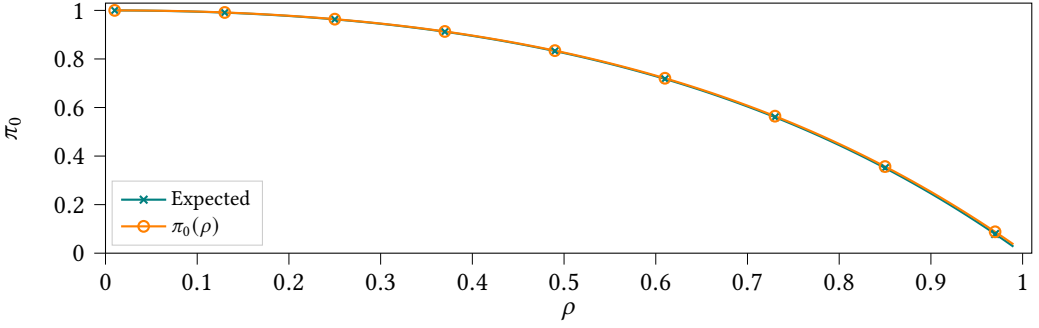
\begin{figure}
    \tikzsetnextfilename{poly}

\begin{tikzpicture}
\definecolor{color0}{rgb}{0.905882352941176,0.16078431372549,0.541176470588235}
\definecolor{color1}{rgb}{0.850980392156863,0.372549019607843,0.00784313725490196}
\definecolor{color2}{rgb}{0.458823529411765,0.43921568627451,0.701960784313725}
\definecolor{color3}{rgb}{0.105882352941176,0.619607843137255,0.466666666666667}
\definecolor{color4}{rgb}{0.4,0.650980392156863,0.117647058823529}
\begin{groupplot}[
group style={
  group size=1 by 1,
  horizontal sep=0.25cm,
  vertical sep=0.25cm,
},
height=0.35\textwidth,
width=\textwidth,
xmax=1.01,
ymax=1.03,
axis on top,
xmin=0, ymin=0,
legend cell align={left},
legend style={
  fill opacity=0.8,
  draw opacity=1,
  text opacity=1,
  at={(0.17,0.03)},
  nodes={scale=0.8, transform shape},
  anchor=south east,
  draw=white!80!black,
},
tick align=outside,
tick pos=left,
x grid style={white!69.0196078431373!black},
xtick style={color=black},
y grid style={white!69.0196078431373!black},
ytick style={color=black},
xlabel={$\rho$},
ylabel={$\pi_0$},
cycle list name=exotic,
]
\nextgroupplot[]
\addplot+ [thick, mark repeat=12, mark size=2, mark=x] table [x index=0, y index=1] {figures/poly.csv};
\addlegendentry{Expected}
\addplot+ [thick, mark repeat=12, mark size=2, mark=o] table [x index=0, y index=2] {figures/poly.csv};
\addlegendentry{$\pi_0(\rho)$}
\end{groupplot}
\end{tikzpicture}

 \caption{Comparison between the expected and the polynomial approximation of the distribution $\pi_0$ for varying system loads $\rho$.}
    \label{fig:poly}
\end{figure}

We calculate $\pi_0$ using the former method for different $\rho$ and $N=500$. We then fit $f(x) = a x^4 + b x^3 + c x^2 + dx + e$
accordingly. The result of the fit procedure produces the polynomial function:
$$
\pi_0(\rho) = -0.24 \rho^4-0.21 \rho^3-0.55 \rho^2+0.01 \rho^1+1
$$
The proposed polynomial function converges to the expected values of $\pi_0$, as
depicted in \autoref{fig:poly}.

\clearpage
\glsaddall
\section{Glossary}\label{sec:glossary}
\printglossary[type=\acronymtype,title={}]

\end{document}